\documentclass[12pt]{article}
\usepackage{amsmath}
\usepackage{graphicx}
\usepackage{enumerate}
\usepackage{natbib}
\usepackage{url} 

\newcommand{\blind}{1}

\usepackage{color}
\usepackage{geometry}
\usepackage{amsbsy, amsmath,amsfonts}
\usepackage{graphicx,url}
\usepackage{blindtext}
\usepackage{adjustbox}
\usepackage{float}
\restylefloat{table}

\def\E{\mathop{\rm E\,\!}\nolimits}
\def\Var{\mathop{\rm Var}\nolimits}
\def\Cov{\mathop{\rm Cov}\nolimits}

\def\vec{\mathop{\rm vec}\nolimits}

\def\tr{\mathop{\rm tr}\nolimits}

\def\amp{\mathop{\;\:}\nolimits}
\def\argmax{\mathop{\rm argmax}\nolimits}

\newcommand{\bb}{\boldsymbol{b}}
\newcommand{\bc}{\boldsymbol{c}}

\newcommand{\be}{\boldsymbol{e}}

\newcommand{\bg}{\boldsymbol{g}}

\newcommand{\br}{\boldsymbol{r}}

\newcommand{\bv}{\boldsymbol{v}}

\newcommand{\bx}{\boldsymbol{x}}
\newcommand{\by}{\boldsymbol{y}}
\newcommand{\bz}{\boldsymbol{z}}

\newcommand{\bB}{\boldsymbol{B}}

\newcommand{\bD}{\boldsymbol{D}}

\newcommand{\bG}{\boldsymbol{G}}

\newcommand{\bJ}{\boldsymbol{J}}

\newcommand{\bL}{\boldsymbol{L}}
\newcommand{\bM}{\boldsymbol{M}}

\newcommand{\bO}{\boldsymbol{O}}

\newcommand{\bT}{\boldsymbol{T}}

\newcommand{\bV}{\boldsymbol{V}}
\newcommand{\bW}{\boldsymbol{W}}
\newcommand{\bX}{\boldsymbol{X}}
\newcommand{\bY}{\boldsymbol{Y}}

\newcommand{\balpha}{\boldsymbol{\alpha}}
\newcommand{\bbeta}{\boldsymbol{\beta}}
\newcommand{\bgamma}{\boldsymbol{\gamma}}

\newcommand{\bmu}{\boldsymbol{\mu}}
\newcommand{\bnu}{\boldsymbol{\nu}}

\newcommand{\btheta}{\boldsymbol{\theta}}

\newcommand{\bGamma}{\boldsymbol{\Gamma}}

\newcommand{\bSigma}{\boldsymbol{\Sigma}}

\newcommand{\bOmega}{\boldsymbol{\Omega}}

\usepackage{changes}

\usepackage{amsmath}
\usepackage[linesnumbered,ruled]{algorithm2e}

\usepackage{hyperref}
\hypersetup{
    colorlinks=true,
    linkcolor=blue,
    filecolor=magenta,      
    urlcolor=blue,
}

\begin{document}

\def\spacingset#1{\renewcommand{\baselinestretch}%
{#1}\small\normalsize} \spacingset{1}


\if1\blind
{
  \title{\bf A Flexible Quasi-Copula Distribution for Statistical Modeling}
  \author{
    \small\thanks{Address all correspondence to smji@g.ucla.edu.}
    Sarah S. Ji$^1$\thanks{
    The authors gratefully acknowledge USPHS grants  GM141798, AI153044, HG002536 and HG006139 and NSF grants DMS-2054253 and IIS-2205441.}\hspace{.2cm}, Benjamin B. Chu$^2$, 
    \small Hua Zhou$^{1,4}$, Kenneth Lange$^{3,4,5}$\\
    \small$^1$Department of Biostatistics, University of California, Los Angeles, CA\\
    \small$^2$Department of Biomedical Data Science, Stanford University, Stanford, CA\\
    \small$^3$Department of Human Genetics, University of California, Los Angeles, CA\\
    \small$^4$Department of Computational Medicine, University of California, Los Angeles, CA\\
    \small$^5$Department of Statistics, University of California, Los Angeles, CA
    }
  \maketitle
} \fi

\if0\blind
{
  \bigskip
  \bigskip
  \bigskip
  \begin{center}
    {\LARGE\bf A Flexible Quasi-Copula Distribution for Statistical Modeling}
\end{center}
  \medskip
} \fi

\bigskip
\begin{abstract}
Copulas, generalized estimating equations, and generalized linear mixed models promote the analysis of grouped data where non-normal responses are correlated. Unfortunately, parameter estimation remains challenging in these three frameworks. Based on prior work of Tonda, we derive a new class of probability density functions that allow explicit calculation of moments, marginal and conditional distributions, and the score and observed information needed in maximum likelihood estimation. We also illustrate how the new distribution flexibly models longitudinal data following a non-Gaussian distribution. Finally, we conduct a tri-variate genome-wide association analysis on dichotomized systolic and diastolic blood pressure and body mass index data from the UK-Biobank, showcasing the modeling prowess and computational scalability of the new distribution.
\end{abstract}

\noindent%
\vfill

\newpage
\spacingset{1.0} 


\newpage

\section{Introduction}

The analysis of correlated data is stymied by the lack of flexible multivariate distributions with fixed margins. Once one ventures beyond the confines of multivariate Gaussian distributions, analysis choices are limited. \citet{liang1986longitudinal} launched the highly influential method of generalized estimating equations (GEEs). This advance allows generalized linear models (GLMs) to accommodate the correlated traits encountered in panel and longitudinal data and effectively broke the stranglehold of Gaussian distributions in analysis. The competing method of statistical copulas introduced earlier by Sklar is motivated by the same consideration \citep{sklar1959fonctions}. Finally, generalized linear mixed models (GLMMs) \citep{breslow1993approximate, zeger1991generalized} attacked the same problem. GLMMs are effective tools for modeling overdispersion and capturing the correlations of multivariate discrete data.

However, none of these three modeling approaches is a panacea. GEEs lack a well-defined likelihood, and estimation searches can fail to converge. For copula models, likelihoods exist, but are unwieldy, particularly for discrete outcomes. Copula calculations scale extremely poorly in high dimensions. Computing with GLMMs is problematic since their densities have no closed form and require evaluation of multidimensional integrals. Gaussian quadratures scale exponentially in the dimension of the parameter space. Markov Chain Monte Carlo (MCMC) can be harnessed in Bayesian versions of GLMMs, but even MCMC can be costly.  For these reasons alone, it is worth pursuing alternative modeling approaches.

This brings us to an obscure paper by the Japanese mathematical statistician Tonda. Working within the framework of Gaussian copulas \citep{song2009joint} and generalized linear models, Tonda introduces a device for relaxing independence assumptions while preserving computable likelihoods \citep{tonda2005class}. He succeeds brilliantly except for the presence of an annoying constraint on the parameter space of the new distribution class. The fact that his construction perturbs marginal distributions is forgivable. The current paper has several purposes. First, by adopting a slightly different working definition, we show how to extend his construction to lift the awkward parameter constraint. Our new definition allows explicit calculation of (a) moments, (b) marginal and conditional distributions, and (c) the score and observed information of the loglikelihood and allows (d) generation of random deviates. Tonda tackles item (a), omits items (b) and (c), and mentions item (d) only in passing. For maximum likelihood estimation (MLE), he relies on a non-standard derivative-free algorithm \citep{ohtaki1999globally} that scales poorly in high dimensions. We present two gradient-based algorithms designed for high-dimensional MLEs. The first is a block ascent algorithm that updates fixed effects by Newton's method and updates variance components by a minorization-maximization (MM) algorithm. The second is a standard quasi-Newton algorithm that updates fixed effects and variance components jointly. 

In contrast to other multivariate outcome models, our loglikelihoods contain no determinants or matrix inverses. These features resolve computational bottlenecks in  parameter estimation. We advocate gradient based estimation methods that avoid computationally intensive second derivatives. Approximate Hessians can be  computed after estimation to provide asymptotic standard errors and confidence intervals. The range of potential applications of our quasi-copula model is enormous. Panel, longitudinal, time series, and all of GLM modeling stand to benefit. In addition to relaxing independence assumptions, our models offer a simple way to capture over-dispersion. Our simulation studies and real data examples highlight not only the virtues of the quasi-copula model but also its limitations. For reasons to be explained, we find that the model reflects reality best when the size of the independent sampling units is low or the correlations between responses within a unit are small.

In subsequent sections, we begin by introducing the quasi-copula model and studying its statistical properties. Then we illustrate how the model is used in practice to 
analyze correlated non-Gaussian variables. Getting the correlation structure of the variables right is a key step in modeling both longitudinal and multi-trait data. Next, we discuss the details of parameter estimation that enable model fitting. Finally, we present analysis results on both simulated and real data. These results showcase the speed and flexibility of the quasi-copula model. In particular, our genome-wide association (GWAS) example demonstrates the scalability of the quasi-copula model. 
The difficulty of multivariate GWAS with mixed outcomes served as the original motivation for this paper.

\section{The Quasi-Copula Model}

\subsection{Notation}

For the record, here are some notational conventions used in the sequel. All vectors and matrices appear in boldface. The entries of the vector $\bf 0$ consist of 0's, and the standard basis vector $\be_i$ has all entries 0 except a 1 in entry $i$. The $^\top$ superscript indicates a vector or matrix transpose. The Euclidean norm of a vector $\bx$ is denoted by $\|\bx\|$, and the spectral norm of a matrix $\bM = (m_{ij})$ is $\|\bM\|  =  \sup_{\bx \ne {\bf 0}}\frac{\|\bM\bx\|}{\|\bx\|}$. For a smooth real-valued function $f(\bx)$, we write its gradient (column vector of partial derivatives) as $\nabla f(\bx)$, its first differential (row vector of partial derivatives) as $df(\bx)=\nabla f(\bx)^\top$, and its second differential (Hessian matrix) as $d^2f(\bx) = d \,\nabla f(\bx)$. If $g(\bx)$ is vector-valued with $i$th component $g_i(\bx)$, then the differential (Jacobi matrix) $dg(\bx)$ has $i$th row $dg_i(\bx)$. The transpose $dg(\bx)^\top$ is the gradient of $g(\bx)$. 
Differentials $dg(\bx)$ can be constructed from directional derivatives $d_{\bv}g(\bx)=\lim_{t \to  0} \frac{g(\bx+t\bv)-g(\bx)}{t}$.

\subsection{Definitions}

Consider $d$ independent random variables $X_1,\ldots,X_d$ with densities $f_i(x_i)$ relative to measures $\alpha_i$, with means $\mu_i$, variances $\sigma_i^2$, third central moments $c_{i3}$, and fourth central moments
$c_{i4}$.  Let $\bGamma =(\gamma_{ij})$ be an $d \times d$ positive semidefinite matrix, and $\alpha$ be the product measure $\alpha_1 \times \cdots \times \alpha_d$. Inspired by \cite{tonda2005class}, we let 
$\bD$ be the diagonal matrix with $i$th diagonal entry $\sigma_i$ and consider the nonnegative function
\begin{eqnarray*}
1 + \frac{1}{2}(\bx-\bmu)^\top \bD^{-1}\bGamma\bD^{-1}(\bx-\bmu).
\end{eqnarray*}
Its average value is
\begin{eqnarray*}
&  & \int \prod_{i=1}^d f_i(x_i)\Big[1 + \frac{1}{2}(\bx-\bmu)^\top \bD^{-1} \bGamma \bD^{-1}(\bx-\bmu)\Big] 
d\alpha(\bx)\\
& = & 1 + \frac{1}{2}\sum_i \sum_j \E\Big[\frac{(x_i-\mu_i)(x_j-\mu_j)}{\sigma_i\sigma_j}
\Big]\gamma_{ij} \\
& = & 1 + \frac{1}{2}\sum_i \gamma_{ii}.
\end{eqnarray*}
It follows that the function 
\begin{eqnarray}
\!\! g(\bx) & \!\!\!\! = \!\!\!\! &  \Big[1 + \frac{1}{2}\tr(\bGamma)\Big]^{-1}
\prod_{i=1}^d f_i(x_i)\Big[1 + \frac{1}{2}(\bx-\bmu)^\top \bD^{-1}\bGamma \bD^{-1}
(\bx-\bmu)\Big] \label{copula_density}
\end{eqnarray}
is a probability density with respect to the measure $\alpha$. Detailed derivations of Tonda's approximation are found in supplemental Section \ref{sec:tonda}. The virtue of the density is that it overcomes the independence restriction 
and steers the sample matrix of the residuals toward the target
covariance matrix $\bGamma$. For example, if $X_1$ is Gaussian and $X_2$ is Bernoulli, then the distribution \eqref{copula_density}  allows one to induce correlation between the two components (one continuous and the other discrete) of the binary random vector $X = (X_1, X_2)^\top$. Note that $g(\bx)$ is technically not a copula because it only approximately preserves the marginal distributions $f_i(x_i)$. Later, we will see that $g(\bx)$ tends to inflate marginal variances. This can be a blessing rather than curse if the true marginal distributions are inflated compared to the assumed marginal distributions.

\subsection{Moments}

Let $\bY=(Y_1,\ldots,Y_d)^\top$ be a random vector distributed as $g(\bx)$.  To calculate the mean of $Y_k$, note that our independence assumption implies
\begin{eqnarray*}
&  & \int (x_k-\mu_k) g(\bx) \alpha(\bx) \\
& = &  \Big[1 + \frac{1}{2}\tr(\bGamma)\Big]^{-1}
\frac{1}{2}\sum_i \sum_j \E\Big[(x_k-\mu_k)\frac{(x_i-\mu_i)(x_j-\mu_j)}{\sigma_i \sigma_j}\Big]\gamma_{ij} \\
& = & \Big[1 + \frac{1}{2}\tr(\bGamma)\Big]^{-1}\frac{c_{k3}\gamma_{kk}}{2\sigma_k^2}.
\end{eqnarray*}
Hence,  if $\kappa_{k3}$ is the skewness of $X_k$, then
\begin{eqnarray*}
\E(Y_k) & = &\mu_k +\Big[1 + \frac{1}{2}\tr(\bGamma)\Big]^{-1}
\frac{c_{k3}\gamma_{kk}}{2\sigma_k^2} \\
& = & \mu_k +\Big[1 + \frac{1}{2}\tr(\bGamma)\Big]^{-1}\frac{\sigma_k \kappa_{k3}\gamma_{kk}}{2}\\
& = & \mu_k +\frac{\sigma_k \kappa_{k3}\gamma_{kk}}{2} +O(\|\bGamma\|^2)
\end{eqnarray*}
for any matrix norm $\|\bGamma\|$. The mean $\E(Y_k)$ is close to $\mu_k$ when the diagonal entries of $\bGamma$ and, hence $\|\bGamma\|$ itself, are small. 

To calculate the covariance matrix of $\bY$, note that
\begin{eqnarray*}
&  & \int (x_k-\mu_k)(x_l-\mu_l) g(\bx) d \alpha(\bx) \\
& = & \Big[1+ \frac{1}{2}\tr(\bGamma)\Big]^{-1}1_{\{k=l\}}\sigma_k^2 
 +\Big[1+ \frac{1}{2}\tr(\bGamma)\Big]^{-1} \\
&  & \times \frac{1}{2}\sum_i \sum_j  
\E\Big[(x_k-\mu_k)(x_l-\mu_l)\frac{(x_i-\mu_i)(x_j-\mu_j)}{\sigma_i\sigma_j}\Big]\gamma_{ij}.
\end{eqnarray*}
The indicated expectations relative to $\prod_{i=1}^d f_i(x_i)$ reduce to
\begin{eqnarray*}
&  & \E[(x_k-\mu_k)(x_l-\mu_l)(x_i-\mu_i)(x_j-\mu_j)]\\
& = & \begin{cases} c_{k4} & k=l=i=j \\
\sigma_{k}^2 \sigma_{i}^2 & k=l \ne i = j \\
 \sigma_{k}^2\sigma_{l}^2 & k = i \ne l = j \\
\sigma_{k}^2\sigma_{l}^2 & k = j \ne l = i \\
0 & \text{otherwise .} \end{cases}
\end{eqnarray*}
When $k=l$ and $\kappa_{k4}$ is the kurtosis of $X_k$,
\begin{eqnarray*}
&  & \int (x_k-\mu_k)^2 g(\bx) d\alpha(\bx) \\
& = &  \Big[1+ \frac{1}{2}\tr(\bGamma)\Big]^{-1} \Big[\sigma_k^2 
+ \frac{1}{2}\frac{c_{k4} \gamma_{kk}}{\sigma_k^2}+\frac{1}{2}\sigma_k^2 \sum_{i \ne k}  \gamma_{ii}\Big] \\
& = & 
\Big[1+ \frac{1}{2}\tr(\bGamma)\Big]^{-1} \sigma_k^2 \Big[1
+ \frac{\kappa_{k4} \gamma_{kk}}{2}+\frac{1}{2} \sum_{i \ne k}  \gamma_{ii}\Big] \\
& = & \sigma_k^2
+ \frac{\sigma_k^2 \kappa_{k4} \gamma_{kk}}{2}+\frac{\sigma_k^2}{2} \sum_{i \ne k}  \gamma_{ii} + O(\|\bGamma\|^2)\\
& = & \sigma_k^2 \Big[1+\frac{(\kappa_{k4}-1)\gamma_{kk}}{2} + \frac{1}{2} \sum_i \gamma_{ii} \Big]+O(\|\bGamma\|^2)\\
& = & \sigma_k^2 \Big[1+\frac{(\kappa_{k4}-1)\gamma_{kk}}{2}\Big]+O(\|\bGamma\|^2). 
\end{eqnarray*}
Because $[\E(Y_k-\mu_k)]^2 = O(\|\bGamma\|^2)$, we find that
\begin{eqnarray*}
\Var(Y_k) & = & \E[(Y_k-\mu_k)^2] - [\E(Y_k-\mu_k)]^2\\
& = &  \sigma_k^2\Big[1+\frac{(\kappa_{k4}-1)\gamma_{kk}}{2}\Big]+O(\|\bGamma\|^2).
\end{eqnarray*}
Because the kurtosis $\kappa_{k4} \ge 1$, the multiplier $\kappa_{k4}-1$ of $\gamma_{kk}$ is nonnegative, and the variance is inflated for $\|\bGamma\|$ small. 

When $k\ne l$, 
\begin{eqnarray*}
\int (x_k-\mu_k)(x_l-\mu_l) g(\bx) d\alpha(\bx)
& = &  \Big[1+ \frac{1}{2}\tr(\bGamma)\Big]^{-1}
\frac{1}{2}2\sigma_k\sigma_l\gamma_{kl} \\
& = & \sigma_k\sigma_l\gamma_{kl}+O(\|\bGamma\|^2).
\end{eqnarray*}
Hence, the covariance and correlation satisfy
\begin{eqnarray*}
\Cov(Y_k,Y_l) & = & \Cov(Y_k-\mu_k,Y_l-\mu_l) \\
& = & \E[(Y_k-\mu_k)(Y_l-\mu_l)] - \E(Y_k-\mu_k)\E(Y_l-\mu_l) \\
& = & \sigma_k\sigma_l\gamma_{kl}+O(\|\bGamma\|^2)\\
\text{\rm Corr}(Y_k,Y_l) & = & \frac{\gamma_{kl}}
{\sqrt{1+\frac{(\kappa_{k4}-1)\gamma_{kk}}{2}+O(\|\bGamma\|^2)}
\sqrt{1+\frac{(\kappa_{l4}-1)\gamma_{ll}}{2}+O(\|\bGamma\|^2)}}.
\end{eqnarray*}
As a check, the quantities $\E(Y_k)$, $\Var(Y_k)$, and $\Cov(Y_k,Y_l)$ reduce to the correct values $\mu_k$, $\sigma_k^2$, and 0, respectively, when $\bGamma = {\bf 0}$. 


\subsection{Marginal and Conditional Distributions}

Let $S$ be a subset of $\{1,\ldots,d\}$ with complement $T$. To
simplify notation, suppose $S = \{1,2,\ldots, s\}$. Now write
\begin{eqnarray*}
\bY & = & \begin{pmatrix} \bY_{\!S} \\\bY_{\!T} \end{pmatrix}, \:\:
\br \amp = \amp \begin{pmatrix} \br_S \\\br_T \end{pmatrix}, \:\:
\bGamma \amp = \amp \begin{pmatrix} \bGamma_S & \bGamma_{ST}
\\ \bGamma_{ST}^\top & \bGamma_T \end{pmatrix}, \:\:
\alpha \amp = \amp \alpha_S \times \alpha_T,
\end{eqnarray*}
where $\br$ is the vector $\bD^{-1}(\bY-\bmu)$ of standardized residuals. The marginal density of $\bY_{\!S}$ is
\begin{eqnarray*}
&  & \Big[1+ \frac{1}{2}\tr(\bGamma)\Big]^{-1}\prod_{i \in S} f_i(y_i)
\int \prod_{i \in T} f_i(y_i)\Big[1+\frac{1}{2}\br^\top \bGamma \br\Big]\, d \alpha_T(\by_T)\\
& = & \Big[1+ \frac{1}{2}\tr(\bGamma)\Big]^{-1}\prod_{i \in S} f_i(y_i)
\Big[1+\frac{1}{2}\br_S^\top \bGamma_S \br_S + \frac{1}{2}\tr(\bGamma_T)\Big]. 
\end{eqnarray*}
To derive the conditional density of $\bY_{\!S}$ given by $\bY_{\!T}$, we divide the joint density by the marginal density of $\bY_{\!T}$. This action produces the conditional density
\begin{eqnarray*}
d_S\prod_{i \in S} f_i(y_i)\Big[1+\frac{1}{2}\br^\top \bGamma \br\Big] 
\end{eqnarray*}
with normalizing constant $d_S=\Big[1+\frac{1}{2}\br_T^\top \bGamma_T \br_T + \frac{1}{2}\tr(\bGamma_S)\Big]^{-1}$. From this density, our well-rehearsed arguments lead to the conditional mean
\begin{eqnarray*}
\E(Y_k \mid \bY_{\!T}) = \mu_k+d_S \Big[\frac{c_{k3}\gamma_{kk}}{2\sigma_k^2} + \frac{1}{\sigma_k}\sum_{j\in T}r_j\gamma_{jk}\Big]
\amp = \amp \mu_k+\frac{c_{k3}\gamma_{kk}}{2\sigma_k^2} +O(\|\bGamma\|^2)
\end{eqnarray*}
for $k \in S$. The corresponding conditional variance is
\begin{eqnarray*}
\Var(Y_k \mid \bY_{\!T}) & = & \sigma_k^2+\frac{1}{2}\Big(\frac{c_{k4}}{\sigma_k^2}-\sigma_k^2\Big)\gamma_{kk}+ \sum_{j \in T}\frac{c_{k3}r_j\gamma_{kj}}{\sigma_k}+O(\|\bGamma\|^2).
\end{eqnarray*}
and the corresponding conditional covariances are 
\begin{eqnarray*}
\Cov(Y_k, Y_l \mid \bY_{\!T}) & = & \sigma_k \sigma_l \gamma_{kl}+O(\|\bGamma\||^2)
\end{eqnarray*}
for $k \in S$, $l \in S$, and $k \ne l$. It is noteworthy that to 
order $O(\|\bGamma\|^2)$, the conditional and marginal means agree,
and the conditional and marginal covariances agree.

\subsection{Generation of Random Deviates}

To generate a random vector from the density (\ref{copula_density}), we first sample $Y_1$ from its marginal density
\begin{eqnarray*}
\Big[1+ \frac{1}{2}\tr(\bGamma)\Big]^{-1} f_1(y_1)
\Big(1 + \frac{\gamma_{11}}{2} r_1^2 + \frac{1}{2} \sum_{j=2}^d \gamma_{jj}\Big),
\end{eqnarray*}
and then sample the subsequent components $Y_i$ from their conditional distributions $Y_i \mid Y_1, \ldots, Y_{i-1}, \forall i \in [1, d]$. If we denote the set $\{1,\ldots,i-1\}$ by $[i-1]$, then the conditional density of 
$Y_i$ given the previous components is
\begin{eqnarray*}
d_{[i-1]}^{-1}f_i(y_i)
\Big[d_{[i-1]} +  r_i \sum_{j=1}^{i-1} r_j \gamma_{ij}  + \frac{\gamma_{ii}}{2} (r_i^2-1) \Big],
\end{eqnarray*}
where $d_{[i-1]}=1 +  \frac{1}{2} r_{[i-1]}^\top \bGamma_{[i-1]} 
r_{[i-1]}+ \frac{1}{2} \sum_{j=i}^d \gamma_{jj}$. 

When the densities $f_i(y_i)$ are discrete, each stage of sampling is straightforward. Consider any random variable $Z$ with nonnegative integer values, discrete density $p_i=\Pr(Z=i)$, and mean $\nu$. The inverse method of random sampling reduces to a sequence of comparisons. We partition the interval $[0,1]$ into subintervals with the $i$th subinterval of length $p_i$. To sample $Z$, we draw a uniform random deviate $U$ from $[0,1]$ and return the deviate $j$ determined by the conditions $\sum_{i=1}^{j-1} p_i \le U < \sum_{i=1}^j p_i$. 
The process is most efficient when the largest $p_i$ occur first. This suggests that we let $k$ denote the least integer $\lfloor \nu \rfloor$ and rearrange the probabilities in the order 
$p_k, p_{k+1},p_{k-1},p_{k+2},p_{k-2},\ldots$ This
tactic is apt put most of the probability mass first and render sampling efficient. 

When the densities $f_i(y_i)$ are continuous, each stage of sampling is probably best performed by inverse transform sampling. This requires
calculating distribution functions and forming their inverses, either
analytically or by Newton's method. The required distribution functions
assume the form
\begin{eqnarray*}
\int_{-\infty}^x \! f(y)[a_0+a_1(y-\mu)+a_2(y-\mu)^2]\,dy & \!\!\!= \!\!\! &
\int_{-\infty}^x \! f(y)[b_0+b_1 y+b_2y^2]\,dy.
\end{eqnarray*}
The integrals $\int_{-\infty}^x f(y)y^j\,dy$ are available as
special functions for Gaussian, beta, and gamma densities $f(y)$.
For instance, if $\phi(y)=\frac{1}{2\pi}e^{-y^2/2}$ is the
standard normal density and $\Phi(x)$ is the standard normal
distribution, then
\begin{eqnarray*}
\int_{-\infty}^x  y \phi(y) \, dy  & = & -\phi(x) \;\; \text{and} \;\;
\int_{-\infty}^x y^2 \phi(y) \, dy  =  \Phi(x) - x\phi(x).
\end{eqnarray*}
To avoid overburdening the text with classical mathematics, we omit further details. Additional derivations can be found in supplemental section \ref{sec:supp_gen_random_deviates}.

\section{Model Choices for Specific Applications}

We now work through the details of how the quasi-copula model is structured in practice. Our examples of longitudinal data analysis and multivariate genome-wide association are typical of the challenges posed by non-continuous data and mixtures of continuous, binary, and count data. 

\subsection{Model for Longitudinal Data}\label{sec:longitudinal_model}

The longitudinal setting involves $n$ independent subjects. The response vector $\by_i$ for subject $i$ consists of $d_i$ traits values and $p$ covariates (features) $\bX_i \in \mathbb{R}^{d_i \times p}$ per trait. The component $y_{ij}$ represents the measured trait of subject $i$ at time $j$. The data matrix $\bX_i$ may include both time-varying features (for example medication use) and time-invariant features (for example gender). Linear mixed models \citep{verbeke1997linear,fitzmaurice2012applied} are a sensible modeling choice when the trait values in $\by_i$ are continuous. When measurements are discrete, the quasi-copula model often runs orders of magnitude faster than generalized linear mixed models and leads to more pluasible loglikelihoods. Sections \ref{sec:simulation_studies} and \ref{sec:nhanes_example} cover both simulated and real longitudinal data. 

In the quasi-copula model of longitudinal data, the random vector $Y = (Y_1,...,Y_{d_i})$ follows the quasi-copula density appearing in equation \eqref{copula_density}. Marginal means are linked to covariates through
\begin{eqnarray}\label{eq:longitudinal_mean}
    \bmu_i(\bbeta) &=& g^{[-1]}(\bX_i\bbeta), \quad \bbeta \in \mathbb{R}^p,
\end{eqnarray}
where the inverse link function $g^{[-1]}(z)$ is applied component-wise. Although
it is not strictly necessary, we assume that each of the $d_i$ measurements follows the same base distribution. This simplifying assumption holds, for example, with repeated blood-pressure measurements. 

Because each subject $i$ can be measured at a different number $d_i$ of time points, the choice of the covariance matrices $\bGamma_i$ is constrained. We explore three structured options:
\begin{eqnarray*}
    \bGamma_i &= & \sum_{j=1}^m \theta_j \bOmega_{ij} \label{VC}\\ 
    \bGamma_i &= & \sigma^2 \times 
    \begin{bmatrix} 
        1 & \rho & \rho^2 & \rho^3 & ...  &\rho^{d_i - 1}\\ \rho & 1 & \rho & \rho^2 & ... \\ & & ... & & \\ & &...& \rho & 1 & \rho \\ \rho^{d_i - 1} & \rho^{d_i - 2} & ...& \rho^2 & \rho & 1
    \end{bmatrix}\label{AR1}\\ 
    \bGamma_i &= & \sigma^2 \times \Big[ \rho {\bf 1}_{d_{i}}{\bf 1}_{d_{i}}^\top + (1 - \rho) \boldsymbol{I}_{d_i} \Big] .\label{CS}
\end{eqnarray*}
These traditional choices define the variance component (VC) model, the auto-regressive (AR1) model, and the compound symmetric (CS) model. The $\theta_i \ge 0$ must be estimated in the VC model, while $\sigma^2 \ge0$ and $\rho$ must be estimated in the AR1 and CS models. Section \ref{sec:structured_covariance_estimation} describes how estimation is performed. The covariance parameters supplement the regression coefficients $\bbeta$ of the base distributions. 

\subsection{Model for Unstructured Multivariate Data}\label{sec:multivariate_model}

This multivariate model involves  $n$ independent samples exhibiting $d$ responses and $p$ covariates. Each component of the quasi-copula response $Y = (Y_1,...,Y_d)$ is allowed to have a different base distribution $f_j(y)$ and a different inverse link $g_j^{[-1]}(z)$. In contrast to the longitudinal model, we postulate a matrix $\bB=
(\bbeta_1 \, \ldots \, \bbeta_d)$ of regression coefficients that capture the unique impacts of the $p$ covariates on the $d$ responses. Means are linked to covariates via $\bmu_i = g^{[-1]}(\bB^\top\bx_i)$. If we define
\begin{eqnarray*}
    \vec(\bB) & = & \begin{bmatrix}
        \bbeta_1\\ \vdots \\ \bbeta_d
    \end{bmatrix}_{pd \times 1}\text{ and } 
    \bX_i \amp = \amp
    \begin{bmatrix}
    \bx_i^\top & & 0 \\
    & \ddots & \\
    0 & & \bx_i^\top
    \end{bmatrix}_{d \times pd}, 
\end{eqnarray*}
then we can apply the previous notation  
\begin{eqnarray}\label{eq:multivariate_mean}
    \bmu_i = g^{[-1]}[\bX_i \vec(\bB)].
\end{eqnarray}
linking  means to covariates. This notational change allows us to estimate a parameter vector $\bbeta$, with the understanding that $\bbeta$ has length $p$ for longitudinal data and length $pd$ for multivariate data.  

Because the covariance matrices $\bGamma_i$ are constant, one can estimate an single unstructured covariance matrix
\begin{eqnarray*}
    \bGamma &=& \bL\bL^\top \amp\in\amp \mathbb{R}^{d \times d},
\end{eqnarray*}
where $\bL$ is the lower-triangular Cholesky decomposition of $\bGamma$. In the interests of parsimony, one could replace $\bL$ by a low-rank matrix, but we will
not pursue this suggestion further. In summary, the multivariate quasi-copula model is parametrized by $\vec(\bB) \in \mathbb{R}^{dp}$  mean-effect parameters in addition to the $d(d+1)/2$ nonzero parameters of the Cholesky decomposition $\bL=(\ell_{ij})$. The requirement $\ell_{ii}\ge 0$ for all $i$ is the only constraint on the parameter space.

\subsection{Genome-Wide Association Studies}

The ability to analyze multiple responses simultaneously are particularly useful in genome-wide association studies (GWAS). Here $n$ subjects are measured on $q$ non-genetic covariates (for example sex, lifestyle, and ancestry) and $p$ single-nucleotide polymorphisms (SNPs), where $p \sim 10^6$. Together the covariates influence a small set of $d$ traits (phenotypes). Although there is no restriction on $d$, the phenotypes should be meaningfully correlated. In practice, $d\le 10$ usually holds. The goal is to select a small subset of SNPs that influence the $d$ traits holistically \citep{galesloot2014comparison}. Most methods for multivariate GWAS are based on Gaussian approximations\citep{chu2023multivariate,zhou2014efficient}. The normality assumption excludes dichotomous and count data that are ubiquitous in genetic studies. The quasi-copula framework allows one to conduct GWAS on multiple correlated phenotypes that may be continuous, binary, or discrete. 

In GWAS each SNP is examined separately. If $\balpha \in \mathbb{R}^d$ represents the mean effect of a SNP on each of $d$ phenotypes, then the null hypothesis
\begin{align}\label{eq:qc_null_hypothesis}
    H_0: \alpha_1 = ... = \alpha_d = 0
\end{align}
is pertinent. To test the hypothesis \eqref{eq:qc_null_hypothesis}, a likelihood ratio tests (LRT) is certainly possible. This involves maximizing the loglikelihood  under the null $H_0$ and comparing it to the likelihood under the alternative $H_a$ where $\balpha$ can vary. If $\mathcal{L}_0$ and $\mathcal{L}_a$ denote the respective maximum loglikelihoods, then the likelihood ratio statistic follows the distribution
\begin{eqnarray}
2(\mathcal{L}_a - \mathcal{L}_0) \sim \chi^2_{d}.
\end{eqnarray} 
SNPs can be selected by applying the Bonferroni's correction to the resulting p-values. For GWAS with human subjects, the stringent cutoff p-value $5 \times 10^{-8}$ is usually applied.

Although this strategy is conceptually simple, it requires fitting $p \sim 10^6$ alternative models. The sizes of modern  datasets generally prohibit this brute-force approach.  To alleviate the computational burden, we calculate the LRT on only the most promising SNPs. One can screen for the top SNPs by examining the gradient $\nabla_{\balpha} \mathcal{L} \in \mathbb{R}^d$ under the maximum likelihood estimates of the null model estimates $(\hat{\bbeta}, \hat{\bL})$. The $\ell_1$ norm $\|\nabla_{\balpha}\mathcal{L}\|_1$ quantifies the signal strength of the SNP under consideration. We will derive the gradient $\nabla_{\balpha}\mathcal{L}$ later.         Computing the gradient under the null is much faster than fitting a full likelihood model under the alternative. We show in supplemental Section \ref{sec:multivariate_gwas_extra_sims1} that ordering SNPs by $\|\nabla_{\balpha}\mathcal{L}\|_1$ is strongly correlated with ordering them by $-\log_{10}(\text{p-value})$. Algorithm \eqref{alg:adhoc_LRT} summarizes our fast multivariate GWAS procedure. 

\begin{algorithm}[H] \SetCustomAlgoRuledWidth{0.45\textwidth}  
\caption{Fast likelihood-ratio tests for multivariate GWAS}\label{alg:adhoc_LRT}
\KwData{phenotypes $\bY \in \mathbb{R}^{n \times d}$, covariates $\bX \in \mathbb{R}^{n \times q}$, genotypes $\bG \in \mathbb{R}^{n \times p}$}
$\mathcal{L}_0 \gets$ fit($\bY, \bX$) \quad\quad\quad \CommentSty{\#\# maximum loglikelihood under $H_0$}\\
\For{$j = 1, 2, ..., p$}{
    $r_j \gets \|\nabla_{\balpha}\mathcal{L}\|_1$ \quad\quad \CommentSty{\#\# normed 
    gradient of SNP $j$ under $H_0$}
}
Sort from largest to smallest $(r_{[1]}, ..., r_{[p]}) \gets \text{sort}(r_1,...,r_p)$\\
\For{$j = 1,2,...,p$}{
    $\bg_j \gets$ $j$th SNP genotypes corresponding to $r_{[j]}$\\
    $\mathcal{L}_a \gets \text{fit}(\bY, \bX, \bg_j)$ \ \CommentSty{\#\# maximum 
    loglikelihood under $H_a$}\\
    LRT: $2(\mathcal{L}_a - \mathcal{L}_0) \sim \chi^2_d$\\
    \textbf{terminate:} if the p-value is larger than a preset level
}
\end{algorithm}

Supplemental Figure \ref{fig:QQ_plot} displays a QQ plot for Algorithm \ref{alg:adhoc_LRT}, showing that the resulting p-values are valid.

\section{Parameter Estimation}

Throughout this section, the vector $\bbeta$ denotes the mean effects tied to the base distributions through equations \eqref{eq:longitudinal_mean} and \eqref{eq:multivariate_mean}. The vector $\btheta$ summarizes the covariance parameters determining the covariance matrices $\bGamma_i$. For example, in the longitudinal AR1 model, $\btheta = (\sigma^2, \rho)$, and in the multivariate model, $\btheta = \text{vech}(\bL)$, where $\text{vech}(\bL)$ captures the lower triangle of the Cholesky decomposition $\bL$. 

\subsection{Mean Components}

Consider $n$ independent realizations $\by_i$ from the quasi-copula density (\ref{copula_density}). Each of these may be of a different
dimension $d_{i}$ and possess a different mean vector $\bmu_i(\bbeta)$, covariance matrice $\bGamma_i(\btheta)=[\gamma_{ijk}(\btheta)]$, and component densities $f_{ij}(y_{ij} \mid \bbeta)$. If $\br_i(\bbeta)$ denotes the vector $\bD_i^{-1}(\by_{i}-\bmu_{i})$ of standardized residuals for sampling unit $i$, then the loglikelihood of the sample is
\begin{eqnarray*}
\mathcal{L}(\bbeta,\btheta) & = &
 - \sum_{i=1}^n \ln \Big\{1\! +\! \frac{1}{2}\tr[\bGamma_{i}(\btheta)]\Big\}+
\sum_{i=1}^n \sum_{j=1}^{d_i} \ln f_{ij}(y_{ij} \mid \bbeta)\\ &  &  + \sum_{i=1}^n
\ln \Big[1\!+\!\frac{1}{2}\br_i(\bbeta)^\top \bGamma_i(\btheta) \br_i(\bbeta)\Big].
\end{eqnarray*}
The score (gradient of the loglikelihood) with respect to $\bbeta$ is clearly
\begin{eqnarray*}
\nabla_{\bbeta} \mathcal{L}(\bbeta,\btheta)
 & \!\!=\!\! & \sum_{i=1}^n \sum_{j =1}^{d_i} \nabla \ln f_{ij}(y_{ij} \mid \bbeta) + \sum_{i=1}^n
\frac{\nabla \br_i(\bbeta)^\top\bGamma_i(\btheta)\br_i(\bbeta)}{1+\frac{1}{2}\br_i(\bbeta)^\top \bGamma_i(\btheta) \br_i(\bbeta)},
\end{eqnarray*}
where $\nabla \br_i(\bbeta)^\top = d \br_i(\bbeta)$ is the differential (Jacobi matrix) of the vector $\br_i(\bbeta)$. 
An easy calculation shows that $\nabla \br_i(\bbeta)$ has entries
\begin{eqnarray*}
\nabla r_{ij}(\bbeta) & = & -\frac{1}{\sigma_{ij}(\bbeta)} \nabla \mu_{ij}(\bbeta)- \frac{1}{2} \frac{y_{ij}-\mu_{ij}(\bbeta)}{\sigma_{ij}^3(\bbeta)} \nabla \sigma_{ij}^2(\bbeta).
\end{eqnarray*}
The Hessian (second differential) of the loglikelihood with respect to $\bbeta$ is
\begin{eqnarray*}
    d_{\bbeta}^2 \mathcal{L}(\bbeta,\btheta) 
    & = & \sum_{i=1}^n \sum_{j=1}^{d_i} d^2 \ln f_{ij}(y_{ij} \mid \bbeta) - \sum_{i=1}^n\frac{[\nabla \br_i(\bbeta)\bGamma_i \br_i(\bbeta)]
    [\nabla \br_i(\bbeta)\bGamma_i \br_i(\bbeta)]^\top}
    {\Big[1+\frac{1}{2}\br_i(\bbeta)^\top \bGamma_i \br_i(\bbeta)\Big]^2} \\
    && + \sum_{i=1}^n \frac{\nabla \br_i(\bbeta)\bGamma_i d\br_i(\bbeta)}{1 + \frac{1}{2}\br_i(\bbeta)^\top\bGamma_i\br_i(\bbeta)} + \sum_{i=1}^n \sum_{j=1}^{d_i} \frac{\be_j^\top\bGamma_i\br_i(\bbeta)d^2r_{ij}(\bbeta)}{1 + \frac{1}{2}\br_i(\bbeta)\bGamma_i\br_i(\bbeta)},
\end{eqnarray*}
where 
\begin{eqnarray*}
    d^2 r_{ij}(\bbeta)
    &=& -\frac{1}{\sigma_{ij}(\bbeta)}d^2\mu_{ij}(\bbeta) + \frac{1}{2}\frac{1}{\sigma_{ij}^3(\bbeta)}\nabla \sigma_{ij}^2(\bbeta) d\mu_{ij}(\bbeta) -\frac{1}{2}\frac{y_{ij} - \mu_{ij}(\bbeta)}{\sigma_{ij}^3(\bbeta)}d^2\sigma_{ij}^2(\bbeta) \\
    && +\frac{1}{2}\left[\frac{1}{\sigma_{ij}^3(\bbeta)}\nabla\mu_{ij}(\bbeta) + \frac{3}{2}\frac{y_{ij}-\mu_{ij}(\bbeta)}{\sigma_{ij}^5(\bbeta)}\nabla\sigma_{ij}^2(\bbeta)\right]d\sigma_{ij}^2(\bbeta).
\end{eqnarray*}
Calculation of $\nabla r_{ij}$ and $d^2 r_{ij}$ requires computing $\nabla \mu_{ij}(\bbeta), d^2 \mu_{ij}(\bbeta), \nabla \sigma^2_{ij}(\bbeta)$, and $d^2 \sigma^2_{ij}(\bbeta)$ by repeated application of the chain rule. A full example appear in supplemental Section \ref{sec:nabla_r}.

\color{black}
In searching the likelihood surface, it is often beneficial to approximate the
observed information by a positive definite matrix.  This suggests replacing the observed information matrix $-d^2 \ln f_{ij}(y_{ij} \mid \bbeta)$ by the expected information matrix $\bJ_{ij}(\bbeta)$ under the base model and dropping indefinite matrices in the exact Hessian. These steps give the approximate Hessian
\begin{eqnarray*}
d_{\bbeta}^2\mathcal{L} & \approx &- \sum_{i=1}^n \sum_{j=1}^{d_i} \bJ_{ij}(\bbeta)
 -\sum_{i=1}^n\frac{[\nabla \br_i(\bbeta)\bGamma_i(\btheta) \br_i(\bbeta)]
[\nabla \br_i(\bbeta)\bGamma_i(\btheta) \br_i(\bbeta)]^\top}
{\Big[1+\frac{1}{2}\br_i(\bbeta)^\top \bGamma_i \br_i(\bbeta)\Big]^2},
\end{eqnarray*}
which is clearly negative semidefinite. These formulas provide the ingredients for implementing Newton's method or a quasi-Newton method for updating $\bbeta$. 

\subsection{Derivatives Pertinent to VC Models}\label{sec:structured_covariance_estimation}

Maximization of the loglikelihood depends on the derivatives of the covariance matrices $\bGamma_i(\btheta)$ through the functions $\tr(\bGamma_i)$ and $\br_i(\bbeta)^\top\bGamma_i \br_i(\bbeta)$. Here we present details of the longitudinal VC model and relegate others to supplement Section \ref{sec:grad_of_theta}. Recall that the longitudinal VC model involves the decomposition
\begin{eqnarray*}
    \bGamma_i (\btheta) &=& \sum_{j=1}^m \theta_{j} \bOmega_{ij}
\end{eqnarray*}
of $\bGamma_i$ into a linear combination of known covariance matrices $\bOmega_{ij}=(\omega_{ijkl})$ against unknown nonnegative variance components $\btheta=(\theta_1,...,\theta_m)$. Assuming there are no shared mean and covariance parameters, 
\begin{eqnarray*}
\tr(\bGamma_i) & = & \sum_{j=1}^m\theta_j\tr(\bOmega_{ij}) \amp = \amp 
\btheta^\top\bc_i \\
\br_i(\bbeta)^\top\bGamma_i\br_i(\bbeta) & = & \sum_{j=1}^m \theta_j \br_i(\bbeta)^\top\bOmega_{ij}\br_i(\bbeta) \amp = \amp \btheta^\top\bb_i ,
\end{eqnarray*}
in obvious notation. The part of the loglikelihood relevant to estimation of $\btheta$ can be expressed as
\begin{eqnarray*}
    \mathcal{L}(\btheta) 
    &=& \sum_{i=1}^n\ln(1 + \btheta^\top\bb_i) - \sum_{i=1}^n\ln(1 + \btheta^\top\bc_i).
\end{eqnarray*}
Consequently, the score and Hessian with respect to $\btheta$ are 
\begin{eqnarray*}
    \nabla_{\btheta} \mathcal{L}(\bbeta,\btheta) &=& \sum_{i=1}^n \frac{\bb_i}{1 + \btheta^\top\bb_i} - \sum_{i=1}^n\frac{\bc_i}{1 + \btheta^\top\bc_i}\\
    d_{\btheta}^2 \mathcal{L}(\bbeta,\btheta) &=& -\sum_{i=1}^n \frac{\bb_i\bb_i^\top}{(1 + \btheta^\top\bb_i)^2} + \sum_{i=1}^n \frac{\bc_i\bc_i^\top}{(1 + \btheta^\top\bc_i)^2}.
\end{eqnarray*}

\subsection{Derivatives of an Unstructured Covariance Matrix}

If we parametrize $\bGamma$ using the Cholesky factor $\bGamma = \bL\bL^\top$ as suggested in Section \ref{sec:multivariate_model}, then the  loglikelihood function can be written as
\begin{eqnarray*}
    \mathcal{L}(\bbeta, \bL) &=& - \sum_{i=1}^n \ln \Big[1\! +\! \frac{1}{2}\tr(\bL\bL^\top)\Big]+
    \sum_{i=1}^n \sum_{j=1}^{d} \ln f_{ij}(y_{ij} \mid \bbeta, \bgamma)\\ &  &  + \sum_{i=1}^n
    \ln \left[1\!+\!\frac{1}{2}\br_i(\bbeta)^\top \bL\bL^\top \br_i(\bbeta)\right].
\end{eqnarray*}
The directional derivatives  
\begin{eqnarray*}
d_{\bV} \frac{1}{2}\tr(\bL\bL^\top) & = & \tr(\bL\bV^\top) \\
d_{\bV}\frac{1}{2}\br_i(\bbeta)^\top \bL\bL^\top \br_i(\bbeta)
& = & \br_i(\bbeta)^\top \bL\bV^\top \br_i(\bbeta)
\end{eqnarray*}
follow from the standard rules of differentiation \citep{lange2024tutorial,magnus2019matrix}. For coding purposes it is easier to invoke reverse-mode automatic differentiation tools such as \texttt{Enzyme.jl} \citep{NEURIPS2020_9332c513}. Indeed,
if we write
\begin{eqnarray*}    
    \bgamma &\equiv& \begin{bmatrix}
        \vec(\bB) \\ \text{vech}(\bL)
    \end{bmatrix} \amp=\amp 
    \begin{bmatrix}
        \bbeta \\ \btheta
    \end{bmatrix}
    \amp\in\amp \mathbb{R}^{pd + d(d+1)/2},
\end{eqnarray*}
then the loglikelihood can be viewed as a vector-input scalar-output function $\mathcal{L}(\bgamma)$. At some sacrifice of computational speed, automatic differentiation will evaluate $\nabla \mathcal{L}(\bgamma)$ and $d^2\mathcal{L}(\bgamma)$ at a current parameter vector $\bgamma$. These derivatives enable implementation of Newton's method or a quasi-Newton method for fitting the multivariate quasi-copula model. 

\subsection{Nuisance Parameter Estimation}

Many distributions are parametrized by additional nuisance parameters that also require estimation. In general, the strategy for estimating these is similar to the strategy for estimating the mean or variance parameters. For brevity, we illustrate this procedure concretely for the Gaussian and negative binomial distributions in 
supplemental Sections \ref{sec:neg_bin_fit} and \ref{sec:gaussian_quasi_copulas}.

\color{black}
\subsection{Initialization}
Most optimization algorithms benefit from good starting values. The obvious candidate for $\bbeta$ is the maximum likelihood estimate delivered by the base model. For structured covariance models, we use an MM algorithm \citep{lange2016mm} to initialize variance components. This process is described in supplemental Section \ref{sec:mm_for_variance_component_estim}. Under the CS and AR1 models, we initialize the variance component $\sigma^2$ by the crude estimate from the MM algorithm treating $\rho = 0$. For the unstructured covariance model, we initialize $\bL$ by the Cholesky decomposition of sample correlation matrix of $\bY$. 

\color{black}
\section{Results}

\subsection{Simulation Studies under the Longitudinal 
Model} \label{sec:simulation_studies}

To assess estimation accuracy of the quasi-copula model, we first present simulation studies for the Poisson and negative binomial base distributions with log link function, under the VC parameterization of $\bGamma_i$. Additional simulation studies with different base distributions under the AR(1), CS and VC parameterizations of $\bGamma_i$ are included in supplemental Section \ref{sec:supp_more_longitudinal_sims}.

In each simulation scenario, the non-intercept entries of the  predictor matrix $\bX_i$ are independent standard normal deviates. True regression coefficients $\bbeta_{\text{true}} \sim \text{Uniform}(-0.2, 0.2)$. For the negative binomial base, all dispersion parameters are $r_{\text{true}} = 10$. Each simulation scenario was run on 100 replicates for each sample size $n \in \{100, 1000, 10000\}$ and number of observations $d_i \in \{2, 5, 10, 15, 20, 25\}$ per independent sampling unit. 

\begin{figure}
    \footnotesize
    \begin{minipage}[h]{0.49\linewidth}
        \begin{center}
        \includegraphics[width=1\linewidth]{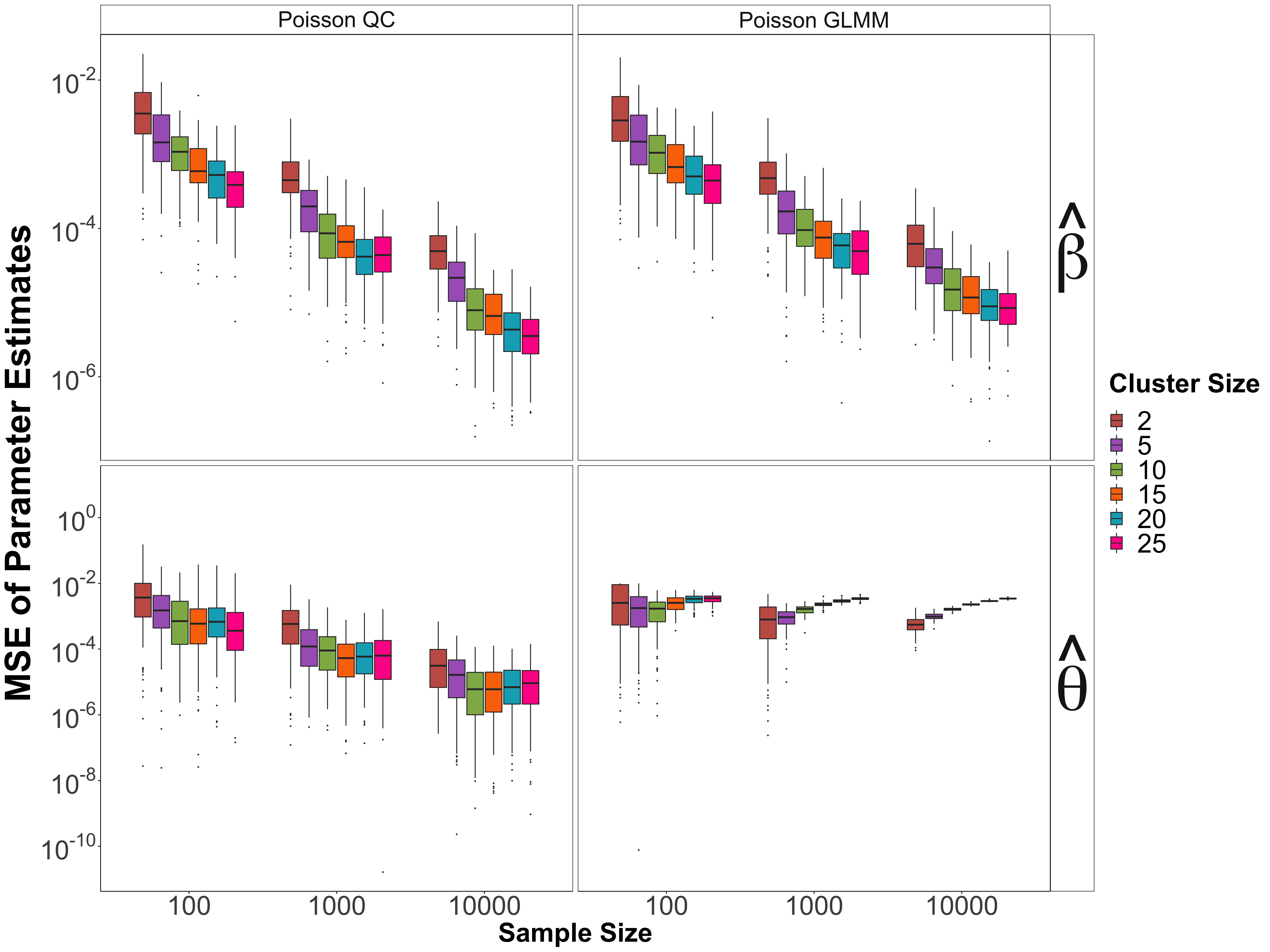} 
        \end{center} 
    \end{minipage}
    \hfill
    \vspace{0.2 cm}
    \begin{minipage}[h]{0.49\linewidth}
        \begin{center}
        \includegraphics[width=1\linewidth]{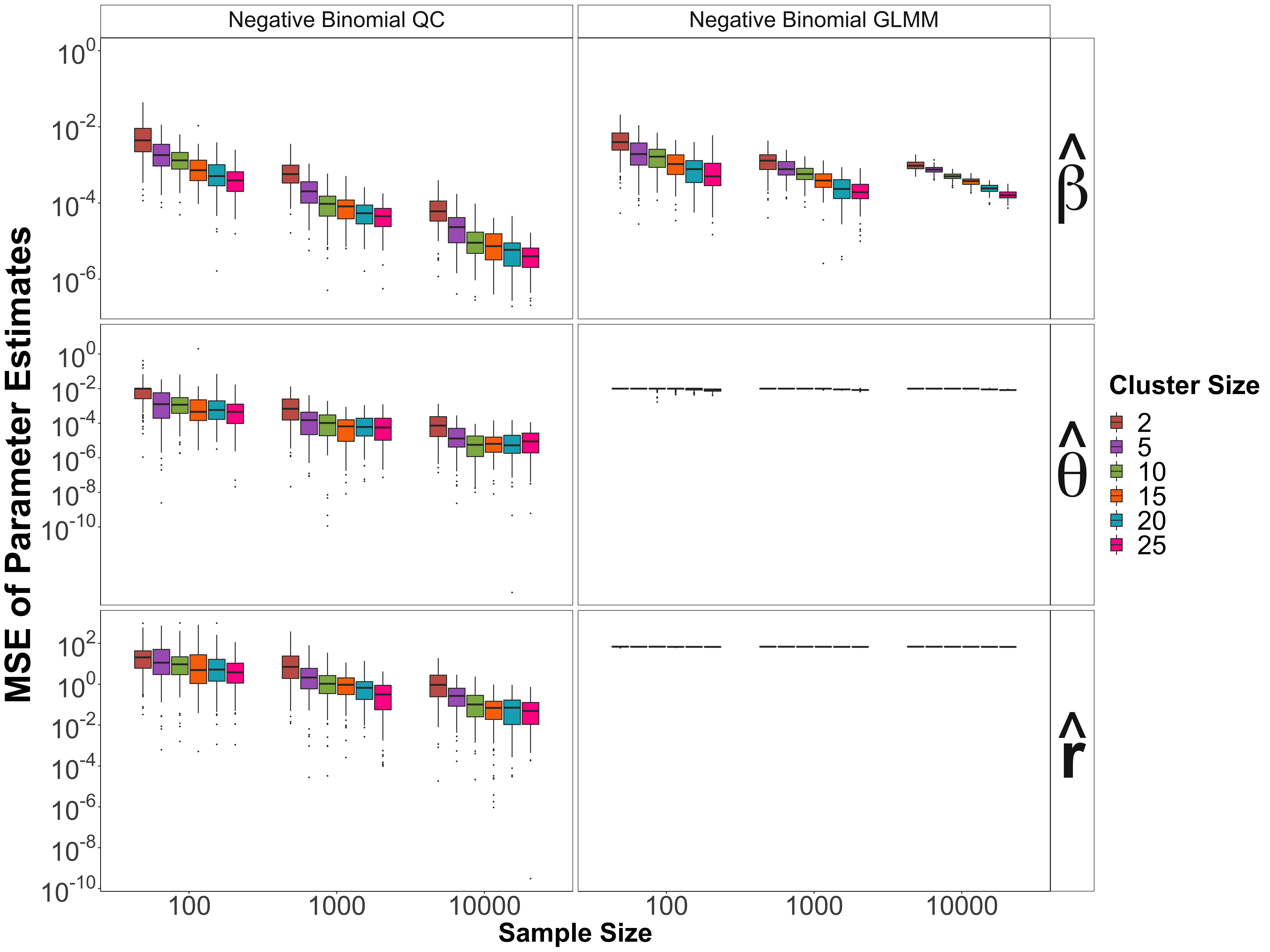} 
        \end{center}
    \end{minipage}
    \vfill
    \vspace{0.2 cm}
    \begin{minipage}[h]{0.49\linewidth}
        \begin{center}
        \includegraphics[width=1\linewidth]{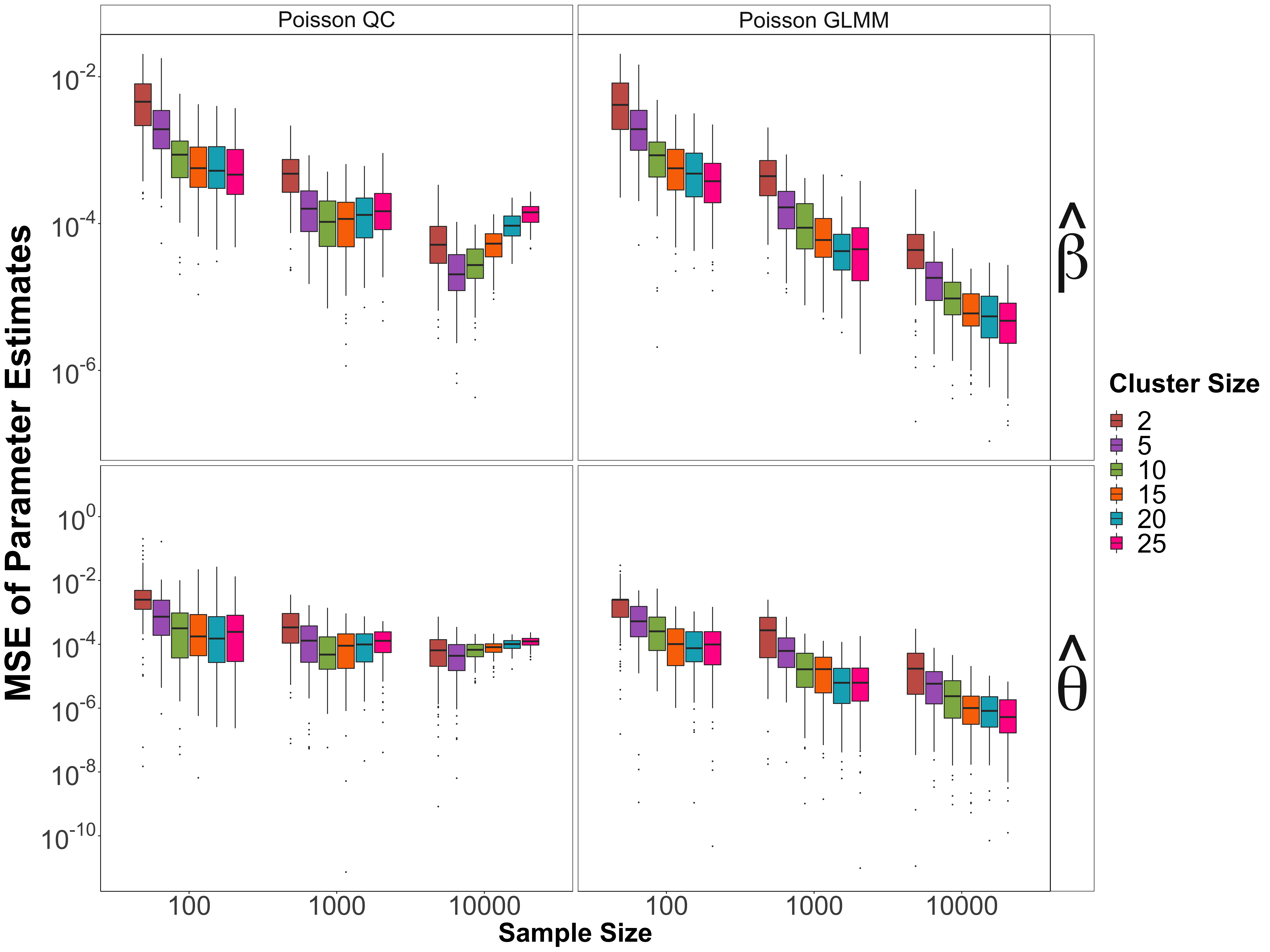} 
        \end{center}
    \end{minipage}
    \hfill
    \begin{minipage}[h]{0.49\linewidth}
        \begin{center}
        \includegraphics[width=1\linewidth]{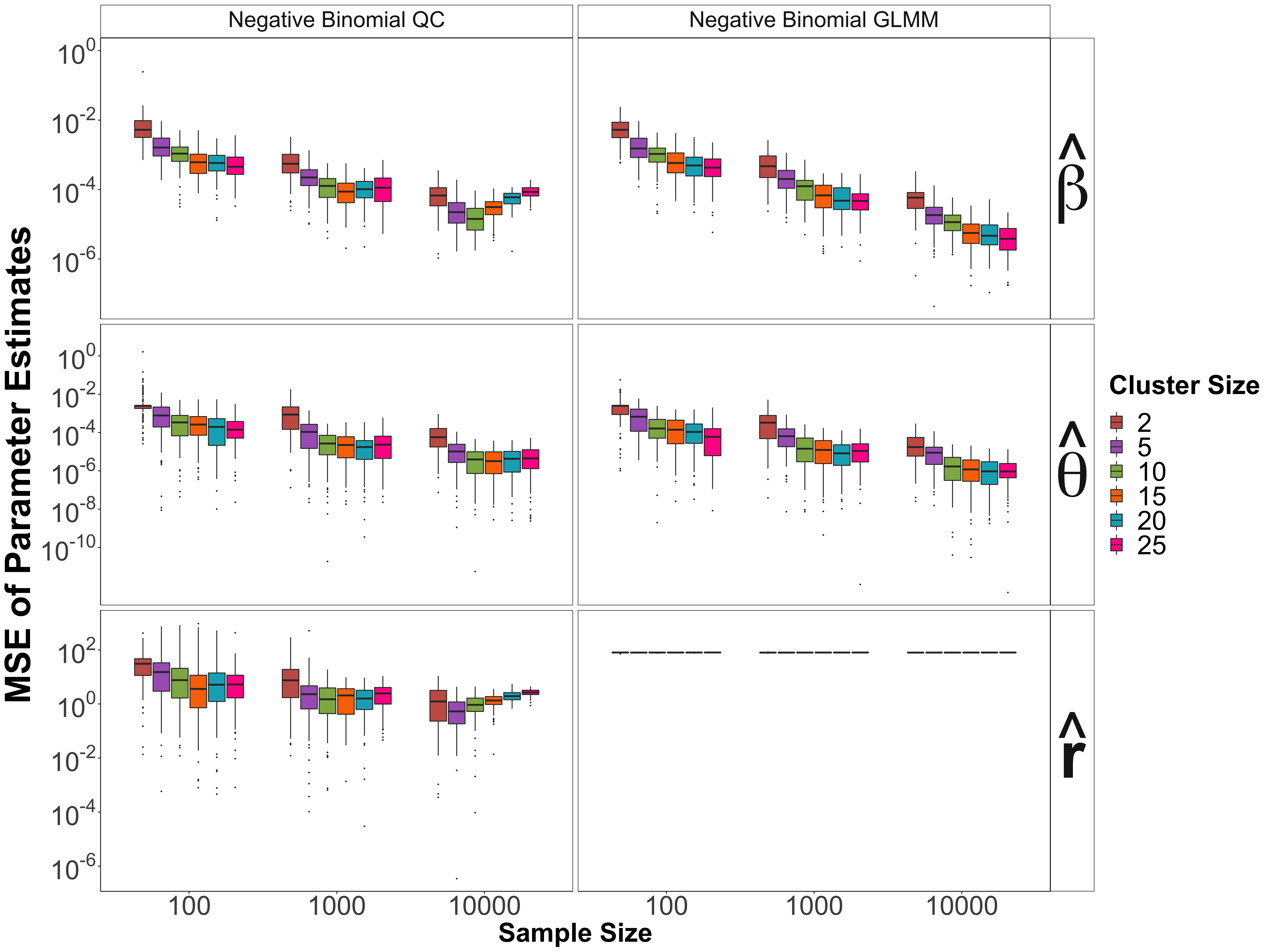} 
        \end{center}
    \end{minipage}
    \caption{Simulation study under the longitudinal model. Top panel features MSE for $\bbeta$ and $\btheta$ under Simulation I setting with Poisson base (left) and Negative binomial base (right). The bottom panel features MSE of $\bbeta$ and $\btheta$ under Simulation II setting with Poisson base (left) and Negative binomial base (right). Here cluster size refers to $d_i$, the number of observations per sample. }
    \label{fig:MSE_of_poisson_and_NB}
\end{figure}

Under the VC parameterization of $\bGamma_i,$ the choice $\bGamma_{i,\text{true}} = \theta_{\text{true}} \times {\bf 1}_{d_i}{\bf 1}_{d_i}^\top$ allows us to compare to the random intercept GLMM fit using {\rm MixedModels.jl}. When the random effect term is a scalar, {\rm MixedModels.jl} uses Gaussian quadrature for parameter estimation. We compare estimates and run-times to the random intercept GLMM fit of {\rm MixedModels.jl} with 25 Gaussian quadrature points. We conduct simulation studies under two scenarios (simulation I and II). In simulation I, it is assumed that the data are generated by the quasi-copula model with $\theta_{\text{true}} = 0.1$, and in simulation II, it is assumed that the true distribution is the random intercept GLMM with $\theta_{\text{true}} = 0.05$.\newline
\newline
\textbf{Simulation I:} In this scenario, we simulate datasets under the quasi-copula model as outlined in Section 5 and compare MLE fits under the quasi-copula model and GLMM. Top panel of Figure \ref{fig:MSE_of_poisson_and_NB} helps us assess estimation accuracy and how well the GLMM density approximates the quasi-copula density. As anticipated, the MSE's across all base distributions decrease as sample size increases. For data simulated under the quasi-copula model, quasi-copula mean squared errors (MSE) are generally lower than GLMM MSE's. GLMM estimated variance components are often zero and stay relatively constant across sample sizes. This confirms the fact that the two models are different in how they handle random effects, particularly with larger sampling units ($d_i > 2$). \newline
\newline
\textbf{Simulation II:} In the second simulation scenario, we generate data\-sets under the random intercept Poisson GLMM and compare MLE fits delivered by the two models. Bottom panel of Figure \ref{fig:MSE_of_poisson_and_NB} now shed light on how well the quasi-copula density approximates the GLMM density under different magnitudes of the variance components. As expected, MSE's under GLMM analysis are now generally lower than those under quasi-copula analysis. For the Poisson and negative binomial base distributions with $\theta_{\text{true}} = 0.05$, the bottom panel of Figure \ref{fig:MSE_of_poisson_and_NB} indicate biases for the quasi-copula estimates of $(\bbeta, \btheta)$ for larger sampling units $(d_i > 2)$ up to sample size $n = 10,000$ similar to the bias observed for $\hat{\btheta}$ in the Poisson GLMM example on top panel.  

\subsection{Run Times}

Run times under simulation I and II are comparable. Table \ref{tab:table1} presents average run times and their standard errors in seconds for $100$ replicates under simulation II with  $\theta_{\text{true}} = 0.01$. All computer runs were performed on a standard 2.3 GHz Intel i9 CPU with 8 cores. Runtimes for the quasi-copula model are presented given multi-threading across 8 cores. We note the current version of {\rm MixedModels.jl} does not allow for multi-threading across multiple cores. 

\begin{table}[H]
    \centering
    
    \caption{Run times and (standard error of run times) in seconds based on $100$ replicates under simulation II with Poisson and negative binomial (NB) Base, $\theta_{\text{true}} = 0.01,$ sampling unit size $d_i$ and sample size $n$.}
    \begin{adjustbox}{width=\textwidth}
    \begin{tabular}{|c|c|c|c|c| c|}
    \hline
        \textbf{n} & $\mathbf{d_i}$ & \textbf{Poisson QC time} & \textbf{Poisson GLMM time} & \textbf{NB QC time} & \textbf{NB GLMM time}  \\ \hline
    100 & 2 & 0.021 ($<$0.001) & 0.022 (0.003) & 0.125 (0.008) & 0.037 (0.003) \\ \hline
    100 & 5 & 0.020 ($<$0.001) & 0.045 (0.003) & 0.095 (0.005) & 0.068 (0.004) \\ \hline
    100 & 10 & 0.023 (0.001) & 0.080 (0.004) & 0.105 (0.004) & 0.187 (0.011) \\ \hline
    100 & 15 & 0.024 (0.001) & 0.148 (0.006) & 0.105 (0.004) & 0.282 (0.017) \\ \hline
    100 & 20 & 0.025 (0.001) & 0.186 (0.007) & 0.112 (0.002) & 0.394 (0.017) \\ \hline
    100 & 25 & 0.026 ($<$0.001) & 0.265 (0.009) & 0.119 (0.003) & 0.461 (0.019) \\ \hline
    1000 & 2 & 0.025 ($<$0.001) & 0.192 (0.007) & 0.163 (0.009) & 0.365 (0.013) \\ \hline
    1000 & 5 & 0.030 ($<$0.001) & 0.516 (0.016) & 0.167 (0.004) & 0.857 (0.033) \\ \hline
    1000 & 10 & 0.035 (0.001) & 1.011 (0.022) & 0.243 (0.003) & 1.972 (0.050) \\ \hline
    1000 & 15 & 0.040 ($<$0.001) & 1.402 (0.030) & 0.303 (0.002) & 2.854 (0.064) \\ \hline
    1000 & 20 & 0.042 ($<$0.001) & 1.887 (0.036) & 0.371 (0.002) & 3.722 (0.077) \\ \hline
    1000 & 25 & 0.051 (0.001) & 2.531 (0.046) & 0.435 (0.002) & 4.815 (0.089) \\ \hline
    10000 & 2 & 0.128 (0.001) & 1.896 (0.032) & 1.169 (0.040) & 3.902 (0.079) \\ \hline
    10000 & 5 & 0.154 (0.001) & 4.333 (0.075) & 1.375 (0.020) & 8.598 (0.140) \\ \hline
    10000 & 10 & 0.232 (0.002) & 9.545 (0.143) & 2.154 (0.007) & 20.499 (0.303) \\ \hline
    10000 & 15 & 0.272 (0.002) & 14.844 (0.249) & 2.78 (0.007) & 29.003 (0.465) \\ \hline
    10000 & 20 & 0.336 (0.002) & 21.423 (0.356) & 3.314 (0.007) & 42.952 (0.679) \\ \hline
    10000 & 25 & 0.429 (0.003) & 29.324 (0.528) & 4.111 (0.011) & 54.676 (0.861) \\ \hline
    \end{tabular}
    \label{tab:table1}
    \end{adjustbox}
\end{table}

Because the quasi-copula loglikelihoods contain no determinants or matrix inverses, our software experiences less pronounced increases in computation time as sample and sampling unit sizes grow compared to GLMM implemented in {\rm MixedModels.jl}. Run times for the quasi-copula model are faster than those of {\rm MixedModels.jl} for discrete outcomes (Table \ref{tab:table1}, Supplementary Table \ref{tab:bernoulli_time}) and slower for Gaussian distributed outcomes (Supplementary Table \ref{tab:gaussian_time}). This general trend also holds on a per core basis. This discrepancy is hardly surprising since {\rm MixedModels.jl} takes into account the low-rank structure of the covariance matrix $\bOmega_i$ in the random intercept linear mixed model (LMM). This tactic reduces the computational complexity per sample from $\bO(d_i^3)$ to $\bO(d_i^2)$. More detailed comparisons appear in supplemental Section \ref{sec:additional_runtimes}. 

Finally, we study the negative binomial base distribution for longitudinal model in more depth. We compared our negative binomial fits with those delivered by the three popular R packages for GLMM estimation in supplementary Section \ref{sec:neg_bin_r_compare}. Within Julia, {\rm MixedModels.jl} explicitly warns the user against fitting GLMM's with unknown dispersion parameter $r$. Our software updates $r$ iteratively by Newton's method, holding the other parameters $(\bbeta, \btheta)$ fixed, see supplemental Section \ref{sec:neg_bin_fit} for more estimation details. 

\subsection{GWAS Simulations}\label{sec:multivariate_sims}

Simulations under the multivariate quasi-copula model
demonstrate the potential of Algorithm \eqref{alg:adhoc_LRT} in GWAS.  Specifically, we simulated $n=5000$ subjects, $p=15$ covariates, $q=1000$ indepedent SNPs, and $d=4$ correlated responses. The simulated covariates $\bX \in \mathbb{R}^{n \times p}$ have entries drawn from $N(0, 1)$. A column of 1's is appended to $\bX$ to accommodate an intercept. The true effect sizes are randomly sampled from a $\text{Uniform}(0, 0.5)$ distribution. The number of minor alleles for each SNP follows a $\text{Binomial}(2, \rho)$ distribution with $\rho = 0.3$. For simplicity, the true covariance is simulated under the AR1 model
\begin{eqnarray*}
    \bGamma = \begin{pmatrix}
        1 & 0.5 & 0.25 & 0.125 \\
        0.5 & 1 & 0.5 & 0.25\\
        0.25 & 0.5 & 1 & 0.5\\
        0.125 & 0.25 & 0.5 & 1
    \end{pmatrix}.
\end{eqnarray*}
Under this setup we explore 3 simulation scenarios:\\
\\
\textbf{Simulation III:} Here we assume each response $\by_i \in \mathbb{R}^d$ is an independent sample from the quasi-copula model. Each of the $d$ components of $\by_i$ is randomly chosen from the Gaussian, Bernoulli, and Poisson base distributions. The variance of the Gaussian base is set at $\sigma^2 = 0.5$. \\
\\
\textbf{Simulation IV:} Here the responses $\by_i$ are also generated from the quasi-copula model, but now all $d$ components $y_{ij}$ have a Bernoulli base. This scenario is appropriate given multiple correlated case-control responses. \\
\\
\textbf{Simulation V:} Here the responses follow a multivariate Gaussian $\by_i = N(\bB^\top\bx_i, \bGamma)$ distribution. This choice allows us to assess whether quasi-copula fitting correctly collapses to the underlying base model.\\
\\
To assess power, we randomly choose $k=10$ causal SNPs. If $\balpha \in \mathbb{R}^d$ denotes the effect of a causal SNP on the $d$ phenotypes, then we impose the constraint $\sum_{i=1}^d \alpha_i = s$, where $s$ varies between 0 and 1. Thus, a causal SNP influences each of the $d$ responses, but the magnitudes of its effect are both correlated and random. 

\begin{figure}
    \centering
    \includegraphics[width=\textwidth]{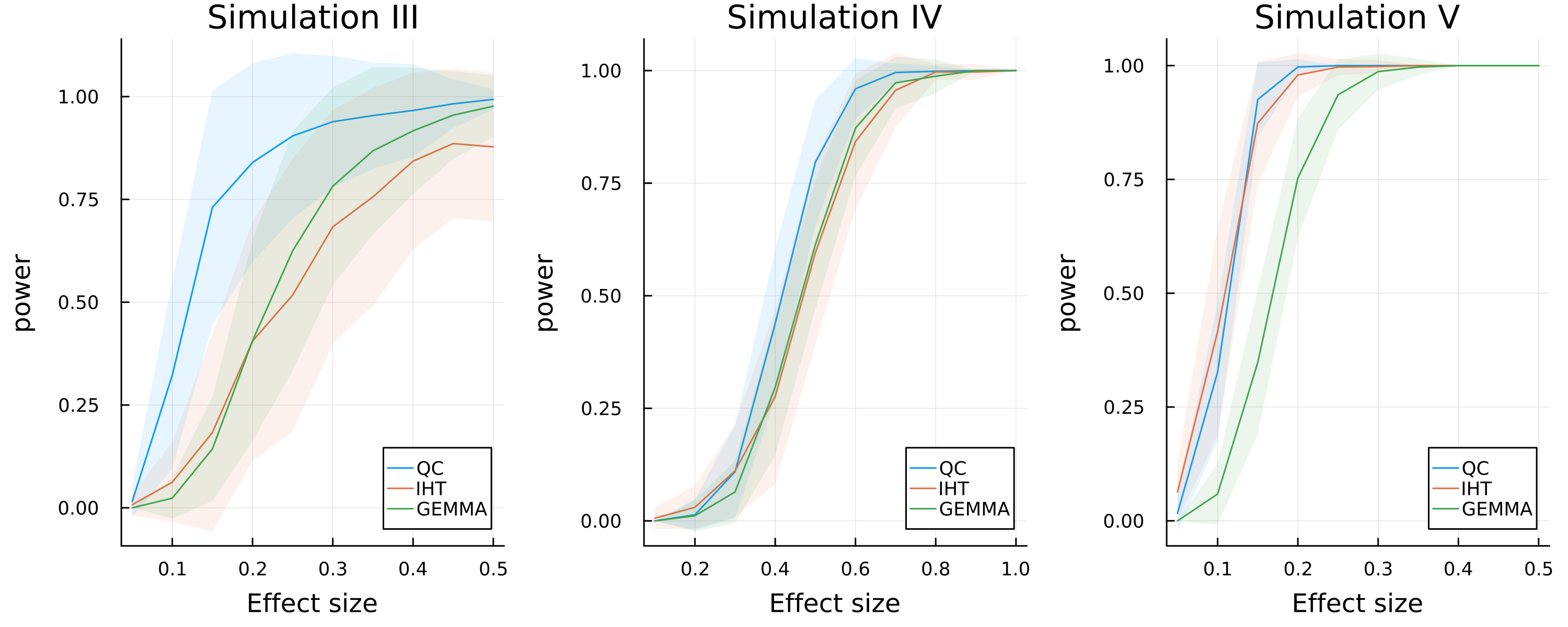}
    \caption{Power simulation for the proposed multivariate GWAS routine in Algorithm \eqref{alg:adhoc_LRT}. Here QC denotes quasi-copula, IHT denotes iterative hard thresholding, a penalized sparse regression method \citep{chu2023multivariate}, and GEMMA implements a multivariate linear-mixed model \citep{zhou2014efficient}.}
    \label{fig:multivariate_gwas_power}
\end{figure}

Over 100 replicates, we compare the power of Algorithm \eqref{alg:adhoc_LRT} against a penalized regression (IHT) algorithm \citep{chu2023multivariate} and the multivariate linear mixed model \citep{zhou2014efficient}. As shown in Figure \ref{fig:multivariate_gwas_power}, we achieve better power than both IHT and GEMMA when the data generative process follows the quasi-copula model (simulation III and IV). When the responses are purely Gaussian (simulation V), quasi-copula fitting offers comparable power to IHT. Finally, supplemental Section \ref{sec:multivariate_gwas_extra_sims2} demonstrates that the quasi-copula model produces valid p-values. 

\color{black}

\subsection{Longitudinal Analysis of the NHANES Data}\label{sec:nhanes_example}

For many repeated measurement problems, a simple random intercept model is sufficient to account for correlations between different responses on the same subject. To illustrate this point and the performance of the quasi-copula model, we now turn to a bivariate example from the NHANES I Epidemiologic Followup Study (NHEFS) dataset \citep{cohen1987plan}. In this example, we group the data by subject ID and jointly model the number of cigarettes smoked per day in 1971 and the number of cigarettes smoked per day in 1982 as a bivariate outcome. For fixed effects, we include an intercept and control for sex, age in 1971, and the average price of tobacco in the state of residence. The average price of tobacco is a time-dependent covariate that is adjusted for inflation using the 2008 U.S. consumer price index (CPI). Participants with missing responses or predictors were excluded from the model cohort. A total of $n = 1537$ NHANES I participants constitute the cohort. 

Table \ref{table4} compares the estimates, loglikelihoods and run times in seconds of the longitudinal regression model with Poisson, negative binomial, and Bernoulli base distributions under the random intercept model of {\rm QuasiCopula.jl} and {\rm MixedModels.jl}. For the Bernoulli base distribution, we transformed each count outcome to a binary indicator with value $1$ if the number of cigarettes smoked per day is greater than the sample average and value $0$ otherwise. 

Because overdispersion is a feature of this dataset, the Poisson base distribution represents a case of model misspecification; the negative binomial base distribution is a better choice for analysis. Under the Poisson base distribution, the quasi-copula model inflates the variance component to account for the overdispersion. Under the negative binomial base distribution, both {\rm QuasiCopula.jl} and {\rm MixedModels.jl} estimate the variance component to be $0$. This suggests that no additional overdispersion exists in the data. The estimates for $\bbeta$ under the quasi-copula model with Poisson base are closer to the more realistic estimates under the negative binomial base than those of GLMM. The maximum loglikelihood of the quasi-copula model is lower than that of GLMM for the Poisson base and higher than that of GLMM for the negative binomial and Bernoulli bases. Run times favor the quasi-copula model.

\begin{table}
    \centering
    \begin{adjustbox}{width=\textwidth}
    \begin{tabular}{|c|c|c|c|c|c|c|}
    \hline
        \textbf{Parameter} & \textbf{QC Poisson} & \textbf{GLMM Poisson} & \textbf{QC NB} & \textbf{GLMM NB} & \textbf{QC Bernoulli} & \textbf{GLMM Bernoulli}\\ \hline
    $\beta_\text{Intercept}$ & 2.509 & 2.039 & 2.580 & 2.580 & -1.768 & -1.411 \\ \hline
    $\beta_\text{sex}$ & -0.210 & -0.225 & -0.187 & -0.187 & -0.793 & -0.761 \\ \hline
    $\beta_\text{age}$ & -0.009 & -0.009 & -0.009 & -0.009 & -0.040 & -0.034 \\ \hline
    $\beta_\text{price}$ & 0.434 & 0.597 &  0.402 &  0.402 & 2.238 & 1.891 \\ \hline
    $\theta$ & 7.080 & 0.458 & 0.0 & 0.0 & 0.666 & 3.461 \\ \hline
    $r$ & - & - & 1.141 & 1.395 & - & - \\ \hline
    \textbf{loglikehood} & -20690.8 & \textbf{-15499.5} & \textbf{-12037.6} & -12047.5 & \textbf{-1938.7} & -1980.9 \\ \hline
    \textbf{time (sec)} & 0.268 & 0.749 & 0.160 & 0.978 & 0.109 & 1.030 \\ \hline
    \end{tabular}
    \end{adjustbox}
    \caption{Random intercept MLEs, loglikelihoods, and run times for for the NHEFS data under the quasi-copula (QC) model and GLMM. All $n = 1537$ sampling units are of size $d_i = 2$.}
    \label{table4}
\end{table}

\subsection{Multivariate GWAS on UK Biobank Data}\label{sec:gwas_example}

We also conducted a 3-trait analysis of hypertension related phenotypes from the second release of the UK-Biobank \citep{sudlow2015uk}. The underlying traits, average systolic blood pressure (SBP), average diastolic blood pressure (DBP), and body mass index (BMI), are correlated, heritable, and the subject of previous association studies \citep{chu2023multivariate}. Although the traits are continuous, we dichotomize both SBP and DBP to illustrate a multivariate analysis with correlated non-continuous responses. Following the clinical definition of stage 2 hypertension \citep{whelton20182017}, we set SBP to 1 if a patient's  average SBP is $\ge 140$ mm Hg and DBP to 1 if a patient's average DBP is $\ge 90$ mm Hg. Otherwise, these traits are set to 0.

\begin{figure}
    \centering
    \includegraphics[width=\linewidth]{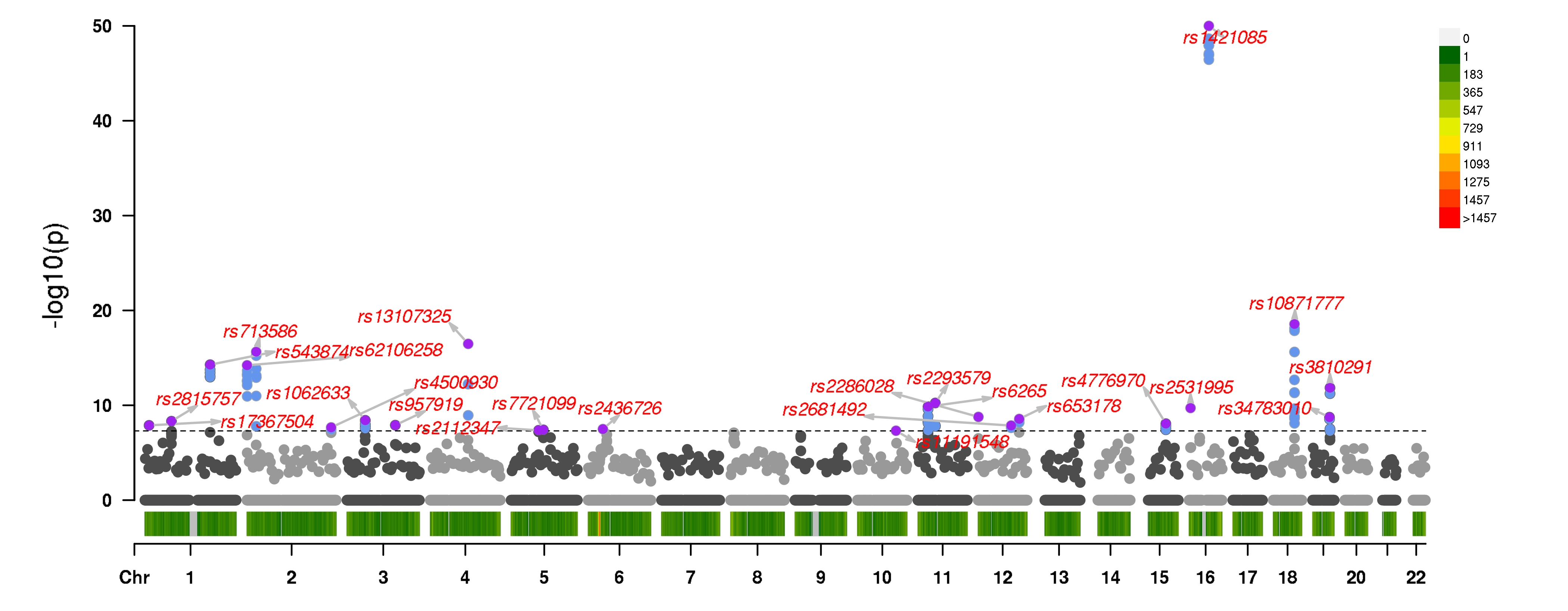}
    \caption{A 3-trait multivariate GWAS on BMI, dichotomized SBP, and dichotomized DBP. The black horizontal dotted line indicates the the genome-wide threshold of $5 \times 10^{-8}$. The most significant SNP within a 1Mb window is labeled and colored purple. All other significant SNPs are colored blue and unlabeled. The legend on the right shows chromosome density.}
    \label{fig:manhattan}
\end{figure}

After quality control (supplemental Section \ref{sec:UKB_qc}), the data includes $p=470,228$ autosomal SNPs and a subset of $n=80,000$ subjects without missing phenotypes.
We split this subset of data into 483 contiguous blocks, each containing roughly contiguous 1000 SNPs, and ran Algorithm $\eqref{alg:adhoc_LRT}$ in parallel across them. Each job finished in less than a day. Altogether 617 SNPs pass our threshold for likelihood ratio testing.
Figure \ref{fig:manhattan} depicts our results in a
GWAS Manhattan plot \citep{yin2021rmvp}.

After pruning secondary but significant SNPs within 1Mb windows, we uncover roughly 24 association hotspots. The strongest signals come from previously known associations with BMI, such as {rs1421085} on chromosome 16, {rs10871777} on chromosome 18, and {rs13107325} on chromosome 4. These SNPs are known to be associated with BMI independently of SBP and DBP \citep{chu2023multivariate}.  
Because dichotomizing a trait loses information, we expect most discoveries to be associated with BMI. However, we were able to discover SNPs such as {rs17367504} on chromosome 1, {rs2681492} on chromosome 12, and {rs653178} on chromosome 12 previously known to be associated with SBP and DBP independently of BMI. Interestingly, five SNPs {rs4500930, rs7721099, rs2293579, rs11191548}, and {rs34783010} missed in our previous analysis are known to be associated with BMI  \citep{macarthur2017new}. It is possible our previous analysis discovered nearby proxies instead. While our current analysis is hardly definitive, it demonstrates that multi-trait GWAS is indeed possible taking proper account of inter-trait correlations. 

\color{black}
\section{Discussion}

We propose a new model for analyzing multivariate data based on Tonda's Gaussian copula approximation. Our quasi-copula model enables the analysis of correlated responses and handles random effects needed in applications such as panel and repeated measures data. The quasi-copula model trades Tonda's awkward parameter space constraint for a simple normalizing constant. This allows one to engage in full likelihood analysis under a tractable probability density function with no implicit integrations or matrix inverses. The quasi-copula model is relatively easy to fit and friendly to likelihood ratio hypothesis testing. 

For maximum likelihood estimation, we recommend a combination of two numerical methods. The first is a block ascent algorithm that alternates between updating the mean parameters $\bbeta$ by a version of Newton's method and updating the variance parameters by a minorization-maximization (MM) algorithm. The second method jointly updates $\bbeta$ and the variances components by a standard quasi-Newton algorithm. The MM algorithm converges quickly to a neighborhood of the MLE but then slows. In contrast, the quasi-Newton struggles at first and then converges quickly. Thus, we start with the block ascent algorithm and then switch to the quasi-Newton algorithm. Both algorithms and their combination are available in our {\rm QuasiCopula.jl} Julia package. 

On balance our numerical tests suggest limitations of the quasi-copula model in handling strongly correlated responses and large sampling units. The presence and size of the normalizing constant $1+\tr(\bGamma)$ in the quasi-copula density may well be the culprit. When the true distribution follows the random intercept GLMM, the quasi-copula estimates are most accurate for small sampling units. When sampling units are large, the quasi-copula estimates are reasonably accurate for smaller magnitudes of variance components. In actual practice many statisticians simply assume the validity of their underlying statistical model. We sorely need good methods for assessing model appropriateness. In the meanwhile, the quasi-copula model offers another avenue for analysis of correlated data. In our opinion, its speed and versatility make up for it defects. We hope other statisticians will agree and assist in probing its properties and applying it to challenging datasets.

\section*{Acknowledgements}

We thank Tetsuji Tonda for his pioneering contributions and Seyoon Ko for help in
parallelizing our Julia code.

\section*{Disclosure statement}
The authors have no conflicts of interest.

\section{Web Resources}

Our software \texttt{QuasiCopula.jl} is freely available to the scientic community through the OpenMendel \citep{zhou2020openmendel} platform.

\noindent
\textbf{Project name}: QuasiCopula.jl\\
\textbf{Project home page}: \url{https://github.com/OpenMendel/QuasiCopula.jl}\\
\textbf{Supported operating systems}: Mac OS, Linux, Windows\\
\textbf{Programming language}: Julia 1.6+\\
\textbf{License}: MIT\\
All commands needed to reproduce the following results are available at \url{https://github.com/sarah-ji/QuasiCopula-reproducibility}

\bibliographystyle{apalike}
\bibliography{Bibliography-MM-MC}

\newpage

\section{Supplemental Materials}

\subsection{Tonda's Approximation Details}\label{sec:tonda}
Let $\bx$ be a random vector with exponential density $f(\bx \mid \bnu)=e^{\bT(\bx)^\top \bnu-A(\bnu)}$.
Note that $\bT(\bx)$ has mean $\bmu(\bnu)= \nabla A(\bnu)$ and covariance matrix $d^2A(\bnu)$. Let us shift $\bnu$ by adding a random Gaussian $\bz$ with mean $\boldsymbol{0}$ and covariance $\bSigma$.
The new density $E[e^{\bT(\bx)^\top (\bnu+\bz)-A(\bnu+\bz)}]$ can be approximated by
expanding the integrand to second order around $\bz=\boldsymbol{0}$ and integrating. This yields
\begin{eqnarray*}
\E[e^{\bT(\bx)^\top (\bnu+\bz)-A(\bnu+\bz)}] & \approx &
\E\Big(e^{\bT(\bx)^\top (\bnu)-A(\bnu)}\{1+[\bT(\bx)-\nabla A(\bnu)]^\top\bz\\
&  & +\frac{1}{2}\bz^\top[\bT(\bx)-\nabla A(\bnu)][\bT(\bx)-\nabla A(\bnu)]^\top\bz\\
&  & -\frac{1}{2}\bz^\top d^2 A(\bnu)\bz\}\Big) \\
& = & e^{\bT(\bx)^\top (\bnu)-A(\bnu)}\Big(1+
\frac{1}{2}\tr\{[\bT(\bx)-\nabla A(\bnu)][\bT(\bx)-\nabla A(\bnu)]^\top\bSigma\}\\
&  & -\frac{1}{2}\tr[d^2 A(\bnu)\bSigma]\Big) \\
& = & e^{\bT(\bx)^\top (\bnu)-A(\bnu)}\Big(1+
\frac{1}{2}\tr\{[\bT(\bx)-\bmu(\bnu)][\bT(\bx)-\bmu(\bnu)]^\top \bSigma\}\\
&  & -\frac{1}{2}\tr[d^2 A(\bnu)\bSigma]\Big) \\
& = & e^{\bT(\bx)^\top (\bnu)-A(\bnu)}\Big\{1+
\frac{1}{2}\bW^\top \sqrt{d^2A(\bnu)}\bSigma\sqrt{d^2A(\bnu)}\bW \\
&  & -\frac{1}{2}\tr[\sqrt{d^2A(\bnu)}\bSigma\sqrt{d^2A(\bnu)}]\Big\} ,
\end{eqnarray*}
where $\bW$ is the standardized version $[\bT(\bx)-\bmu(\bnu)]d^2A(\bnu)^{-1/2}$ of the base sufficient statistic $\bT(\bx)$. The condition $1- -\frac{1}{2}\tr[\sqrt{d^2A(\bnu)}\bSigma\sqrt{d^2A(\bnu)}]>0$ is sufficient but not necessary for the approximate density to be nonnegative. When this condition holds, the approximate density has mass 1. In our quasi-copula density, we drop the offending term $ -\frac{1}{2}\tr[\sqrt{d^2A(\bnu)}\bSigma\sqrt{d^2A(\bnu)}]$, replace $\sqrt{d^2A(\bnu)}\bSigma\sqrt{d^2A(\bnu)}$ by $\bGamma$, assume $T(\bx)=\bx$, and normalize.

\subsection{Generate Random Deviates}\label{sec:supp_gen_random_deviates}

We can construct the $d$ dimensional multivariate vector, $\textbf{y}$ from the multivariate density $g_\textbf{y}(\textbf{y})$ element wise using conditional densities. We recognize the joint density can be represented as a product of conditional densities: 
\begin{eqnarray*}
g_\textbf{y}(\textbf{y}) &=& g_{y_1}(y_1) \times 
g_{y_2 | y_1}(y_2 | y_1) \times ... \times g_{y_d | y_1, ..., y_{d-1}}(y_d | y_1, ..., y_{d-1})\end{eqnarray*}

Thus we can first sample $y_1$ from its marginal density $g_{y_1}(y_1),$ and then sample $y_2$ from the conditional density $g_{y_2 | y_1}(y_2 | y_1).$ The resulting set is a sample from the joint density of $g_{y_1, y_2}(y_1, y_2).$ Continuing this process for all $n$ values of the multivariate vector, $\textbf{y},$ we can sample from it's joint density $g_\textbf{y}(\textbf{y}).$ First we derive the form of the marginal densities $g_{y_1}(y_1),$ and then show the derivation of the conditional density $g_{y_d | y_1, ..., y_{d-1}}(y_d | y_1, ..., y_{d-1}).$

\subsubsection{Marginal Distribution}
For every univariate base distribution, the required  probability density functions (PDFs) $g_{y}(y)$ are of the same form, where $c_0, c_1$ and $c_2$ are constants that depend on the parameters of the specified base distribution $f_{y}(y).$ 
\begin{eqnarray}
g_{y}(y) &=& c f_y(y)[a_0 + a_1(y-\mu) + a_2(y-\mu)^2]\\ &=& c f_y(y) \times \Big[c_0 + c_1 y + c_2 y^2\Big],
\end{eqnarray}
We can re-arrange the PDF to derive the constants $c_0, c_1, c_2$ in the marginal PDF  $g_{y}(y)$ as follows:
\begin{eqnarray*}
	g_{y}(y) &=& \left(1 + \frac 12 \text{tr} \boldsymbol{\Gamma} \right)^{-1} f_{y}(y) \Big[1 + \frac{\gamma_{11}}{2} \Big(\frac{y - \mu}{\sigma}\Big)^2 + \frac{1}{2} \sum_{j=2}^d \gamma_{jj}\Big]\\
	&=& \left(1 + \frac 12 \text{tr} \boldsymbol{\Gamma} \right)^{-1} f_{y}(y) \Big[1 + \frac{\gamma_{11}}{2} \Bigg(\frac{y^2 - 2 y \mu + \mu^2}{\sigma^2}\Bigg) + \frac{1}{2} \sum_{j=2}^d \gamma_{jj}\Big]\\
	&=& \left(1 + \frac 12 \text{tr} \boldsymbol{\Gamma} \right)^{-1} f_{y}(y) \Big[1 + \frac{\gamma_{11}}{2} \Bigg(\frac{y^2}{\sigma^2}\Bigg) + \frac{\gamma_{11}}{2} \Bigg(\frac{-2 y \mu}{\sigma^2}\Bigg) + \frac{\gamma_{11}}{2} \Bigg(\frac{\mu^2}{\sigma^2}\Bigg)  + \frac{1}{2} \sum_{j=2}^d \gamma_{jj}\Big]\\
	&=& \left(1 + \frac 12 \text{tr} \boldsymbol{\Gamma} \right)^{-1} f_{y}(y) \Bigg[\Bigg(1 + \frac{\gamma_{11}}{2} \Big(\frac{\mu^2}{\sigma^2}\Big)  + \frac{1}{2} \sum_{j=2}^d \gamma_{jj} \Bigg) + \Bigg( \frac{\gamma_{11}}{2} \Big(\frac{-2\mu}{\sigma^2} \Big)\Bigg) y  + \Bigg( \frac{\gamma_{11}}{2} \Big(\frac{1}{\sigma^2}\Big) \Bigg) y^2 \Bigg]\\
	&=& c \times f_{y}(y) \Big[\Bigg(c_0\Bigg) + \Bigg(c_1\Bigg) y + \Bigg(c_2\Bigg) y^2 \Big], \label{eq:y1-marginal}
\end{eqnarray*}
\begin{itemize}
\item $c = \left[1 + \frac 12 \text{tr}(\boldsymbol{\Gamma}) \right]^{-1},$ for all base distributions $f_{y}(y).$
\item $c_0 = \Bigg(1 + \frac{\gamma_{11}}{2} \Big(\frac{\mu^2}{\sigma^2}\Big)  + \frac{1}{2} \sum_{j=2}^d \gamma_{jj} \Bigg),$
    \item $c_1 = \Bigg( \frac{\gamma_{11}}{2} \Big(\frac{-2\mu}{\sigma^2} \Big)\Bigg),$
    \item $c_2 = \Bigg( \frac{\gamma_{11}}{2} \Big(\frac{1}{\sigma^2}\Big) \Bigg)$
\end{itemize}

Thus, the required marginal Cumulative Distribution Function (CDF)  $G_y(x)$ takes the following form. We will derive the CDF by finding the appropriate scaled cumulative distributions of the three terms. 
\begin{eqnarray*}
G_y(x) &=& \int_{0}^{\infty} g_{y}(y) \,dy\\
&=& c \int_{-\infty}^x f(y)[c_0 + c_1 y + c_2 y^2]\,dy\\
&=& c\times c_0\int_{-\infty}^x f_{y}(y)\,dy\\ &+& c\times c_1\int_{-\infty}^x yf_{y}(y) \,dy\\ &+& c\times c_2\int_{-\infty}^x y^2 f_{y}(y) \,dy\\
&=& \textbf{term1} + \textbf{term2} + \textbf{term3} 
\end{eqnarray*}
The first term is a scalar multiple of the base distribution CDF, $F_y(y)$, and $d_1, d_2$ are normalizing constants for random variables $v_1, v_2$ from named distributions $f_{v_1}(v_1), f_{v_2}(v_2)$ with CDFs $F_{v_1}(x)$ and $F_{v_2}(x)$ in terms 2 and 3, respectively. 

\begin{itemize}
    \item $\textbf{term1} = c \times c_0 \int_{-\infty}^x f_y(y) \,dy = c\times (c_0) \times F_{y}(x), $
    \item $\textbf{term2} = c \times c_1 \int_{-\infty}^x y*f_y(y) \,dy = c \times c_1 \times d_1 \times \int_{-\infty}^x f_{v_1}(y) \,dy = c\times (c_1) \times d_1 \times F_{v_1}(x)$
    \item $\textbf{term3} = c \times c_2 \int_{-\infty}^x y^2 f_y(y)\,dy = c \times c_2 \times d_2 \times \int_{-\infty}^x f_{v_2}(y) \,dy =  c\times (c_2) \times d_2 \times F_{v_2}(x)$
\end{itemize}

For every base distribution, to satisfy properties of a proper distribution function we require $$c \times [c_0 + c_1 \times d_1 + c_2 \times d_2] = 1.$$

\subsubsection{Conditional Distribution}
Let $\mathbf{y_{[i-1]}}$ indicate elements $y_1,..., y_{i-1}, \forall i \in [1, d].$ Then the conditional density of $y_i$ given the previous components $\mathbf{y_{[i-1]}}$ is:

\begin{eqnarray*}
&&g_{y_i | \mathbf{y_{[i-1]}}}(y_i| \mathbf{y_{[i-1]}}) \\
&=& d_{[i-1]}^{-1}f_i(y_i)
\Big[d_{[i-1]} +  r_i \sum_{j=1}^{i-1} r_j \gamma_{ij}  + \frac{\gamma_{ii}}{2} (r_i^2-1) \Big]\\
&=& d_{[i-1]}^{-1}f_i(y_i)
\Big[d_{[i-1]} +  \Big(\frac{y_i - \mu_i}{\sigma_i} \Big) \sum_{j=1}^{i-1} r_j \gamma_{ij}  + \frac{\gamma_{ii}}{2} (r_i^2-1) \Big]\\
&=& d_{[i-1]}^{-1}f_i(y_i) \Big[\Big(d_{[i-1]}^{-1} - \frac{\gamma_{ii}}{2}\Big) +\Big( \frac{\sum_{j=1}^{i-1} r_j \gamma_{ij}}{\sigma_i}\Big)y_i - \mu_i \Big(\frac{\sum_{j=1}^{i-1} r_j \gamma_{ij}}{\sigma_i}\Big) + \frac{\gamma_{ii}}{2} \frac{(y_i - \mu_i)^2}{\sigma_i^2} \Big]\\
&=& d_{[i-1]}^{-1}f_i(y_i) \Big[\Big(d_{[i-1]}^{-1} - \frac{\gamma_{ii}}{2} - \mu_i \Big(\frac{\sum_{j=1}^{i-1} r_j \gamma_{ij}}{\sigma_i}\Big)\Big) + \Big( \frac{\sum_{j=1}^{i-1} r_j \gamma_{ij}}{\sigma_i}\Big)y_i + \frac{\gamma_{ii}}{2} \frac{(y_i - \mu_i)^2}{\sigma_i^2} \Big]\\
&=& d_{[i-1]}^{-1}f_i(y_i) \Big[\Big(d_{[i-1]}^{-1} - \frac{\gamma_{ii}}{2} - \mu_i \Big(\frac{\sum_{j=1}^{i-1} r_j \gamma_{ij}}{\sigma_i}\Big) + \frac{\gamma_{ii}}{2} \frac{\mu_i^2}{\sigma_i^2}\Big)\\ &+& \Big( \frac{\sum_{j=1}^{i-1} r_j \gamma_{ij}}{\sigma_i} + \frac{\gamma_{ii}}{2} \Big(\frac{-2 \mu_i}{\sigma_i^2}\Big)\Big) y_i\\ &+& \Big(\frac{\gamma_{ii}}{2} \Big(\frac{1}{\sigma_i^2}\Big)\Big) y_i^2 \Big]\\
&=& c f_i(y_i) \Big[ c_0 + c_1 y_i + c_2 y_i^2\Big]
\end{eqnarray*}
where $d_{[i-1]}=1 +  \frac{1}{2} r_{[i-1]}^\top \bGamma_{[i-1]} 
r_{[i-1]}+ \frac{1}{2} \sum_{j=i}^d \gamma_{jj}$. 
\begin{itemize}
\item $c = d_{[i-1]}^{-1}$
\item $c_0 = \Big(d_{[i-1]}^{-1} - \frac{\gamma_{ii}}{2} - \mu_i \Big(\frac{\sum_{j=1}^{i-1} r_j \gamma_{ij}}{\sigma_i}\Big) + \frac{\gamma_{ii}}{2} \frac{\mu_i^2}{\sigma_i^2}\Big)$
    \item $c_1 = \Big( \frac{\sum_{j=1}^{i-1} r_j \gamma_{ij}}{\sigma_i} + \frac{\gamma_{ii}}{2} \Big(\frac{-2 \mu_i}{\sigma_i^2}\Big)\Big)$
    \item $c_2 = \Big(\frac{\gamma_{ii}}{2} \Big(\frac{1}{\sigma_i^2}\Big)\Big)$
\end{itemize}
We can construct each conditional density given the previously sampled elements as a combination of three constants, just as in the marginal density.


\subsubsection{Continuous Outcomes}
When marginal densities $f(y)$ are continuous, each stage of sampling is probably best performed by inverse transform sampling. 
\paragraph{Gamma distribution}
This note considers the special case of Gamma base in the copula framework outlined in Ken's notes, where $f_1(y) \sim \Gamma(\alpha, \theta).$ We will simulate $y$ directly from its marginal density, $g_{y}(y),$ which can also be represented as a mixture distribution.
$$y \sim \Gamma(\alpha, \theta); f_{y}(y) = \frac{1}{\Gamma(\alpha) \theta^\alpha} y^{\alpha - 1} e^{\frac{-y}{\theta}}$$

\begin{itemize}
    \item $\mu = E[y] = \alpha \theta,$ and $\sigma^2 = Var(y) = \alpha \theta^2.$
\begin{eqnarray*}
g_{y}(y) &=& \Big[1+ \frac{1}{2}\tr(\bGamma)\Big]^{-1} \Big[ \frac{1}{\Gamma(\alpha) \theta^\alpha} y^{\alpha - 1} e^{\frac{-y}{\theta}} \Big]
\Big(1 + \frac{\gamma_{11}}{2} \Bigg[\frac{(y - \mu)^2}{\sigma^2}\Bigg] + \frac{1}{2} \sum_{j=2}^d \gamma_{jj}\Big)\\
&=& \Big[1+ \frac{1}{2}\tr(\bGamma)\Big]^{-1} \Big[ \frac{1}{\Gamma(\alpha) \theta^\alpha} y^{\alpha - 1} e^{\frac{-y}{\theta}} \Big]
\Bigg( \Big(1 + \frac{1}{2} \sum_{j=2}^d \gamma_{jj} \Big) + \frac{\gamma_{11}}{2} \Bigg[\frac{y^2 - 2 y \mu + \mu^2}{\sigma^2}\Bigg] \Bigg)\\
&=& c \times f_{y}(y) \Big[\Bigg(c_0\Bigg) + \Bigg(c_1\Bigg) y + \Bigg(c_2\Bigg) y_2^2 \Big]
\end{eqnarray*}
\item Here $c_0 = \Bigg(1 + \frac{\gamma_{11}}{2} \Big(\frac{\mu^2}{\sigma^2}\Big)  + \frac{1}{2} \sum_{j=2}^d \gamma_{jj} \Bigg) = \Bigg(1 + \frac{\gamma_{11}}{2} \Big(\alpha\Big)  + \frac{1}{2} \sum_{j=2}^d \gamma_{jj} \Bigg)$
    \item $c_1 = \Bigg( \frac{\gamma_{11}}{2} \Big(\frac{-2\mu}{\sigma^2} \Big)\Bigg) = \Bigg( \frac{\gamma_{11}}{2} \Big(\frac{-2}{\theta} \Big)\Bigg)$
    \item $c_2 = \Bigg( \frac{\gamma_{11}}{2} \Big(\frac{1}{\sigma^2}\Big) \Bigg) = \Bigg( \frac{\gamma_{11}}{2} \Big(\frac{1}{\alpha\theta^2}\Big) \Bigg)$
\end{itemize}

We will use the given information here: $y \sim \Gamma(\alpha,  \theta),$ where $F_{Y}(x) = P(y \leq x)$ to derive the CDF $G_{Y}(x)$ term by term. 
\begin{eqnarray*}
\textbf{term1} &=& c\times (c_0) \int_{0}^x f_{y}(y) dy\\
&=& c\times (c_0) \times F_{Y}(Y = x)\\
&=& c \times \Bigg(1 + \frac{\gamma_{11}}{2} \Big(\alpha\Big)  + \frac{1}{2} \sum_{j=2}^m \gamma_{jj} \Bigg) \times F_{Y}(Y = x)
\end{eqnarray*}
Define a new random variable $v_1 \sim \Gamma(\alpha + 1, \theta),$ where $F_{v_1}(x) = P(v_1 \leq x).$ \begin{eqnarray*}
\textbf{term2} &=& c\times (c_1) \times \int_{0}^x y f_{y}(y) dy\\
&=& c \times \Bigg( \frac{\gamma_{11}}{2} \Big(\frac{-2}{\theta} \Big)\Bigg) \times \int_{0}^x y f_{y}(y) dy\\
&=& c \times \Bigg( \frac{\gamma_{11}}{2} \Big(\frac{-2}{\theta} \Big)\Bigg) \times \theta \int_{0}^x \frac{y}{\theta} f_{y}(y) dy\\
&=& c\times \Bigg( \frac{\gamma_{11}}{2} \Big(\frac{-2}{\theta} \Big)\Bigg) \times \theta \times \int_{0}^x \Big[ \frac{1}{\Gamma(\alpha) \theta^{\alpha + 1}} y^{(\alpha + 1) - 1} e^{\frac{-y}{\theta}} \Big] dy\\
&=& c\times \Bigg( \frac{\gamma_{11}}{2} \Big(\frac{-2}{\theta} \Big)\Bigg) \times \theta \times \frac{\Gamma(\alpha + 1)}{\Gamma(\alpha)} \times \int_{0}^x \Big[ \frac{1}{\Gamma(\alpha + 1) \theta^{\alpha + 1}} y^{(\alpha + 1) - 1} e^{\frac{-y}{\theta}} \Big] dy\\
&=& c\times \Bigg( \frac{\gamma_{11}}{2} \Big(\frac{-2}{\theta} \Big)\Bigg) \times \theta \times \frac{\Gamma(\alpha + 1)}{\Gamma(\alpha)}  \times F_{v_1}(x)\\
\end{eqnarray*}

Define another random variable $v_2 \sim \Gamma(\alpha + 2, \theta);$ with CDF $F_{v_2}(x) = P(v_2 \leq x).$  
\begin{eqnarray*}
\textbf{term 3} &=&c\times (c_2) \times \int_{0}^x y^2 f_{y}(y)\,dy\\
&=& c \times \Bigg( \frac{\gamma_{11}}{2} \Big(\frac{1}{\alpha\theta^2}\Big) \Bigg) \times \int_{0}^x y^2 f_{y}(y)\,dy\\
&=& c \times \Bigg( \frac{\gamma_{11}}{2} \Big(\frac{1}{\alpha\theta^2}\Big) \Bigg) \times \theta^2 \int_{0}^x \frac{y^2}{\theta^2} f_{y}(y)\,dy\\
&=& c\times \Bigg( \frac{\gamma_{11}}{2} \Big(\frac{1}{\alpha\theta^2}\Big) \Bigg) \times \theta^2 \times \int_{0}^x \Big[ \frac{1}{\Gamma(\alpha) \theta^{\alpha + 2}} y^{(\alpha + 2) - 1} e^{\frac{-y}{\theta}} \Big] dy\\
&=& c\times \Bigg( \frac{\gamma_{11}}{2} \Big(\frac{1}{\alpha\theta^2}\Big) \Bigg) \times \theta^2 \times \frac{\Gamma(\alpha + 2)}{\Gamma(\alpha)} \times \int_{0}^x \Big[ \frac{1}{\Gamma(\alpha + 2) \theta^{\alpha + 2}} y^{(\alpha + 2) - 1} e^{\frac{-y}{\theta}} \Big] dy\\
&=& c\times \Bigg( \frac{\gamma_{11}}{2} \Big(\frac{1}{\alpha\theta^2}\Big) \Bigg) \times \theta^2 \times \frac{\Gamma(\alpha + 2)}{\Gamma(\alpha)} \times F_{v_2}(x)\\
\end{eqnarray*}



\paragraph{Exponential distribution}
Next, we consider when $f_{y}(y) \sim Exponential(\frac{1}{\theta}).$ To find the appropriate CDF function under this exponential base, we make note of the relationship between the exponential and gamma densities. 
$$y \sim Exponetial\Big(\frac{1}{\theta}\Big) \iff y \sim \Gamma(\alpha = 1, \theta \geq 0);$$
$$f_{y}(y) = \frac{1}{\theta} e^{\frac{-y}{\theta}} = \frac{1}{\Gamma(1) \theta^1} y^{1 - 1} e^{\frac{-y}{\theta}} ; y, \theta \geq 0$$

\begin{itemize}
    \item $\mu = E[y] = \theta,$ and $\sigma^2 = Var(y) = \theta^2.$
\begin{eqnarray*}
g_{y}(y) &=& \Big[1+ \frac{1}{2}\tr(\bGamma)\Big]^{-1} \Big[ \frac{1}{\Gamma(1) \theta^1} y^{1 - 1} e^{\frac{-y}{\theta}} \Big]
\Big(1 + \frac{\gamma_{11}}{2} \Bigg[\frac{(y - \mu)^2}{\sigma^2}\Bigg] + \frac{1}{2} \sum_{j=2}^d \gamma_{jj}\Big)\\
&=& \Big[1+ \frac{1}{2}\tr(\bGamma)\Big]^{-1} \Big[ \frac{1}{\Gamma(1) \theta^1} y^{1 - 1} e^{\frac{-y}{\theta}} \Big]
\Bigg( \Big(1 + \frac{1}{2} \sum_{j=2}^d \gamma_{jj} \Big) + \frac{\gamma_{11}}{2} \Bigg[\frac{y^2 - 2 y \mu + \mu^2}{\sigma^2}\Bigg] \Bigg)\\
&=& c \times f_{y}(y) \Big[\Bigg(c_0\Bigg) + \Bigg(c_1\Bigg) y + \Bigg(c_2\Bigg) y_2^2 \Big]
\end{eqnarray*}
    \item $\text{ Here } c_0 = \Bigg(1 + \frac{\gamma_{11}}{2} \Big(\frac{\mu^2}{\sigma^2}\Big)+\frac{1}{2} \sum_{j=2}^d \gamma_{jj}\Bigg) = \Bigg(1 + \frac{\gamma_{11}}{2}+\frac{1}{2} \sum_{j=2}^d \gamma_{jj} \Bigg)$
    \item $c_1 = \frac{\gamma_{11}}{2} \Big(\frac{-2\mu}{\sigma^2} \Big) = \frac{\gamma_{11}}{2} \Big(\frac{-2}{\theta} \Big)$
    \item $c_2 = \frac{\gamma_{11}}{2} \Big(\frac{1}{\sigma^2}\Big) =  \frac{\gamma_{11}}{2} \Big(\frac{1}{\theta^2}\Big) $
\end{itemize}

$y \sim \text{Exponential}(\frac{1}{\theta}) = \Gamma(\alpha = 1, \theta),$ with CDF $F_{Y}(Y = x) = P(Y \leq x)$
\begin{eqnarray*}
\textbf{term1} &=& c\times (c_0) \int_{0}^x f_{y}(y) dy\\
&=& c\times (c_0) \times F_{y}(x)\\
&=& c \times \Bigg(1 + \frac{\gamma_{11}}{2}  + \frac{1}{2} \sum_{j=2}^d \gamma_{jj} \Bigg) \times F_{y}(x)
\end{eqnarray*}
Define a new random variable $v_1 \sim \Gamma(\alpha + 1, \theta) = \Gamma(2, \theta),$ with CDF $F_{v_1}(x) = P(v_1 \leq x).$  
\begin{eqnarray*}
\textbf{term2} &=& c\times (c_1) \times \int_{0}^x y f_{y}(y) dy\\
&=& c \times \Bigg( \frac{\gamma_{11}}{2} \Big(\frac{-2}{\theta} \Big)\Bigg) \times \int_{0}^x y f_{y}(y) dy\\
&=& c\times \Bigg( \frac{\gamma_{11}}{2} \Big(\frac{-2}{\theta} \Big)\Bigg) \times \int_{0}^x \Big[ \frac{1}{\Gamma(1) \theta^1} y^{(1 + 1) - 1} e^{\frac{-y}{\theta}} \Big] dy\\
&=& c\times \Bigg( \frac{\gamma_{11}}{2} \Big(\frac{-2}{\theta} \Big)\Bigg) \times \frac{\theta^2}{\theta} \times \frac{\Gamma(2)}{\Gamma(1)} \times \int_{0}^x \Big[ \frac{1}{\Gamma(2) \theta^{2}} y^{(2) - 1} e^{\frac{-y}{\theta}} \Big] dy\\
&=& c\times \Bigg( \frac{\gamma_{11}}{2} \Big(\frac{-2}{\theta} \Big)\Bigg) \times \theta \times \frac{\Gamma(2)}{\Gamma(1)}  \times F_{v_1}(x)
\end{eqnarray*}
Define another random variable $v_2 \sim \Gamma(1 + 2, \theta) = \Gamma(3, \theta);$ with CDF $F_{v_2}(x) = P(v_2 \leq x).$  
\begin{eqnarray*}
\textbf{term 3} &=&c\times (c_2) \times \int_{0}^x y^2 f_{y}(y)\,dy\\
&=& c \times \Bigg( \frac{\gamma_{11}}{2} \Big(\frac{1}{\theta^2}\Big) \Bigg) \times \int_{0}^x y^2 f_{y}(y)\,dy\\
&=& c\times \Bigg( \frac{\gamma_{11}}{2} \Big(\frac{1}{\theta^2}\Big) \Bigg) \times \int_{0}^x \Big[ \frac{1}{\Gamma(1) \theta^{1}} y^{(1 + 2) - 1} e^{\frac{-y}{\theta}} \Big] dy\\
&=& c\times \Bigg( \frac{\gamma_{11}}{2} \Big(\frac{1}{\theta^2}\Big) \Bigg) \times \frac{\theta^3}{\theta} \times \frac{\Gamma(3)}{\Gamma(1)} \times \int_{0}^x \Big[ \frac{1}{\Gamma(3) \theta^{3}} y^{(3) - 1} e^{\frac{-y}{\theta}} \Big] dy\\
&=& c\times \Bigg( \frac{\gamma_{11}}{2} \Big(\frac{1}{\theta^2}\Big) \Bigg) \times \theta^2 \times \frac{\Gamma(3)}{\Gamma(1)} \times F_{v_2}(x)\\
\end{eqnarray*}

\paragraph{Beta distribution}
Next, we consider when $f_{y}(y) \sim \text{Beta}(\alpha, \beta).$ 
$$y \sim f(y; \alpha, \beta) = \frac{1}{B(\alpha, \beta)}
 y^{\alpha - 1} (1 - y)^{\beta - 1} \\
 = \frac{\Gamma(\alpha + \beta)}{\Gamma(\alpha) \Gamma(\beta)}
 y^{\alpha - 1} (1 - y)^{\beta - 1}, \quad y \in [0, 1]$$

\begin{itemize}
    \item $\mu = E[y] = \frac{\alpha}{\alpha + \beta},$ and $\sigma^2 = Var(y) = \frac{\alpha\beta}{(\alpha + \beta)^2 (\alpha + \beta + 1)},$
\begin{eqnarray*}
&&g_{y}(y) \\
&=& \Big[1+ \frac{1}{2}\tr(\bGamma)\Big]^{-1} \Bigg[\frac{\Gamma(\alpha + \beta)}{\Gamma(\alpha) \Gamma(\beta)}
 y^{\alpha - 1} (1 - y)^{\beta - 1}\Bigg]
\Big(1 + \frac{\gamma_{11}}{2} \Bigg[\frac{(y - \mu)^2}{\sigma^2}\Bigg] + \frac{1}{2} \sum_{j=2}^m \gamma_{jj}\Big)\\
&=& \Big[1+ \frac{1}{2}\tr(\bGamma)\Big]^{-1}\Bigg[\frac{\Gamma(\alpha + \beta)}{\Gamma(\alpha) \Gamma(\beta)}
 y^{\alpha - 1} (1 - y)^{\beta - 1}\Bigg]
\Bigg( \Big(1 + \frac{1}{2} \sum_{j=2}^d \gamma_{jj} \Big) + \frac{\gamma_{11}}{2} \Bigg[\frac{y^2 - 2 y \mu + \mu^2}{\sigma^2}\Bigg] \Bigg)\\
&=& c \times f_{y}(y) \Big[\Bigg(c_0\Bigg) + \Bigg(c_1\Bigg) y + \Bigg(c_2\Bigg) y_2^2 \Big]
\end{eqnarray*}
    \item $\text{ Here } c_0 = \Bigg(1 + \frac{\gamma_{11}}{2}\Big(\frac{\mu^2}{\sigma^2}\Big) +\frac{1}{2} \sum_{j=2}^d \gamma_{jj} \Bigg) = \Bigg(1 + \frac{\gamma_{11}}{2}\Bigg(\frac{\alpha (\alpha + \beta + 1)}{\beta} \Bigg) +\frac{1}{2} \sum_{j=2}^d \gamma_{jj} \Bigg)$
    \item $c_1 = \frac{\gamma_{11}}{2} \Big(\frac{-2\mu}{\sigma^2} \Big) = \frac{\gamma_{11}}{2} \Big(\frac{-2(\alpha + \beta) (\alpha + \beta + 1)}{\beta} \Big)$
    \item $c_2 = \frac{\gamma_{11}}{2} \Big(\frac{1}{\sigma^2}\Big) =  \frac{\gamma_{11}}{2} \Big(\frac{(\alpha + \beta)^2(\alpha + \beta + 1)}{\alpha \beta}\Big) $
\end{itemize}

$y \sim \text{Beta}(\alpha, \beta)$ with CDF $F_{Y}(Y = x) = P(Y \leq x)$
\begin{eqnarray*}
\textbf{term1} &=& c\times (c_0) \int_{0}^x f_{y}(y) dy\\
&=& c\times (c_0) \times F_{y}(x)\\
&=& c \times \Bigg(1 + \frac{\gamma_{11}}{2}\Bigg(\frac{\alpha (\alpha + \beta + 1)}{\beta} \Bigg) +\frac{1}{2} \sum_{j=2}^d \gamma_{jj} \Bigg) \times F_{y}(x)
\end{eqnarray*}
Define a new random variable $v_1 \sim \text{Beta}(\alpha + 1, \beta)$ with CDF $F_{v_1}(x) = P(v_1 \leq x).$  
\begin{eqnarray*}
\textbf{term2} &=& c\times (c_1) \times \int_{0}^x y f_{y}(y) dy\\
&=& c \times \frac{\gamma_{11}}{2} \Big(\frac{-2(\alpha + \beta) (\alpha + \beta + 1)}{\beta} \Big) \times \int_{0}^x y f_{y}(y) dy\\
&=& c\times \frac{\gamma_{11}}{2} \Big(\frac{-2(\alpha + \beta) (\alpha + \beta + 1)}{\beta} \Big) \times \int_{0}^x \Bigg[\frac{\Gamma(\alpha + \beta)}{\Gamma(\alpha) \Gamma(\beta)}
 y^{(\alpha+1) - 1} (1 - y)^{\beta - 1}\Bigg] dy\\
 &=& c\times \frac{\gamma_{11}}{2} \Big(\frac{-2(\alpha + \beta) (\alpha + \beta + 1)}{\beta} \Big) \times \frac{\Gamma(\alpha + \beta)}{\Gamma(\alpha)} \frac{\Gamma(\alpha + 1)}{\Gamma(\alpha + \beta + 1)} \times \int_{0}^x f_{v_1}(y) dy\\
&=& c\times \frac{\gamma_{11}}{2} \Big(\frac{-2(\alpha + \beta) (\alpha + \beta + 1)}{\beta} \Big) \times \frac{\Gamma(\alpha + \beta)}{\Gamma(\alpha)} \frac{\Gamma(\alpha + 1)}{\Gamma(\alpha + \beta + 1)} \times F_{v_1}(x)
\end{eqnarray*}
Define another random variable $v_2 \sim \text{Beta}(\alpha + 2, \beta)$ with CDF $F_{v_2}(x) = P(v_2 \leq x).$  
\begin{eqnarray*}
\textbf{term 3} &=&c\times (c_2) \times \int_{0}^x y^2 f_{y}(y)\,dy\\
&=& c \times  \frac{\gamma_{11}}{2} \Big(\frac{(\alpha + \beta)^2(\alpha + \beta + 1)}{\alpha \beta}\Big) \times \int_{0}^x y^2 f_{y}(y)\,dy\\
&=& c\times  \frac{\gamma_{11}}{2} \Big(\frac{(\alpha + \beta)^2(\alpha + \beta + 1)}{\alpha \beta}\Big) \times \int_{0}^x \Bigg[\frac{\Gamma(\alpha + \beta)}{\Gamma(\alpha) \Gamma(\beta)} y^{(\alpha+2) - 1} (1 - y)^{\beta - 1}\Bigg] dy\\
 &=& c\times \frac{\gamma_{11}}{2} \Big(\frac{-2(\alpha + \beta) (\alpha + \beta + 2)}{\beta} \Big) \times \frac{\Gamma(\alpha + \beta)}{\Gamma(\alpha)} \frac{\Gamma(\alpha + 2)}{\Gamma(\alpha + 2 + \beta)} \times \int_{0}^x f_{v_2}(y) dy\\
&=& c\times  \frac{\gamma_{11}}{2} \Big(\frac{(\alpha + \beta)^2(\alpha + \beta + 2)}{\alpha \beta}\Big) \times \frac{\Gamma(\alpha + \beta)}{\Gamma(\alpha)} \frac{\Gamma(\alpha + 2)}{\Gamma(\alpha + 2 + \beta)} \times F_{v_2}(x)
\end{eqnarray*}

\subsubsection{Discrete Outcomes}
When the densities $f_i(y_i)$ are discrete, it may be necessary to compute infinite sums involving these probabilities. For example, such sums occur naturally in numerical algorithms developed for Poisson, Geometric, negative binomial variate generation. From a practical standpoint, it is necessary to truncate these infinite sums
after a finite number of terms. 

Consider any random variable $Z$ with nonnegative integer values, discrete density $p_i=\Pr(Z=i)$, and mean $\nu$. The inverse method of random sampling reduces to a sequence of comparisons. We partition the interval $[0,1]$ into subintervals with the $i$th subinterval of length $p_i$. To sample $Z$, we draw a uniform random deviate $U$ from $[0,1]$ and return the deviate $j$ determined by the conditions $\sum_{i=1}^{j-1} p_i \le U < \sum_{i=1}^j p_i$. There is no need to invoke the distribution of $Z$. The process is most efficient when the largest $p_i$ occur first. This suggests that we let $k$ denote the least integer $\lfloor \nu \rfloor$ and rearrange the probabilities in the order $p_k, p_{k+1},p_{k-1},p_{k+2},p_{k-2},\ldots$ This tactic is apt put most of the probability mass first and render sampling efficient. 

\paragraph{Poisson Distribution}
A Poisson distribution describes the number of independent events occurring within a unit time interval, given the average rate of occurrence $\theta.$

$$y \sim \text{Poisson}(\theta); f_{y}(y) = \frac{\theta^y e^{-\theta y}}{y!}, y = 0,1,2,3,...$$

\begin{itemize}
    \item $\mu = E[y] = \theta = \sigma^2 = Var(y)$
\begin{eqnarray*}
g_{y}(y) &=& \Big[1+ \frac{1}{2}\tr(\bGamma)\Big]^{-1} \Big[ \frac{\theta^y e^{-\theta y}}{y!} \Big]
\Big(1 + \frac{\gamma_{11}}{2} \Bigg[\frac{(y - \mu)^2}{\sigma^2}\Bigg] + \frac{1}{2} \sum_{j=2}^d \gamma_{jj}\Big)\\
&=& \Big[1+ \frac{1}{2}\tr(\bGamma)\Big]^{-1} \Big[ \frac{\theta^y e^{-\theta y}}{y!} \Big]
\Bigg( \Big(1 + \frac{1}{2} \sum_{j=2}^d \gamma_{jj} \Big) + \frac{\gamma_{11}}{2} \Bigg[\frac{y^2 - 2 y \mu + \mu^2}{\sigma^2}\Bigg] \Bigg)\\
&=& c \times f_{y}(y) \Big[\Bigg(c_0\Bigg) + \Bigg(c_1\Bigg) y + \Bigg(c_2\Bigg) y_2^2 \Big]
\end{eqnarray*}
\item Here $c_0 = \Bigg(1 + \frac{\gamma_{11}}{2} \Big(\frac{\mu^2}{\sigma^2}\Big)  + \frac{1}{2} \sum_{j=2}^m \gamma_{jj} \Bigg) = \Bigg(1 + \frac{\gamma_{11}}{2} \Big(1\Big)  + \frac{1}{2} \sum_{j=2}^d \gamma_{jj} \Bigg)$
    \item $c_1 = \Bigg( \frac{\gamma_{11}}{2} \Big(\frac{-2\mu}{\sigma^2} \Big)\Bigg) = \Bigg( \frac{\gamma_{11}}{2} \Big(-2 \Big)\Bigg)$
    \item $c_2 = \Bigg( \frac{\gamma_{11}}{2} \Big(\frac{1}{\sigma^2}\Big) \Bigg) = \Bigg( \frac{\gamma_{11}}{2} \Big(\frac{1}{\theta}\Big) \Bigg)$
\end{itemize}

\paragraph{Binomial Distribution}
A Binomial distribution characterizes the number of successes in a sequence of independent trials. It has two parameters: $N$, the number of trials, and $p$, the probability of success in an individual trial, with the distribution:
\begin{eqnarray*}
    y \sim \text{Binomial}(N, p), \quad f_{y}(y) = \binom{N}{y} p^y (1-p)^{N-y}, \quad y = 0,1,2,...,N
\end{eqnarray*}
\begin{itemize}
    \item $\mu = E[y] = Np; \sigma^2 = Var(y) = Np(1-p)$
    \begin{eqnarray*}
        g_{y}(y) &=& \Big[1+ \frac{1}{2}\tr(\bGamma)\Big]^{-1} \Big[ \binom{N}{y} p^y (1-p)^{N-y} \Big]
        \Big(1 + \frac{\gamma_{11}}{2} \Bigg[\frac{(y - \mu)^2}{\sigma^2}\Bigg] + \frac{1}{2} \sum_{j=2}^m \gamma_{jj}\Big)\\
        &=& \Big[1+ \frac{1}{2}\tr(\bGamma)\Big]^{-1} \Big[ \binom{N}{y} p^y (1-p)^{N-y} \Big]
        \Bigg( \Big(1 + \frac{1}{2} \sum_{j=2}^m \gamma_{jj} \Big) + \frac{\gamma_{11}}{2} \Bigg[\frac{y^2 - 2 y \mu + \mu^2}{\sigma^2}\Bigg] \Bigg)\\
        &=& c \times f_{y}(y) \Big[\Bigg(c_0\Bigg) + \Bigg(c_1\Bigg) y + \Bigg(c_2\Bigg) y_2^2 \Big]
    \end{eqnarray*}
    \item Here $c_0 = \Bigg(1 + \frac{\gamma_{11}}{2} \Big(\frac{\mu^2}{\sigma^2}\Big)  + \frac{1}{2} \sum_{j=2}^m \gamma_{jj} \Bigg) = \Bigg(1 + \frac{\gamma_{11}}{2} \Big(\frac{(Np)^2}{Np(1-p)}\Big)  + \frac{1}{2} \sum_{j=2}^m \gamma_{jj} \Bigg)$
    \item $c_1 = \Bigg( \frac{\gamma_{11}}{2} \Big(\frac{-2\mu}{\sigma^2} \Big)\Bigg) = \Bigg( \frac{\gamma_{11}}{2} \Big(\frac{-2Np}{Np(1-p)} \Big)\Bigg)$
    \item $c_2 = \Bigg( \frac{\gamma_{11}}{2} \Big(\frac{1}{\sigma^2}\Big) \Bigg) = \Bigg( \frac{\gamma_{11}}{2} \Big(\frac{1}{Np(1-p)}\Big) \Bigg)$
\end{itemize}

\paragraph{Geometric Distribution}
A Geometric distribution characterizes the number of failures before the first success in a sequence of independent Bernoulli trials with success rate `p`.
$$y \sim \text{Geometric}(p); f_{y}(y) = (1-p)^{y}p, y = 0, 1, 2, ...$$

\begin{itemize}
    \item $\mu = E[y] = \frac{1}{p}; \sigma^2 = Var(y) = \frac{1-p}{p^2}$
\begin{eqnarray*}
g_{y}(y) &=& \Big[1+ \frac{1}{2}\tr(\bGamma)\Big]^{-1} \Big[(1-p)^{y}p \Big]
\Big(1 + \frac{\gamma_{11}}{2} \Bigg[\frac{(y - \mu)^2}{\sigma^2}\Bigg] + \frac{1}{2} \sum_{j=2}^d \gamma_{jj}\Big)\\
&=& \Big[1+ \frac{1}{2}\tr(\bGamma)\Big]^{-1} \Big[(1-p)^{y}p \Big]
\Bigg( \Big(1 + \frac{1}{2} \sum_{j=2}^d \gamma_{jj} \Big) + \frac{\gamma_{11}}{2} \Bigg[\frac{y^2 - 2 y \mu + \mu^2}{\sigma^2}\Bigg] \Bigg)\\
&=& c \times f_{y}(y) \Big[\Bigg(c_0\Bigg) + \Bigg(c_1\Bigg) y + \Bigg(c_2\Bigg) y_2^2 \Big]
\end{eqnarray*}
\item Here $c_0 = \Bigg(1 + \frac{\gamma_{11}}{2} \Big(\frac{\mu^2}{\sigma^2}\Big)  + \frac{1}{2} \sum_{j=2}^d \gamma_{jj} \Bigg) = \Bigg(1 + \frac{\gamma_{11}}{2} \Big(\frac{\frac{1}{p}}{\frac{1-p}{p^2}}\Big)  + \frac{1}{2} \sum_{j=2}^d \gamma_{jj} \Bigg)$
    \item $c_1 = \Bigg( \frac{\gamma_{11}}{2} \Big(\frac{-2\mu}{\sigma^2} \Big)\Bigg) = \Bigg( \frac{\gamma_{11}}{2} \Big(\frac{\frac{-2}{p}}{\frac{1-p}{p^2}} \Big)\Bigg)$
    \item $c_2 = \Bigg( \frac{\gamma_{11}}{2} \Big(\frac{1}{\sigma^2}\Big) \Bigg) = \Bigg( \frac{\gamma_{11}}{2} \Big(\frac{p^2}{1-p}\Big) \Bigg)$
\end{itemize}

\paragraph{Negative Binomial Distribution}
A negative binomial distribution describes the number of failures before the `r`th success in a sequence of independent Bernoulli trials. It is parameterized by `r`, the number of successes, and `p`, the probability of success in an individual trial.

$$y \sim \text{Negative Binomial}(r, p); f_{y}(y) = \binom{y+r-1}{y} p^r (1 - p)^y, y = 0, 1, 2, ...$$

\begin{itemize}
    \item $\mu = E[y] = \frac{pr}{1-p}; \sigma^2 = Var(y) = \frac{pr}{(1-p)^2}$
\begin{eqnarray*}
g_{y}(y) &=& \Big[1+ \frac{1}{2}\tr(\bGamma)\Big]^{-1} \Big[\binom{y+r-1}{y} p^r (1 - p)^y\Big]
\Big(1 + \frac{\gamma_{11}}{2} \Bigg[\frac{(y - \mu)^2}{\sigma^2}\Bigg] + \frac{1}{2} \sum_{j=2}^d \gamma_{jj}\Big)\\
&=& \Big[1+ \frac{1}{2}\tr(\bGamma)\Big]^{-1} \Big[\binom{y+r-1}{y} p^r (1 - p)^y\Big]
\Bigg( \Big(1 + \frac{1}{2} \sum_{j=2}^d \gamma_{jj} \Big) + \frac{\gamma_{11}}{2} \Bigg[\frac{y^2 - 2 y \mu + \mu^2}{\sigma^2}\Bigg] \Bigg)\\
&=& c \times f_{y}(y) \Big[\Bigg(c_0\Bigg) + \Bigg(c_1\Bigg) y + \Bigg(c_2\Bigg) y_2^2 \Big]
\end{eqnarray*}
\item Here $c_0 = \Bigg(1 + \frac{\gamma_{11}}{2} \Big(\frac{\mu^2}{\sigma^2}\Big)  + \frac{1}{2} \sum_{j=2}^d \gamma_{jj} \Bigg) = \Bigg(1 + \frac{\gamma_{11}}{2} \Big(\frac{\frac{pr}{1-p}}{\frac{pr}{(1-p)^2}}\Big)  + \frac{1}{2} \sum_{j=2}^d \gamma_{jj} \Bigg)$
    \item $c_1 = \Bigg( \frac{\gamma_{11}}{2} \Big(\frac{-2\mu}{\sigma^2} \Big)\Bigg) = \Bigg( \frac{\gamma_{11}}{2} \Big(\frac{\frac{-2pr}{1-p}}{\frac{pr}{(1-p)^2}} \Big)\Bigg)$
    \item $c_2 = \Bigg( \frac{\gamma_{11}}{2} \Big(\frac{1}{\sigma^2}\Big) \Bigg) = \Bigg( \frac{\gamma_{11}}{2} \Big(\frac{(1-p)^2}{pr}\Big) \Bigg)$
\end{itemize}


\newpage
\subsection{Parameter Estimation:}
We extend the Gaussian Base Model to accommodate densities in exponential family of distributions under the generalized linear model (GLM) framework. In this note, we pay close attention to the density-specific quantities which facilitate parameter estimation, and illustrate using the Poisson and Bernoulli density. 



\subsubsection{Fisher Scoring to Estimate Beta}
The loglikelihood is
\begin{eqnarray}\label{eq:supp_logl}
\mathcal{L}(\bbeta) &=&  -\sum_{i=1}^n \ln \Big[1\! +\! \frac{1}{2}\tr(\bGamma_{i})\Big] + \sum_{i=1}^n
\ln \Big\{1\!+\!\frac{1}{2}\br_i(\bbeta)^\top \bGamma_i \br_i(\bbeta)\Big\} \\
&&+ \sum_{i=1}^{n} \sum_{j=1}^{n_{i}} \ln f_{ij}(y_{ij} \mid \bbeta).\nonumber
\end{eqnarray}

For each distribution, the objective function is the loglikelihood \eqref{eq:supp_logl}, and can be viewed as three separate pieces. The last term of the loglikelihood is specific to the hypothesized density, and has first derivative, $\sum_{i=1}^n \sum_j \nabla \ln f_{ij}(y_{ij} \mid \bbeta),$ and second derivative $d^2 L_n(\bbeta),$ that generalize to the exponential family of distributions.

The score (gradient of the loglikelihood) with respect to $\bbeta$ is:
\begin{eqnarray}
\nabla L_n(\bbeta) = \sum_{i=1}^n \sum_j \nabla \ln f_{ij}(y_{ij} \mid \bbeta) + \sum_{i=1}^n
\frac{\nabla r_i(\bbeta)\bGamma_ir_i(\bbeta)}{1+\frac{1}{2}r_i(\bbeta)^\top \bGamma_i r_i(\bbeta)},
\end{eqnarray}

The first term in the gradient, $\sum_{i=1}^n \sum_j \nabla \ln f_{ij}(y_{ij} \mid \bbeta)$, corresponds to the first derivative of the piece of the loglikelihood, specific to the hypothesized density. We can write this first term as a function of $\mathbf{W_{1i}},$ a diagonal matrix of "working weights".
\begin{eqnarray*}
    \sum_{i=1}^n \sum_j \nabla \ln f_{ij}(y_{ij} \mid \bbeta) &=& \sum_{i=1}^n \sum_{j=1}^{d_i} \frac{(y_{ij}-\mu_{ij}) \mu_{ij}'(\eta_{ij})}{\sigma_{ij}^2} \mathbf{x}_{ij}= \sum_{i=1}^n \mathbf{X_i}^\top \mathbf{W_{1i}}(\mathbf{Y_i}-\boldsymbol{\mu_i})
\end{eqnarray*}
\begin{eqnarray*}
\boldsymbol{W_{1i}} = \mathbf{Diagonal\bigg(\frac{g'(X_{i}^\top\boldsymbol{\beta})}{ var(Y_i | \mu_i)}}\bigg)
&=& \begin{pmatrix}
\frac{{\mu_{i1}'(\eta_{i1})}}{{\sigma_{i1}^2}} & 0 & \cdots & 0 \\
0 & \frac{{\mu_{i2}'(\eta_{i2})}}{{\sigma_{i2}^2}} & \cdots & 0 \\
\vdots  & \vdots  & \ddots & \vdots  \\
0 & 0 & \cdots & \frac{{\mu_{i{n_{i}}}'(\eta_{i{n_{i}}})}}{{\sigma_{i{n_{i}}}^2}}
\end{pmatrix}
\end{eqnarray*}

As discussed in the main text, we can use the expected Fisher Information to get an approximation of the Hessian, which is clearly negative semi-definite. 
\begin{eqnarray}
- \sum_{i=1}^n \mathbf{X_i}^\top \mathbf{W_{2i}} \mathbf{X_i} -\sum_{i=1}^n\frac{[\nabla \br_i(\bbeta)^\top\bGamma_i \br_i(\bbeta)]
[\nabla \br_i(\bbeta)^\top\bGamma_i \br_i(\bbeta)]^\top}
{\Big[1+\frac{1}{2}\br_i(\bbeta)^\top \bGamma_i \br_i(\bbeta)\Big]^2}
\end{eqnarray}

Specifically, we approximate the second derivative of the piece of the loglikelihood particular to the hypothesized density, $d^2 L_n(\beta).$ Using the Expected Fisher Information Matrix, we present this term as a function of another diagonal weight matrix, $\mathbf{W_{2i}}.$
\begin{eqnarray*}
    d^2 L_n(\beta) &=& \sum_{i=1}^n \sum_{j=1}^n \frac{[\mu_{ij}'(\eta_{ij})]^2}{\sigma_{ij}^2} \mathbf{x}_{ij} \mathbf{x}_{ij}^\top - \sum_{i=1}^n \sum_{j=1}^n \frac{(y_{ij} - \mu_{ij}) \mu_{ij}''(\eta_{ij})}{\sigma_{ij}^2} \mathbf{x}_{ij} \mathbf{x}_{ij}^\top \\
& & + \sum_{i=1}^n \sum_{j=1}^n\frac{(y_{ij} - \mu_{ij}) [\mu_{ij}'(\eta_{ij})]^2 (d \sigma_{ij}^{2} / d\mu_{ij})}{\sigma_{ij}^4} \mathbf{x}_{ij} \mathbf{x}_{ij}^\top  \\
    \mathbf{FIM}_n(\beta) &=& \mathbf{E} [- d^2 L_n(\beta)] = - \sum_{i=1}^n \sum_{j=1}^{d_i} \frac{[\mu_{ij}'(\eta_{ij})]^2}{\sigma_{ij}^2} \mathbf{x}_{ij} \mathbf{x}_{ij}^\top = - \sum_{i=1}^n \mathbf{X_i}^\top \mathbf{W_{2i}} \mathbf{X_i}.
\end{eqnarray*}

\begin{eqnarray*}
\boldsymbol{W_{2i}} &=& \mathbf{Diagonal\bigg(\frac{g'(X_{i}^\top\boldsymbol{\beta})^2}{var(Y_i | \mu_i)}}\bigg)
= \begin{pmatrix}
\frac{({\mu_{i1}'(\eta_{i1})})^2}{{\sigma_{i1}^2}} & 0 & \cdots & 0 \\
0 & \frac{({\mu_{i2}'(\eta_{i2})})^2}{{\sigma_{i2}^2}} & \cdots & 0 \\
\vdots  & \vdots  & \ddots & \vdots  \\
0 & 0 & \cdots & \frac{({\mu_{i{n_{i}}}'(\eta_{i{n_{i}}})})^2}{{\sigma_{i{n_{i}}}^2}}
\end{pmatrix}
\end{eqnarray*}
The score and approximate Hessian provide the ingredients for a kind of scoring algorithm for improving $\bbeta$ in our model. For each Newton update of the fixed effect parameter, $\bbeta$, in addition to updating the residual vector, $r_i(\bbeta)$, we must also update these weight matrices, $\boldsymbol{W_{1i}}$ and $\boldsymbol{W_{2i}}$, in the update of the Score and Hessian. We can find these quantities easily by making the appropriate calls to the GLM package, GLM.jl.

Let $y_{ij}$ represent the $j^{th}$ outcome for person $i,$ hypothesized to come from a non-normal density in the exponential family of distributions, $f_{ij}(y_{ij} \mid \bbeta)$. For each hypothesized density under the GLM framework, we have mean parameter $\mu_{ij}(\bbeta) =  g^{[-1]}(\eta_{ij}(\bbeta)) =  g^{[-1]}(\mathbf{x_{ij}} \bbeta),$ and variance parameter $\sigma_{ij}^2(\bbeta)$. Using these quantities, we define $r_{ij}(\bbeta), j \in [1, d_i]$ as the $j^{th}$ entry in the standardized residual vector for observation or group $i$.

\begin{eqnarray} 
r_{ij}(\bbeta) =  \sqrt{\tau} (y_{ij} - \mu_{ij}(\bbeta)) = \frac{(y_{ij}-\mu_{ij}(\bbeta))}{\sqrt{\sigma_{ij}^2(\bbeta)}} \in \mathbb{R}
\end{eqnarray}

Let $\nabla r_{i}(\bbeta) \in \mathbb{R}^{d_i \times p}$ denote the matrix of differentials of all $d_i$ observations for the $i^{th}$ individuals standardized residual vector $r_i(\bbeta)$.  This quantity is important for our score and hessian computation, which really helps the optimization algorithm to speed up convergence.
\begin{center}
\begin{tabular}{ccccccc}
$\nabla r_{i}(\bbeta)^\top$ & = $\begin{pmatrix}
\nabla r_{i1}(\bbeta) & \nabla r_{i2}(\bbeta)& ... & \nabla r_{id_i}(\bbeta)&\\
\end{pmatrix}$
\end{tabular}
\end{center}
For each of the $j \in [1, d_i]$ observations for the $i^{th}$ individual, $\nabla r_{i}(\bbeta)^\top$ can be formed column by column, where $\nabla r_{ij}(\bbeta)$ denotes the $j^{th}$ column. $\nabla \mu_{ij}(\bbeta)$ and $\nabla \sigma_{ij}^2(\bbeta)$ respectively reflect the derivative of the mean and variance of the hypothesized density, with respect to $\bbeta.$
\begin{eqnarray}
\nabla r_{ij}(\bbeta) &=&  -\frac{1}{\sigma_{ij}(\bbeta)} \nabla \mu_{ij}(\bbeta)- \frac{1}{2} \frac{y_{ij}-\mu_{ij}(\bbeta)}{\sigma_{ij}^3(\bbeta)} \nabla \sigma_{ij}^2(\bbeta)  \in \mathbb{R}^p
\end{eqnarray}
$$
\nabla \mu_{ij}(\bbeta) = \frac{\partial \mu_{ij}(\bbeta)}{\partial\eta_{ij}(\bbeta)} * \frac{\partial \eta_{ij}(\bbeta)}{\partial\bbeta} = \begin{pmatrix}
\frac{\partial \mu_{ij}(\bbeta)}{\partial\eta_{ij}(\bbeta)} * \frac{\partial \mathbf{x_{ij}} \bbeta}{\partial \beta_{1}}\\
\vdots\\
\frac{\partial \mu_{ij}(\bbeta)}{\partial\eta_{ij}(\bbeta)}* \frac{\partial \mathbf{x_{ij}} \bbeta}{\partial \beta_{p}} 
\end{pmatrix} = \frac{\partial \mu_{ij}(\bbeta)}{\partial\eta_{ij}(\bbeta)} * \begin{pmatrix}
x_{{ij}_1}\\
x_{{ij}_2}\\
\vdots\\
x_{{ij}_p} 
\end{pmatrix} = \frac{\partial \mu_{ij}(\bbeta)}{\partial\eta_{ij}(\bbeta)} * \mathbf{x_{ij}} \in \mathbb{R}^{p}
$$
$$\nabla \sigma_{ij}^2(\bbeta) = \frac{ \partial \sigma_{ij}^2(\bbeta)}{ \partial \mu_{ij}(\bbeta)} \frac{\partial \mu_{ij}(\bbeta)}{\partial \eta_{ij}(\bbeta)} \frac{\partial \eta_{ij}(\bbeta)}{\partial \bbeta} =\frac{ \partial \sigma_{ij}^2(\bbeta)}{ \partial \mu_{ij}(\bbeta)} \frac{\partial \mu_{ij}(\bbeta)}{\partial \eta_{ij}(\bbeta)}  * \mathbf{x_{ij}} \in \mathbb{R}^{p}$$

For the Gaussian base model, since the identity function is the appropriate canonical link, we have that $\mu_{ij}(\bbeta) = \eta_{ij}(\bbeta) =\mathbf{ x_{ij}} \bbeta = x_{{ij}_1} * \beta_1 + ... + x_{{ij}_p} *\beta_p$ where $\mathbf{x_{ij}}$ denotes the vector of $p$ covariate values for $j^{th}$ measurement of the $i^{th}$ person. 

$$
\nabla \mu_{ij}(\bbeta) =  \frac{\partial \eta_{ij}(\bbeta)}{\partial\bbeta} =  \begin{pmatrix}
\frac{\partial \mathbf{x_{ij}}  \bbeta }{\partial \beta_{1}}\\
\vdots\\
\frac{\partial \mathbf{x_{ij}} \bbeta }{\partial \beta_{p}} 
\end{pmatrix} = \begin{pmatrix}
X_{{ij}_1}\\
X_{{ij}_2}\\
\vdots\\
X_{{ij}_p} 
\end{pmatrix} = \mathbf{x_{ij}} \in \mathbb{R}^{p}
$$

$$\nabla \sigma_{ij}^2(\bbeta) = \frac{ \partial \sigma_{ij}^2(\bbeta)}{ \partial \mu_{ij}(\bbeta)} \frac{\partial \mu_{ij}(\bbeta)}{\partial \eta_{ij}(\bbeta)} \frac{\partial \eta_{ij}(\bbeta)}{\partial \bbeta} = 0 * 1 * \mathbf{x_{ij}} = \textbf{0} \in \mathbb{R}^{p}$$

In the table below, we derive the same quantities for the Normal, Poisson, Bernoulli and negative binomial distributions, under the appropriate canonical link function. The details of the derivation for the above table is below.
\begin{center}
\begin{adjustbox}{width=1\textwidth}
\begin{tabular}{ |c|c|c|c|c|c| } 
\hline
Distribution & $g(\mu_{ij}(\bbeta)) = \eta_{ij}(\bbeta)$ & $\mu_{ij}(\bbeta) \in \mathbb{R}$ & $\sigma_{ij}^2(\bbeta) \in \mathbb{R}$ & $\nabla \mu_{ij}(\bbeta)  \in \mathbb{R}^{p}$ & $\nabla \sigma_{ij}^2(\bbeta) \in \mathbb{R}^{p}$ \\
\hline
Normal & Identity Link & $\eta_{ij}(\bbeta)$  & $\sigma_{ij}^2$ & $\mathbf{x_{i}}$& $\textbf{0}$ \\\hline
Poisson & Log Link& $e^{\eta_{ij}(\bbeta)}$ & $\mu_{ij}(\bbeta)$ & $e^{\eta_{ij}(\bbeta)} * \mathbf{x_{i}}$ & $e^{\eta_{ij}(\bbeta)}* \mathbf{x_{ij}}$\\\hline
Bernoulli & Logit Link & $\frac{e^{\eta_{ij}(\bbeta)}}{1+ e^{\eta_{ij}(\bbeta)}}$ & $\frac{e^{\eta_{ij}(\bbeta)}}{(1+ e^{\eta_{ij}(\bbeta)})^2}$& $\frac{ e^{\eta_{ij}(\bbeta)})}{(1 + e^{\eta_{ij}(\bbeta)})^2} * \mathbf{x_{ij}} $ & $\frac{e^{\eta_{ij}(\bbeta)}(1 -  e^{\eta_{ij}(\bbeta)})^2}{(1 + e^{\eta_{ij}(\bbeta)})^2} * \mathbf{x_{i}}$\\\hline
 Negative Binomial & Log Link & $e^{\eta_{ij}(\bbeta)}$ & $e^{\eta_{ij}(\bbeta)} * (1 + \frac{e^{\eta_{ij}(\bbeta)}}{r}) $ & $e^{\eta_{ij}(\bbeta)} * \mathbf{x_{i}}$
 & $(\frac{e^{\eta_{ij}(\bbeta)}}{r} + (1 + \frac{e^{\eta_{ij}(\bbeta)}}{r})) * e^{\eta_{ij}(\bbeta)} * \mathbf{x_{i}}$\\
\hline
\end{tabular}
\end{adjustbox}
\end{center}

\subsubsection{MM Algorithm for the  VC Model Parameters}\label{sec:mm_for_variance_component_estim}

One can construct an iterative MM algorithm for updating $\btheta$ holding $\bbeta$ fixed. There exists a substantial literature on the MM principle for optimization \citep{lange2000optimization,lange2016mm,zhou2019mm}. The idea in maximization is to concoct a surrogate function $g(\btheta \mid \btheta_r)$ that is easy to maximize and hugs the objective $f(\btheta)$ tightly. Here $\btheta_r$ is the current value of $\btheta$. Construction of the surrogate is guided by two minorization requirements:
\begin{eqnarray*}
f(\btheta) &\ge& g(\btheta \mid \btheta_r) \quad \forall \:\: \btheta \quad \text{(dominance condition)} \\
f(\btheta_r) &=& g(\btheta_r \mid \btheta_r) \quad \text{(tangent condition)}.
\end{eqnarray*}
The next iterate is determined by
$\btheta_{r+1} = \argmax \, g(\btheta \mid \btheta_r)$.
The MM principle guarantees that $f(\btheta_{r+1})\ge f(\btheta_r)$, with strict inequality being the rule. In practice, minorization is carried out piecemeal on a sum of terms defining the objective.

To update the variance components $\btheta = \{\theta_k, k \in [1, m]\},$ the relevant part of the loglikelihood is
\begin{eqnarray}
f(\btheta) & = &
\sum_{i=1}^n \ln(1+\btheta^\top\bb_i) - \sum_{i=1}^n \ln(1+\btheta^\top \bc_i) 
\end{eqnarray}
by defining the vectors $\bb_i$ and $\bc_i$ with nonnegative components
\begin{eqnarray*}
\bb_{ik} & = & \frac{1}{2}r_i(\bbeta)^\top \bOmega_{ik}
r_i(\bbeta) \\
\bc_{ik} & = & \frac{1}{2}\tr(\bOmega_{ik}).
\end{eqnarray*} 
Jensen's inequality gives the minorization for the first term
\begin{eqnarray*}
\sum_{i=1}^n \ln(1+\btheta^\top\bb_i) & \ge &
\sum_{i=1}^n \frac{1}{1+\btheta_r^\top\bb_i} \ln \left(\frac{1+\btheta_r^\top\bb_i}{1}1\right)\\
&  & +  \sum_{i=1}^n\sum_j
\frac{\theta_{rj}b_{ij}}{1+\btheta_r^\top\bb_i} \ln \left(\frac{1+\btheta_r^\top\bb_i}
{\theta_{rj}b_{ij}} \theta_jb_{ij}\right). 
\end{eqnarray*}
For the second term, we capitalize on the convexity of $-\ln(s)$. The supporting hyperplane inequality implies the linear minorization
\begin{eqnarray*}
-\sum_{i=1}^n\ln(1+\btheta^\top \bc_i) & \ge &
-\sum_{i=1}^n \frac{1}{1+\btheta_r^\top\bc_i}(1+\btheta^\top\bc_i-1 - \btheta_r^\top\bc_i). 
\end{eqnarray*}
The sum of these two minorizations constitutes the overall minorization
$g(\btheta \mid \btheta_r)$. The stationary condition $\nabla g(\btheta \mid \btheta_r) = {\bf 0}$ can be solved to yield the updates
\begin{eqnarray*}
\theta_{r+1,j} & = & \theta_{rj} \frac{\sum_{i=1}^n
\frac{b_{ri}}{1+\btheta_r^\top\bb_i}}{\sum_{i=1}^n \frac{c_{ij}}{1+\btheta_r^\top \bc_i}}.
\end{eqnarray*}
Note that the update $\theta_{r+1,j}$ remains nonnegative if $\theta_{rj}$ is nonnegative and equals 0 if and only if $\theta_{rj}=0$. However, convergence of $\theta_{rj}$ to 0 is possible. More importantly, the MM updates drive the loglikelihood uphill. 

\subsubsection{Quasi-Newton Algorithm} \label{sec:grad_of_theta}

Alternatively, we can estimate the mean and variance parameters joinly using the Quasi-Newton algorithm.

\paragraph{Score and Hessian}
For the AR(1) model, $\btheta = \{\sigma^2, \rho\},$ the score (gradient of loglikelihood function) is
\begin{eqnarray*}
	\nabla_{\sigma^2}\mathcal{L} &=& - \sum_{i=1}^n \frac{\frac{d_{i}}{2}}{1 + \frac{d_{i} \sigma^2}{2}} + \sum_{i=1}^n \frac{\frac{1}{2}  r_i(\bbeta)^\top \bV_i(\rho) r_i(\bbeta)}{1 + \frac{\sigma^2}{2}  r_i(\bbeta)^\top \bV_i(\rho) r_i(\bbeta)}\\
	\nabla_{\rho} \mathcal{L}&=& \sum_{i=1}^n \frac{1}{1 + \frac{\sigma^2}{2}  r_i(\bbeta)^\top \bV_i(\rho) r_i(\bbeta)} * \frac{\sigma^2}{2}  r_i(\bbeta)^\top \nabla \bV_i(\rho) r_i(\bbeta).
\end{eqnarray*}
The approximate Hessian is
\begin{eqnarray*}
	d^2_{\sigma^2}
	\mathcal{L}&=& \sum_{i=1}^n \frac{(\frac{d_{i}}{2})^2}{(1 + \frac{d_{i}}{2}\sigma^2)^2} - \sum_{i=1}^n \frac{(\frac{1}{2} r_i(\bbeta)^\top \bV_i(\rho) r_i(\bbeta)
	)^2}{(1 + \frac{\sigma^2}{2}  r_i(\bbeta)^\top \bV_i(\rho) r_i(\bbeta))^2}\\
	d^2_{\rho}\mathcal{L} &=& \sum_{i=1}^n  \frac{1}{1 + \frac{\sigma^2}{2}  r_i(\bbeta)^\top \bV_i(\rho) r_i(\bbeta)} * \frac{\sigma^2}{2}  r_i(\bbeta)^\top d^2 \bV_i(\rho) r_i(\bbeta)\\
	&-& \sum_{i=1}^n \frac{1}{(1 + \frac{\sigma^2}{2}  r_i(\bbeta)^\top \bV_i(\rho) r_i(\bbeta))^2} * \Big( \frac{\sigma^2}{2}  r_i(\bbeta)^\top \nabla \bV_i(\rho) r_i(\bbeta) \Big)^2,
	\end{eqnarray*}
	
For the CS model, $\btheta = (\sigma^2, \rho),$ and the gradient is
\begin{eqnarray*}
	\nabla_{\sigma^2}\mathcal{L} &=& - \sum_{i=1}^n \frac{\frac{d_{i}}{2}}{1 + \frac{d_{i} \sigma^2}{2}} + \sum_{i=1}^n \frac{\frac{1}{2}  r_i(\bbeta)^\top \bV_i(\rho) r_i(\bbeta)}{1 + \frac{\sigma^2}{2}  r_i(\bbeta)^\top \bV_i(\rho) r_i(\bbeta)}\\
	\nabla_{\rho} \mathcal{L}&=& \sum_{i=1}^n \frac{1}{1 + \frac{\sigma^2}{2}  r_i(\bbeta)^\top \bV_i(\rho) r_i(\bbeta)} * \frac{\sigma^2}{2}  r_i(\bbeta)^\top \nabla \bV_i(\rho) r_i(\bbeta).
\end{eqnarray*}
The approximate Hessian is
\begin{eqnarray*}
	d^2_{\sigma^2}
	\mathcal{L}&=& \sum_{i=1}^n \frac{(\frac{d_{i}}{2})^2}{(1 + \frac{d_{i}}{2}\sigma^2)^2} - \sum_{i=1}^n \frac{(\frac{1}{2} r_i(\bbeta)^\top \bV_i(\rho) r_i(\bbeta)
	)^2}{(1 + \frac{\sigma^2}{2}  r_i(\bbeta)^\top \bV_i(\rho) r_i(\bbeta))^2}\\
	d^2_{\rho}\mathcal{L} &=& - \sum_{i=1}^n \frac{1}{(1 + \frac{\sigma^2}{2}  r_i(\bbeta)^\top \bV_i(\rho) r_i(\bbeta))^2} * \Big( \frac{\sigma^2}{2}  r_i(\bbeta)^\top \nabla \bV_i(\rho) r_i(\bbeta) \Big)^2,
	\end{eqnarray*}
where $\nabla \bV_i(\rho)$ and $d^2 \bV_i(\rho)$ are, respectively, the element-wise first and second derivatives of the matrix $\bV_i(\rho)$ with respect to $\rho$.

For the VM model the gradient is
\begin{eqnarray*}
	\nabla_{\boldsymbol{\theta}} f(\btheta) &=& \sum_{i=1}^n \frac{1}{(1+\btheta^\top\bb_i)} * \bb_i - \sum_{i=1}^n \frac{1}{1+\btheta^\top \bc_i} * \bc_i.
\end{eqnarray*}
The approximate Hessian is
\begin{eqnarray*}
	d^2_{\boldsymbol{\theta},\boldsymbol{\theta}} f(\btheta) 
	&=& - \sum_{i=1}^n \frac{1}{(1+\btheta^\top\bb_i)} * \bb_i \bb_i^\top + \sum_{i=1}^n \frac{1}{1+\btheta^\top \bc_i} * \bc_i \bc_i^\top.
\end{eqnarray*}

\subsubsection{Nuisance parameter estimation for Negative Binomial base}\label{sec:neg_bin_fit}

To estimate the nuisance parameter $r$ in a Negative Binomial model, we use maximum likelihood. Because we are dealing with 1 parameter optimization, Newton's method is a good candidate due to its quadratic rate of convergence. The full loglikelihood is
\begin{align*}
    -\sum_{i=1}^n\ln\left(1+\frac{1}{2}\tr(\bGamma_i)\right)+ \sum_{i=1}^n\sum_{i=j}^{d_i}\ln f_{ij}(y_{ij} \ | \ \bbeta) + \sum_{i=1}^n \ln\left(1 + \frac{1}{2}r_i(\bbeta)^\top\bGamma_ir_i(\bbeta)\right)
\end{align*}
where only the 2nd and 3rd term depends on $r$. First consider the 2nd term. Because $\mu_{ij} = \frac{r(1-p_{ij})}{p_{ij}}, p_{ij} = \frac{r}{r+\mu_{ij}}$, the 2nd term of the loglikelihood is
\begin{align*}
    &\sum_{i=1}^n\sum_{j=1}^{d_i} \ln \left[\binom{y_{ij}+r-1}{y_{ij}} p_{ij}^r (1-p_{ij})^{y_{ij}} \right]\\
    =& \sum_{i=1}^n\sum_{j=1}^{d_i}\ln{\binom{y_{ij}+r-1}{y_{ij}}} + r\ln\left(\frac{r}{\mu_{ij}+r}\right) + y_{ij}\ln\left(\frac{\mu_{ij}}{\mu_{ij}+r}\right)\\
    =& \sum_{i=1}^n\sum_{j=1}^{d_i}\ln((y_{ij}+r-1)!)-\ln(y_{ij}!) - \ln((r-1)!) + r\ln(r) - (r+y_{ij})\ln(\mu_{ij}+r) + y_{ij}\ln(\mu_{ij})
\end{align*}
Let $\Psi^{(0)}$ be the digamma function and $\Psi^{(1)}$ the trigamma function, then the first and second derivative is
\begin{align*}
    &\sum_{i=1}^n\sum_{j=1}^{d_i} \Psi^{(0)}(y_{ij}+r) - \Psi^{(0)}(r) + 1 + \ln(r) - \frac{r + y_{ij}}{\mu_{ij}+r} - \ln(\mu_{ij}+r),\\
    &\sum_{i=1}^n\sum_{j=1}^{d_i}\Psi^{(1)}(y_{ij}+r) - \Psi^{(0)}(r) + \frac{1}{r} - \frac{2}{\mu_{ij} + r} + \frac{r + y_{ij}}{(\mu_{ij} + r)^2}.
\end{align*}
Now consider the 3rd term of the full loglikelihood. First recall 
\begin{align*}
    \bD_i 
    &= diagonal(\sqrt{var(\by_i)})\\
    var(y_{ij}) 
    &= \frac{r(1-p_{ij})}{p^2_{ij}} = \frac{e^{\eta_{ij}}(e^{\eta_{ij}}+r)}{r}.
\end{align*}
Repeated application of the chain rule leads to
\begin{align*}
    \frac{d}{dr}\ln\left(1 + \frac{1}{2}r_i(\bbeta)^\top\bGamma_ir_i(\bbeta)\right)
    &= \sum_{i=1}^n\frac{r_i(\bbeta)^\top\bGamma_i dr_i}{1 + \frac{1}{2}r_i(\bbeta)^\top\bGamma_ir_i(\bbeta)}\\
    \frac{d^2}{dr^2}\ln\left(1 + \frac{1}{2}r_i(\bbeta)^\top\bGamma_ir_i(\bbeta)\right)
    &= \sum_{i=1}^n\frac{-[r_i(\bbeta)^\top\bGamma_i dr_i]^2}{[1 + \frac{1}{2}r_i(\bbeta)^\top\bGamma_ir_i(\bbeta)]^2} + \frac{d\br(\bbeta)^\top\bGamma_i dr_i(\bbeta) + \br(\bbeta)^\top\bGamma_i dr_i^2(\bbeta)}{1 + \frac{1}{2}r_i(\bbeta)^\top\bGamma_ir_i(\bbeta)}
\end{align*}
where
\begin{align*}
    r_i(\bbeta) 
    &= \bD_i^{-1}(\by_i - \bmu_i)\\
    dr_i(\bbeta)
    &= -\bD_i^{-1}d\bD_i\bD_i^{-1}(\by_i-\bmu_i)\\
    dr_i^2(\bbeta)
    &= [2\bD_i^{-1}d\bD_i\bD_i^{-1}d\bD_i\bD_i^{-1} - \bD_i^{-1} d^2\bD_i\bD_i^{-1}](\by_i - \bmu_i)\\
    d\bD_i
    &= diagonal\left(\frac{d}{dr}\sqrt{\frac{e^{\eta_{ij}}(e^{\eta_{ij}}+r)}{r}}\right) = diagonal\left(\frac{-e^{2\eta_{ij}}}{2r^{1.5}\sqrt{e^{\eta_{ij}}(e^{\eta_{ij}}+r)}}\right)\\
    d^2\bD_i
    &= diagonal\left(\frac{e^{3\eta}}{4r^{1.5}(e^\eta(e^\eta+r))^{1.5}}+ \frac{3e^{2\eta}}{4r^{2.5}(e^\eta(e^\eta+r))^{0.5}}\right).
\end{align*}
Note we used the identity $df(\bX)^{-1} = -f(\bX)^{-1}df(\bX)f(\bX)^{-1}$ for obtaining $dr_i(\bbeta)$ and for obtaining $d\br^2_i(\bbeta)$, chain rule implies
\begin{align*}
    &d(f(\bX)^{-1}df(\bX)f(\bX)^{-1})\\
    =& [-f(\bX)^{-1}df(\bX)f(\bX)^{-1}]df(\bX)f(\bX)^{-1} + f(\bX)^{-1} d(df(\bX)f(\bX)^{-1})\\
    =& -f(\bX)^{-1}df(\bX)f(\bX)^{-1}df(\bX)f(\bX)^{-1} + \\
    & f(\bX)^{-1}\left[df(\bX)(-f(\bX)^{-1}df(\bX)f(\bX)^{-1}) + d^2f(\bX)f(\bX)^{-1}\right]\\
    =& -2f(\bX)^{-1}df(\bX)f(\bX)^{-1}df(\bX)f(\bX)^{-1} + f(\bX)^{-1}d^2f(\bX)f(\bX)^{-1}
\end{align*}
In summary, we update the nuisance parameter $r$ using Newton's update
\begin{align*}
    r_{n+1} = r_n - \frac{\frac{d}{dr}L(r \ | \ \bmu, \bGamma, \by) }{\frac{d^2}{dr^2}L(r \ | \ \bmu, \bGamma, \by) }
\end{align*}
where 
\begin{eqnarray*}
    \frac{d}{dr} L(r \ | \ \bmu, \bGamma, \by)
    &=& \sum_{i=1}^n\sum_{j=1}^{d_i} \Psi^{(0)}(y_{ij}+r) - \Psi^{(0)}(r) + 1 + \ln(r) - \frac{r + y_{ij}}{\mu_{ij}+r} - \ln(\mu_{ij}+r)\\
    &&+ \sum_{i=1}^n\frac{r_i(\bbeta)^\top\bGamma_i dr_i}{1 + \frac{1}{2}r_i(\bbeta)^\top\bGamma_ir_i(\bbeta)} \\
    \frac{d^2}{dr^2} L(r \ | \ \bmu, \bGamma, \by)
    &=& \sum_{i=1}^n\sum_{j=1}^{d_i}\Psi^{(1)}(y_{ij}+r) - \Psi^{(0)}(r) + \frac{1}{r} - \frac{2}{\mu_{ij} + r} + \frac{r + y_{ij}}{(\mu_{ij} + r)^2}\\
    &&-\sum_{i=1}^n\frac{[r_i(\bbeta)^\top\bGamma_i dr_i]^2}{[1 + \frac{1}{2}r_i(\bbeta)^\top\bGamma_ir_i(\bbeta)]^2} + \frac{dr_i(\bbeta)^\top\bGamma_i dr_i(\bbeta) + r_i(\bbeta)^\top\bGamma_i dr_i^2(\bbeta)}{1 + \frac{1}{2}r_i(\bbeta)^\top\bGamma_ir_i(\bbeta)}
\end{eqnarray*}
For stability, we need to (1) perform line-search and (2) set the second derivative equal to 1 if it is negative. By default, we allow for a maximum of $10$ block  iterations; In each block iteration, we allow for a maximum of $15$ iterations for the quasi-newton update of $\bbeta$ and $\btheta,$ and a maximum of $10$ newton iterations for the update of $r.$

\subsubsection{Compound Symmetric Covariance:}
Under the Compound Symmetric (CS) parameterization of $\bGamma_i$,
\begin{eqnarray*}
\bGamma_i &=& \sigma^2 \times \Big[ \rho {\bf 1}_{d_{i}}{\bf 1}_{d_{i}}^\top + (1 - \rho) \boldsymbol{I_{d_i}} \Big]\\
&=& \sigma^2 \times \bV_i(\rho)
\end{eqnarray*}
\paragraph{Bounding Correlation Parameter}
To ensure that the covariance matrix $\bGamma_i$ is positive semi-definite, we will focus on $\bV_i(\rho)$ and use an eigenvalue argument to bound $\rho \in (-\frac{1}{d_i - 1}, 1)$. Let $\bv$ be a vector of dimension $d_i$ such that $<\bv, \bv> = 1.$ We will find the conditions on $\rho$ such that $\bv^\top \bV_i(\rho) \bv \geq 0.$

\begin{eqnarray*}
\bv^\top \bV_i(\rho) \bv &=& \bv^\top \Big[ \rho {\bf 1}_{d_{i}}{\bf 1}_{d_{i}}^\top + (1 - \rho) \boldsymbol{I_{d_i}} \Big] \bv\\
&=& \rho \bv^\top {\bf 1}_{d_{i}}{\bf 1}_{d_{i}}^\top \bv + (1 - \rho) \bv^\top \bv\\
&=& \rho ({\bf 1}_{d_{i}}^\top \bv)^2 + 1 - \rho\\
&=& \rho \Big( ({\bf 1}_{d_{i}}^\top \bv)^2 - 1 \Big) + 1\\
& \geq & 0
\end{eqnarray*}
Now solving for $\rho$ and using the Cauchy-Schwartz Inequality, we get 
\begin{eqnarray*}
\rho & \geq & \frac{-1}{\Big( ({\bf 1}_{d_{i}}^\top \bv)^2 - 1 \Big) } \\
& \geq & \frac{-1}{({\bf 1}_{d_{i}}^\top {\bf 1}_{d_{i}}) * (\bv^\top \bv) - 1}\\
&=& \frac{-1}{d_i - 1}
\end{eqnarray*}
In summary, the CS longitudinal model has an additional constraint on the parameter space: $\rho \ge \frac{-1}{d_i - 1}$. 

\subsubsection{Gradients and Hessians of residual function}\label{sec:nabla_r}

Here we give an example to calculate the explicit expressions for $\nabla r_{ij}(\bbeta)$ and $\nabla^2 r_{ij}(\bbeta)$. For simplicity, we consider the longitudinal Bernoulli model  with logit link function. In the main text, we saw that the chain and product rules gives the following expressions
\begin{eqnarray*}
    r_{ij}(\bbeta) &=& \frac{y_{ij} - \mu_{ij}}{\sqrt{\sigma_{ij}^2(\bbeta)}}, \quad (r_{ij}(\bbeta) \in \mathbb{R} \text{ denotes sample $i$'s residual at time $j$}, \br_i(\bbeta) \in \mathbb{R}^{d_i})\\
    \nabla r_{ij}(\bbeta) &=& -\frac{1}{\sqrt{\sigma^2_{ij}(\bbeta)}}\nabla \mu_{ij}(\bbeta) - \frac{1}{2}\frac{y_{ij} - \mu_{ij}(\bbeta)}{\sigma_{ij}^3(\bbeta)}\nabla \sigma_{ij}^2(\bbeta)\\
    && \left(\nabla r_{ij}(\bbeta) \in \mathbb{R}^{p} \text{ is $j$th column of }\nabla\br_i(\bbeta)\in \mathbb{R}^{p \times d_i}\right)\\
    \nabla^2 r_{ij}(\bbeta)
    &=& \frac{-1}{\sqrt{\sigma_{ij}^2(\bbeta)}}\nabla^2\mu_{ij}(\bbeta) + \frac{1}{2}\frac{1}{\sigma_{ij}^3(\bbeta)}\nabla \sigma_{ij}^2(\bbeta) \nabla\mu_{ij}(\bbeta)^\top - \left(\frac{1}{2}\frac{y_{ij} - \mu_{ij}(\bbeta)}{\sigma_{ij}^3(\bbeta)}\nabla^2\sigma_{ij}^2(\bbeta)\right)\\
    && \qquad - \nabla\left\{\frac{1}{2}\frac{y_{ij}-\mu_{ij}(\bbeta)}{(\sigma_{ij}^2(\bbeta))^{3/2}}\right\}\nabla \sigma^2_{ij}(\bbeta)^\top\\
    &=& \frac{-1}{\sqrt{\sigma_{ij}^2(\bbeta)}}\nabla^2\mu_{ij}(\bbeta) + \frac{1}{2}\frac{1}{\sigma_{ij}^3(\bbeta)}\nabla \sigma_{ij}^2(\bbeta) \nabla\mu_{ij}(\bbeta)^\top - \left(\frac{1}{2}\frac{y_{ij} - \mu_{ij}(\bbeta)}{\sigma_{ij}^3(\bbeta)}\nabla^2\sigma_{ij}^2(\bbeta)\right) \\
    && \qquad -\frac{1}{2}\left\{\frac{-1}{(\sigma_{ij}^2(\bbeta))^{3/2}}\nabla\mu_{ij}(\bbeta) - \frac{3}{2}\frac{y_{ij}-\mu_{ij}(\bbeta)}{(\sigma_{ij}^2(\bbeta))^{5/2}}\nabla\sigma_{ij}^2(\bbeta)\right\}\nabla\sigma_{ij}^2(\bbeta)^\top\\
\end{eqnarray*}
    To calculate these quantities, we need expressions for $\nabla \mu_{ij}(\bbeta) \in \mathbb{R}^{p}, \nabla^2 \mu_{ij}(\bbeta) \in \mathbb{R}^{p\times p}, \nabla \sigma_{ij}^2(\bbeta) \in \mathbb{R}^{p},$ and $\nabla^2 \sigma_{ij}^2(\bbeta) \in \mathbb{R}^{p \times p}$ (note these gradients are evaluated with respect to $\bbeta$). Since $\bmu_i = \bg^{[-1]}(\boldsymbol{\eta}_i) =  \bg^{[-1]}(\bX_{i}\bbeta)$, by chain rule
\begin{align*}
    \nabla \mu_{ij}(\bbeta) &= \frac{\partial \mu_{ij}}{\partial \eta_{ij}}\frac{\partial \eta_{ij}}{\partial \bbeta} = \frac{\partial \mu_{ij}}{\partial \eta_{ij}}\bx_{ij} \quad (\bx_{ij} \in \mathbb{R}^p \text{ are covariates for sample $i$ at time $j$})\\
    \nabla^2 \mu_{ij}(\bbeta) &= \frac{\partial }{\partial \bbeta}\left( \frac{\partial \mu_{ij}}{\partial \eta_{ij}}\bx_{ij}\right) = \frac{\partial^2 \mu_{ij}^2}{\partial \eta_{ij}^2}\frac{\partial \eta_{ij}}{\partial \bbeta}\bx_{ij} =  \frac{\partial^2 \mu_{ij}^2}{\partial \eta_{ij}^2}\bx_{ij}\bx_{ij}^\top \in \mathbb{R}^{p \times p}
\end{align*}
Here $\frac{\partial\mu_{ij}}{\partial \eta_{ij}} = \frac{\partial [g^{[-1]}(\bx_{ij}^\top\bbeta)]}{\partial [\bx_{ij}^\top\bbeta]} \in \mathbb{R}$ is just the derivative of the inverse link function evaluated at the linear predictor $\eta_{ij} = \bx_{ij}^\top\bbeta \in \mathbb{R}$. This is implemented for various link functions as \href{https://github.com/JuliaStats/GLM.jl/blob/master/src/glmtools.jl#L176}{mueta} in GLM.jl which we call internally. Following its definition, we also implement \texttt{mueta2} for evaluating $\frac{\partial^2\mu_{ij}}{\partial \eta_{ij}^2} \in \mathbb{R}.$\\
\\
To compute $\nabla \sigma_{ij}^2(\bbeta)$ and $\nabla^2 \sigma_{ij}^2(\bbeta)$, note variance is typically a function of the mean, that is,  
\begin{align*}
    \nabla \sigma^2_{ij}(\bbeta)
    &= \frac{\partial \sigma_{ij}^2}{\partial\mu_{ij}}\frac{\partial\mu_{ij}}{\partial\eta_{ij}}\frac{\partial\eta_{ij}}{\partial\bbeta} = \frac{\partial \sigma_{ij}^2}{\partial\mu_{ij}}\frac{\partial\mu_{ij}}{\partial\eta_{ij}} \bx_{ij} \in \mathbb{R}^p\\
    \nabla^2 \sigma^2_{ij}(\bbeta)
    &= \frac{\partial^2 \sigma^2_{ij}}{\partial \mu^2_{ij}}\frac{\partial \mu_{ij}}{\partial \eta_{ij}}\frac{\partial \eta_{ij}}{\partial \bbeta}\frac{\partial\mu_{ij}}{\partial \eta_{ij}}\bx_{ij} + \frac{\partial \sigma^2_{ij}}{\partial \mu_{ij}}\frac{\partial^2 \mu_{ij}}{\partial \eta_{ij}^2}\frac{\partial \eta_{ij}}{\partial \bbeta}\bx_{ij}\\
    &= \left(\frac{\partial^2\sigma^2_{ij}}{\partial\mu_{ij}^2}\left(\frac{\partial\mu_{ij}}{\partial\eta_{ij}}\right)^2 + \frac{\partial\sigma_{ij}^2}{\partial\mu_{ij}}\frac{\partial^2\mu_{ij}}{\partial\eta_{ij}^2}\right)\bx_{ij}\bx_{ij}^\top \in \mathbb{R}^{p \times p}
\end{align*}
Terms $\frac{\partial \sigma^2_{ij}}{\partial \mu_{ij}}$ and $\frac{\partial^2 \sigma^2_{ij}}{\partial \mu^2_{ij}}$ are distribution-dependent, and we implement them as \texttt{sigmamu} and \texttt{sigmamu2} in our software. 

For Bernoulli distribution, if $\mu$ is the mean, then $\mu(1 - \mu)$ is its variance. Given the logit link, we have
\begin{align*}
    \nabla \mu_{ij}
    &= \frac{\partial \mu_{ij}}{\partial \eta_{ij}}\frac{\partial \eta_{ij}}{\partial\bbeta} = \frac{e^{-\eta_{ij}}}{(1 + e^{-\eta_{ij}})^2}\bx_{ij} \quad (\text{expression for } \frac{\partial \mu_{ij}}{\partial \eta_{ij}} \text{ uses the logit link})\\
    \nabla \sigma^2_{ij} 
    &= \frac{\partial \sigma^2_{ij}}{\partial \mu_{ij}}\frac{\partial \mu_{ij}}{\partial \eta_{ij}}\frac{\partial \eta_{ij}}{\partial \bbeta} = \frac{\partial(\mu_{ij}(1 - \mu_{ij}))}{\partial\mu_{ij}}\frac{e^{-\eta_{ij}}}{(1 + e^{-\eta_{ij}})^2}\bx_{ij} = (1 - 2\mu_{ij})\frac{e^{-\eta_{ij}}}{(1 + e^{-\eta_{ij}})^2}\bx_{ij}.
\end{align*}
Differentiating again yields the expressions
\begin{align*}
    \nabla^2 \mu_{ij}^2
    &= \left(\frac{-e^{-\eta_{ij}}}{(1 + e^{-\eta_{ij}})^2} + \frac{2e^{-2\eta_{ij}}}{(1+e^{-\eta_{ij}})^3}\right)\bx_{ij}\bx_{ij}^\top\\
    \nabla^2 \sigma_{ij}^2
    &= -2\bx_{ij}\bx_{ij}^\top.
\end{align*}
For the special Gaussian base, the variance is parametrized by $\tau = \sigma^{-2}$ (see section \ref{sec:gaussian_quasi_copulas}) which is a separate parameter from the mean. Thus, $\frac{\partial \sigma^2}{\partial \mu} = 0$. Given the identity link $\bmu_i(\bbeta) = \boldsymbol{\eta}_i = \bX_i\bbeta$, 
\begin{align*}
    [\nabla \br_{i}(\bbeta)]_j = \frac{-1}{\sigma} \bx_{ij} = -\sqrt{\tau} \bx_{ij}
\end{align*}

\subsection{Special case: Gaussian quasi-copulas}\label{sec:gaussian_quasi_copulas}

This section considers the special case of Gaussian base in the quasi-copula framework, and presents detailed derivations of data generation and estimation methods. The joint density of $\mathbf{y} \in \mathbb{R}^d$ is
\begin{eqnarray}
	\left(c + \frac 12 \text{tr} \boldsymbol{\Gamma} \right)^{-1} \left(\frac{1}{\sqrt{2\pi} \sigma_0}\right)^d e^{-\frac{\|\mathbf{y} - \boldsymbol{\mu}\|_2^2}{2\sigma_0^2}} \left[c + \frac{1}{2\sigma_0^2} (\mathbf{y} - \boldsymbol{\mu})^\top \boldsymbol{\Gamma} (\mathbf{y} - \boldsymbol{\mu}) \right]. \label{eq:gaussian-copula-density}
\end{eqnarray}
The parameter $c \ge 0$ tips the balance between the independent and dependent components.


\subsubsection{Moments}
In the Gaussian case, we have 
\begin{eqnarray*}
	\mathbb{E}(y_i) &=& \mu_i \\
	\mathbf{Var}(y_i) &=& \sigma_0^2 \left( 1 +  \frac{\gamma_{ii}}{c + \frac 12 \text{tr}(\boldsymbol{\Gamma})} \right) \\
	\mathbf{Cov}(y_i, y_j) &=& \sigma_0^2 \frac{\gamma_{ij}}{c + \frac 12 \text{tr}(\boldsymbol{\Gamma})}, \\
	\mathbf{Cor}(y_i, y_j) &=& \frac{\gamma_{ij}}{\sqrt{(c + \frac 12 \text{tr}(\boldsymbol{\Gamma}) + \gamma_{ii})(c + \frac 12 \text{tr}(\boldsymbol{\Gamma}) + \gamma_{jj})}}.
\end{eqnarray*}
In summary, 
$$
\mathbf{Cov} (\mathbf{y}) = \sigma_0^2 \left[ \mathbf{I} + \left(\frac{1}{c + \frac 12 \text{tr} \boldsymbol{\Gamma}} \right) \boldsymbol{\Gamma} \right].
$$
In the special case of $\boldsymbol{\Gamma}=\sigma_1^2 \mathbf{I}$, we have
\begin{eqnarray*}
	\mathbf{Var}(y_i)  &=& \sigma_0^2 \left( 1 + \frac{\sigma_1^2}{c + \frac n2 \sigma_1^2} \right) \mathbf{I} \\
	\mathbf{Cov}(y_i, y_j) &=& 0,  \quad i \ne j.
\end{eqnarray*}
In the regression model, we would keep the variance $\sigma_0^2$ parameter for more flexibility in modeling the variance.

\paragraph{Random number generation}
If we are able to generate a residual vector $\mathbf{R}$ from the (standardized) Gaussian copula model
\begin{eqnarray*}
	\left[1 + \frac 12 \text{tr} (\boldsymbol{\Gamma}) \right]^{-1} \left(\frac{1}{\sqrt{2\pi}}\right)^d e^{-\frac{\|\mathbf{r}\|_2^2}{2}} \left(1 + \frac{1}{2} \mathbf{r}^\top \boldsymbol{\Gamma} \mathbf{r} \right),
\end{eqnarray*}
then $\mathbf{Y} = \sigma_0 \mathbf{R} + \boldsymbol{\mu}$ is a desired sample from density \eqref{eq:gaussian-copula-density}.


To generate a sample from the standardized Gaussian copula model, we first sample $R_1$ from its marginal distribution and then generate remaining components sequentially from the conditional distributions $R_k \mid R_1, \ldots, R_{k-1}$ for $k=2,\ldots,d$.
\begin{itemize}
\item To generate $R_1$ from its marginal density
\begin{eqnarray}
	\left[1 + \frac 12 \text{tr}(\boldsymbol{\Gamma}) \right]^{-1} \frac{1}{\sqrt{2\pi}} e^{- \frac{r_1^2}{2}} \left[1 + \frac{\gamma_{11}}{2} r_1^2 + \frac 12 \sum_{i=2}^d \gamma_{ii}\right], \label{eq:R1-marginal}
\end{eqnarray}
we recognize it as a mixture of three distributions $\text{Normal}(0,1)$, $\sqrt{\chi_3^2}$ and $- \sqrt{\chi_3^2}$ with mixing probabilities $\frac{1 + 0.5 \sum_{i=2}^d \gamma_{ii}}{1 + 0.5 \sum_{i=1}^d \gamma_{ii}}$, $ \frac{0.25 \gamma_{11}}{1 + 0.5 \sum_{i=1}^d \gamma_{ii}}$ and $ \frac{0.25 \gamma_{11}}{1 + 0.5 \sum_{i=1}^d \gamma_{ii}}$ respectively.
\item Next we consider generating $R_2$ from the conditional distribution $R_2 \mid R_1$.  Dividing the marginal distribution of $(R_1, R_2)$ 
\begin{eqnarray*}
	\left[1 + \frac 12 \text{tr}(\boldsymbol{\Gamma}) \right]^{-1} \left( \frac{1}{\sqrt{2\pi}} \right)^2 e^{- \frac{r_1^2 + r_2^2}{2}} \left(1 + \frac{\gamma_{22}}{2} r_2^2 + \gamma_{12} r_1 r_2 + \frac{\gamma_{11}}{2} r_1^2  + \frac 12 \sum_{i=3}^d \gamma_{ii} \right)
\end{eqnarray*}
by the marginal distribution of $R_1$ \eqref{eq:R1-marginal} yields the conditional density
\begin{eqnarray*}
	& & \frac{\frac{1}{\sqrt{2\pi}} e^{-\frac{r_2^2}{2}} \left(1 + \frac{\gamma_{11}}{2} r_1^2  + \frac 12 \sum_{i=3}^d \gamma_{ii} +  \gamma_{12} r_1 r_2  + \frac{\gamma_{22}}{2} r_2^2 \right)}{1 + \frac{\gamma_{11}}{2} r_1^2  + \frac 12 \sum_{i=2}^d \gamma_{ii}},
\end{eqnarray*}
which unfortunately is not a mixture of standard distributions. However we can evaluate its cumulative distribution function (CDF) 
\begin{eqnarray*}
	F(x) = \frac{\left(1 + \frac{\gamma_{11}}{2} r_1^2 + \frac 12 \sum_{i=3}^d \gamma_{ii}\right) \Phi(x) - \gamma_{12} r_1 \phi(x) + \frac{\gamma_{22}}{2} \left[\frac 12 + \frac{\text{sgn}(x)}{2} F_{\chi_3^2}(x^2)\right]}{1 + \frac{\gamma_{11}}{2} r_1^2  + \frac 12 \sum_{i=2}^d \gamma_{ii}}
\end{eqnarray*}
in terms of the density $\phi$ and CDF $\Phi$ of standard normal and the CDF $F_{\chi_3^2}$ of chi-squared distribution with degree of freedom 3. This suggests the inverse CDF approach. To generate one sample from $R_2 \mid R_1$, we draw a uniform variate $U$ and use nonlinear root finding to locate $R_2$ such that $F(R_2)=U$.

\item In general, the conditional distribution $R_k \mid R_1, \ldots, R_{k-1}$ has density
\begin{eqnarray*}
	& & \frac{\frac{1}{\sqrt{2\pi}} e^{-\frac{r_k^2}{2}} \left(1 + \frac 12 \mathbf{r}_{[k-1]}^\top \boldsymbol{\Gamma}_{[k-1],[k-1]} \mathbf{r}_{[k-1]} + \frac 12 \sum_{i=k+1}^n \gamma_{ii} +  (\sum_{i=1}^{k-1} r_i \gamma_{ik}) r_k  + \frac{\gamma_{kk}}{2} r_k^2 \right)}{1 +  \frac 12 \mathbf{r}_{[k-1]}^\top \boldsymbol{\Gamma}_{[k-1],[k-1]} \mathbf{r}_{[k-1]}  + \frac 12 \sum_{i=k}^d \gamma_{ii}}
\end{eqnarray*}
and CDF
\begin{eqnarray*}
	\frac{\left(1 + \frac 12 \mathbf{r}_{[k-1]}^\top \boldsymbol{\Gamma}_{[k-1],[k-1]} \mathbf{r}_{[k-1]} + \frac 12 \sum_{i=k+1}^d \gamma_{ii}\right) \Phi(x) - (\sum_{i=1}^{k-1} r_i \gamma_{ik}) \phi(x) + \frac{\gamma_{kk}}{2} \left[\frac 12 + \frac{\text{sgn}(x)}{2} F_{\chi_3^2}(x^2)\right]}{1 +  \frac 12 \mathbf{r}_{[k-1]}^\top \boldsymbol{\Gamma}_{[k-1],[k-1]} \mathbf{r}_{[k-1]}  + \frac 12 \sum_{i=k}^d \gamma_{ii}}.
\end{eqnarray*}
We apply the inverse CDF approach to sample $R_k$ given $R_1,\ldots,R_{k-1}$.
\end{itemize}

For a general GLM model, we need to sample from conditional densities of form $c f(y) (a_0 + a_1y + a_2y^2)$ where $a_i$, $i=1,2,3$, are constants and $c$ is a normalizing constant. For most continuous distributions, e.g., exponential, gamma, beta, chi-squared, and beta, the CDF can be expressed conveniently using special functions. 

\subsubsection{Parameter Estimation}
Suppose we have $n$ independent realizations $\by_i$ from the quasi-copula density. Each of these may be of different dimensions, $d_{i}$. Assuming the component distribution $\mathbf{y}_i \sim \text{Normal}(\mathbf{X}_i \boldsymbol{\beta}, \sigma_0^2 \mathbf{I}_{d_i})$, the component densities take form
\begin{eqnarray*}
	\ln f_i(\mathbf{y}_i \mid \boldsymbol{\beta}, \sigma_0^2) = - \frac{d_i}{2} \ln 2 \pi - \frac{d_i}{2} \ln \sigma_0^2 - \frac 12 \frac{\|\mathbf{y}_i - \mathbf{X}_i \boldsymbol{\beta}\|_2^2}{\sigma_0^2}
\end{eqnarray*}
and the joint loglikelihood of the sample is
\begin{eqnarray*}
	& & - \sum_i \ln \left(c + \frac{1}{2} \text{tr}(\boldsymbol{\Gamma}_i) \right) - \frac{\sum_i d_i}{2} \ln 2 \pi - \frac{\sum_i d_i}{2} \ln \sigma_0^2 - \frac 12 \frac{\sum_i \|\mathbf{y}_i - \mathbf{X}_i \boldsymbol{\beta}\|_2^2}{\sigma_0^2} \\
	& & + \sum_i \ln \left[c + \frac{1}{2\sigma_0^2} (\mathbf{y}_i - \mathbf{X}_i \boldsymbol{\beta})^\top \boldsymbol{\Gamma}_i  (\mathbf{y}_i - \mathbf{X}_i \boldsymbol{\beta}) \right] \\
	&=& - \sum_i \ln \left(c + \frac{1}{2} \text{tr}(\boldsymbol{\Gamma}_i) \right) - \frac{\sum_i d_i}{2} \ln 2 \pi + \frac{\sum_i d_i}{2} \ln \tau - \frac{\tau}{2} \sum_i \|\mathbf{y}_i - \mathbf{X}_i \boldsymbol{\beta}\|_2^2 \\
	& & + \sum_i \ln \left[c + \frac{\tau}{2} (\mathbf{y}_i - \mathbf{X}_i \boldsymbol{\beta})^\top \boldsymbol{\Gamma}_i  (\mathbf{y}_i - \mathbf{X}_i \boldsymbol{\beta}) \right] 
\end{eqnarray*}
where $\boldsymbol{\Gamma}_i = \sum_{k=1}^m \theta_k \mathbf{V}_{ik}$ are parameterized via variance components $\boldsymbol{\theta} = (\theta_1, \ldots, \theta_m)$. We work with the parameterization $\tau =\sigma_0^{-2}$ because the loglikelihood is concave in $\tau$.

\paragraph{Score and Hessian}
The score (gradient of loglikelihood function) is
\begin{eqnarray*}
	\nabla_{\boldsymbol{\beta}} &=& \sigma_0^{-2} \sum_i \mathbf{X}_i^\top (\mathbf{y}_i - \mathbf{X}_i \boldsymbol{\beta}) - \sum_i \frac{\mathbf{X}_i^\top \boldsymbol{\Gamma}_i (\mathbf{y}_i - \mathbf{X}_i \boldsymbol{\beta})}{c\sigma_0^2 + \frac 12 (\mathbf{y}_i - \mathbf{X}_i \boldsymbol{\beta})^\top \boldsymbol{\Gamma}_i  (\mathbf{y}_i - \mathbf{X}_i \boldsymbol{\beta})} \\
	&=& \tau \sum_i \mathbf{X}_i^\top (\mathbf{y}_i - \mathbf{X}_i \boldsymbol{\beta}) - \tau \sum_i \frac{\mathbf{X}_i^\top \boldsymbol{\Gamma}_i (\mathbf{y}_i - \mathbf{X}_i \boldsymbol{\beta})}{c + \frac{\tau}{2} (\mathbf{y}_i - \mathbf{X}_i \boldsymbol{\beta})^\top \boldsymbol{\Gamma}_i  (\mathbf{y}_i - \mathbf{X}_i \boldsymbol{\beta})} \\
	\nabla_{\tau} &=& \frac{\sum_i d_i}{2\tau} - \frac 12 \sum_i \|\mathbf{y}_i - \mathbf{X}_i \boldsymbol{\beta}\|_2^2 + \sum_i \frac{\frac{1}{2} (\mathbf{y}_i - \mathbf{X}_i \boldsymbol{\beta})^\top \boldsymbol{\Gamma}_i  (\mathbf{y}_i - \mathbf{X}_i \boldsymbol{\beta})}{c + \frac{\tau}{2} (\mathbf{y}_i - \mathbf{X}_i \boldsymbol{\beta})^\top \boldsymbol{\Gamma}_i  (\mathbf{y}_i - \mathbf{X}_i \boldsymbol{\beta})} \\
	\nabla_{c} &=& \sum_i \frac{1}{c + \frac{\tau}{2} (\mathbf{y}_i - \mathbf{X}_i \boldsymbol{\beta})^\top \boldsymbol{\Gamma}_i  (\mathbf{y}_i - \mathbf{X}_i \boldsymbol{\beta})}  \\
	\nabla_{\boldsymbol{\theta}} &=& - \sum_i \left(c + \sum_k \theta_k t_{ik}\right)^{-1} \mathbf{t}_i + \tau \sum_i \left( c + \color{red}\tau\color{black}\sum_k \theta_k q_{ik} \right)^{-1} \mathbf{q}_i
\end{eqnarray*}
where
\begin{eqnarray*}
	t_{ik} &=& \frac 12 \text{tr}(\mathbf{\Omega}_{ik}), \quad \mathbf{t}_i = (t_{i1}, \ldots, t_{im})^\top \\
	q_{ik} &=& \frac{1}{2} (\mathbf{y}_i - \mathbf{X}_i \boldsymbol{\beta})^\top \mathbf{\Omega}_{ik}  (\mathbf{y}_i - \mathbf{X}_i \boldsymbol{\beta}), \quad \mathbf{q}_i = (q_{i1}, \ldots, q_{im})^\top.
\end{eqnarray*}
The Hessian is
\begin{eqnarray*}
	d^2_{\boldsymbol{\beta},\boldsymbol{\beta}} &=& -  \tau \sum_i \mathbf{X}_i^\top \mathbf{X}_i + \sum_i \frac{\tau \mathbf{X}_i^\top \boldsymbol{\Gamma}_i  \mathbf{X}_i}{c + \frac{\tau}{2} (\mathbf{y}_i - \mathbf{X}_i \boldsymbol{\beta})^\top \boldsymbol{\Gamma}_i (\mathbf{y}_i - \mathbf{X}_i \boldsymbol{\beta})}  \\
	& & -  \sum_i \frac{\tau^2 [\mathbf{X}_i^\top \boldsymbol{\Gamma}_i (\mathbf{y}_i - \mathbf{X}_i \boldsymbol{\beta})][\mathbf{X}_i^\top \boldsymbol{\Gamma}_i (\mathbf{y}_i - \mathbf{X}_i \boldsymbol{\beta})]^\top}{\left[c + \frac{\tau}{2} (\mathbf{y}_i - \mathbf{X}_i \boldsymbol{\beta})^\top \boldsymbol{\Gamma}_i (\mathbf{y}_i - \mathbf{X}_i \boldsymbol{\beta})\right]^2} \\
	&\approx& - \tau \sum_i \mathbf{X}_i^\top \mathbf{X}_i -  \tau \sum_i \frac{\tau [\mathbf{X}_i^\top \boldsymbol{\Gamma}_i (\mathbf{y}_i - \mathbf{X}_i \boldsymbol{\beta})][\mathbf{X}_i^\top \boldsymbol{\Gamma}_i (\mathbf{y}_i - \mathbf{X}_i \boldsymbol{\beta})]^\top}{\left[c + \frac{\tau}{2} (\mathbf{y}_i - \mathbf{X}_i \boldsymbol{\beta})^\top \boldsymbol{\Gamma}_i (\mathbf{y}_i - \mathbf{X}_i \boldsymbol{\beta})\right]^2} \\
	d^2_{\boldsymbol{\beta},\tau} &=& \sum_i \mathbf{X}_i^\top (\mathbf{y}_i - \mathbf{X}_i \boldsymbol{\beta}) - \sum_i \frac{\mathbf{X}_i^\top \boldsymbol{\Gamma}_i (\mathbf{y}_i - \mathbf{X}_i \boldsymbol{\beta})}{\left[c + \frac{\tau}{2} (\mathbf{y}_i - \mathbf{X}_i \boldsymbol{\beta})^\top \boldsymbol{\Gamma}_i  (\mathbf{y}_i - \mathbf{X}_i \boldsymbol{\beta})\right]^2} \\
	d^2_{\boldsymbol{\beta},\boldsymbol{\theta}} &=& \sum_i \frac{\mathbf{X}_i^\top \boldsymbol{\Gamma}_i (\mathbf{y}_i - \mathbf{X}_i \boldsymbol{\beta}) \mathbf{q}_i^\top}{\left[c + \frac{\tau}{2} (\mathbf{y}_i - \mathbf{X}_i \boldsymbol{\beta})^\top \boldsymbol{\Gamma}_i  (\mathbf{y}_i - \mathbf{X}_i \boldsymbol{\beta})\right]^2} \\
	d_{\tau, \tau}^2 &=& - \frac{\sum_i d_i}{2 \tau^2} - \sum_i \left[ \frac{\frac{1}{2} (\mathbf{y}_i - \mathbf{X}_i \boldsymbol{\beta})^\top \boldsymbol{\Gamma}_i  (\mathbf{y}_i - \mathbf{X}_i \boldsymbol{\beta})}{c + \frac{\tau}{2} (\mathbf{y}_i - \mathbf{X}_i \boldsymbol{\beta})^\top \boldsymbol{\Gamma}_i  (\mathbf{y}_i - \mathbf{X}_i \boldsymbol{\beta})} \right]^2 \\
	d^2_{\tau,\boldsymbol{\theta}} &=& - \sum_i \frac{\mathbf{X}_i^\top \boldsymbol{\Gamma}_i (\mathbf{y}_i - \mathbf{X}_i \boldsymbol{\beta})}{\left[c + \frac{\tau}{2} (\mathbf{y}_i - \mathbf{X}_i \boldsymbol{\beta})^\top \boldsymbol{\Gamma}_i  (\mathbf{y}_i - \mathbf{X}_i \boldsymbol{\beta})\right]^2} \mathbf{q}_i^\top\\
	d^2_{\boldsymbol{\theta},\boldsymbol{\theta}} &=& \sum_i \left(c + \sum_k \theta_k t_{ik}\right)^{-2} \mathbf{t}_i \mathbf{t}_i^\top - \tau \sum_i \left(c + \sum_k \theta_k q_{ik} \right)^{-2} \mathbf{q}_i \mathbf{q}_i^\top.
\end{eqnarray*}
Note $\mathbb{E} [d^2_{\boldsymbol{\beta},\tau}]$, $\mathbb{E}[d^2_{\boldsymbol{\beta},\boldsymbol{\theta}}]$, and $\mathbb{E}[d^2_{\tau,\boldsymbol{\theta}}]$ are approximately zero. 

\paragraph{MM algorithm}
Because the MM update of $\boldsymbol{\theta}$ and $\tau$ is cheap, we maximize the profiled likelihood. That is, after each Newton update of $\boldsymbol{\beta}$, we update $(\tau, \boldsymbol{\theta})$ conditional on current $\boldsymbol{\beta}$ using the MM algorithm and evaluate the gradient and (approximate) Hessian using the newest $(\tau, \boldsymbol{\theta})$. To update $\tau$ and $\boldsymbol{\theta}$ given $\boldsymbol{\beta}$, the relevant objective function is
\begin{eqnarray*}
	- \sum_i \ln \left(c + \sum_k \theta_k t_{ik}\right) + \frac{\sum_i d_i}{2} \ln \tau - \frac{\sum_i r_i^2}{2} \tau + \sum_i \ln \left(c + \tau \sum_k \theta_k q_{ik}\right),
\end{eqnarray*}
which is minorized by
\begin{eqnarray*}
	&-& \sum_i \sum_k \frac{t_{ik}}{c^{(t)} + \sum_k \theta_k^{(t)} t_{ik}} \theta_k - \sum_i \frac{1}{c^{(t)} + \sum_k \theta_k^{(t)} t_{ik}} c \\
	&+&  \frac{\sum_i d_i}{2} \ln \tau - \frac{\sum_i r_i^2}{2} \tau \\
	&+& \sum_i \sum_k \frac{\tau^{(t)} \theta_k^{(t)} q_{ik}}{c^{(t)} + \tau^{(t)} \sum_k \theta_k^{(t)} q_{ik}} (\ln \tau + \ln \theta_k) \\
	&+& \sum_i \frac{c^{(t)}}{c^{(t)} + \tau^{(t)} \sum_k \theta_k^{(t)} q_{ik}} \ln c \\
	&+& \text{const}.
\end{eqnarray*}
The resultant updates are
\begin{eqnarray}
	\tau^{(t+1)} &=& \frac{\sum_i d_i + 2\sum_i \frac{\tau^{(t)} q_i^{(t)}}{c^{(t)} + \tau^{(t)} q_i^{(t)}}}{\sum_i r_i^2} \nonumber \\
	c^{(t+1)} &=& c^{(t)} \frac{\sum_i \frac{1}{c^{(t)} + \tau^{(t)} q_i^{(t)}}}{\sum_i \frac{1}{c^{(t)} + t^{(t)}}} \\
	\theta_k^{(t+1)} &=& \theta_k^{(t)} \frac{\sum_i \frac{\tau^{(t)} q_{ik}}{c^{(t)} + \tau^{(t)} q_i^{(t)}}}{\sum_i \frac{t_{ik}}{c^{(t)} + t_i^{(t)}}}, \quad k = 1,\ldots,m, \label{eqn:jensen-mm-update}
\end{eqnarray}
where $q_i^{(t)} = \sum_k \theta_k^{(t)} q_{ik}$ and $t_i^{(t)} = \sum_k \theta_k^{(t)} t_{ik}$.

If we opt to use the optimal quadratic minorization
\begin{eqnarray*}
	\ln (1 + x) \ge \ln (1 + x^{(t)}) + (x - x^{(t)}) - \frac{x^2 - x^{2(t)}}{2(1 + x^{(t)})},
\end{eqnarray*}
the minorization function becomes
\begin{eqnarray*}
	&&- \sum_i \sum_k \frac{t_{ik}}{1 + \sum_k \theta_k^{(t)} t_{ik}} \theta_k +  \frac{\sum_i d_i}{2} \ln \tau + \left( \sum_i \sum_k \theta_k q_{ik} - \frac{\sum_i r_i^2}{2} \right) \tau - \\
    && \frac{\tau^2}{2} \sum_i \frac{(\sum_k \theta_k q_{ik})^2}{1 + \tau^{(t)} \sum_k \theta_k^{(t)} q_{ik}} + c^{(t)}.
\end{eqnarray*}
To update $\tau$ given $\theta_k$, let
\begin{eqnarray*}
	a^{(t)} &=& \sum_i \frac{(\sum_k \theta_k^{(t)} q_{ik})^2}{1 + \tau^{(t)} \sum_k \theta_k^{(t)} q_{ik}} \\
	b^{(t)} &=& \sum_i \sum_k \theta_k^{(t)} q_{ik} - \frac{\sum_i r_i^2}{2} \\
	c^{(t)} &=& \frac{\sum_i d_i}{2}
\end{eqnarray*}
then
\begin{eqnarray*}
	\tau^{(t+1)} = \frac{b^{(t)} + \sqrt{b^{2(t)} + 4 a^{(t)} c^{(t)}}}{2 a^{(t)}}.
\end{eqnarray*}
To update $\theta_k$ given $\tau$, we minimize quadratic function
\begin{eqnarray*}
	\frac 12 \boldsymbol{\sigma}^{2T} \mathbf{Q}^\top \mathbf{W^{(t)}} \mathbf{Q}  \boldsymbol{\sigma}^2 - \mathbf{c}^{(t)T} \boldsymbol{\sigma}^2
\end{eqnarray*}
subject to nonnegativity constraint $\theta_k \ge 0$, where $\mathbf{W}^{(t)} = \text{diag}(w_1^{(t)}, \ldots, w_n^{(t)})$ with
\begin{eqnarray*}
	w_i^{(t)} = \frac{\tau^{2(t)}}{1 + \tau^{(t)} \sum_k \theta_k^{(t)} q_{ik}}
\end{eqnarray*}
and $\mathbf{c}^{(t)}$ has entries
\begin{eqnarray*}
	c_k^{(t)} = \tau^{(t)} \sum_i q_{ik} - \sum_i \frac{t_{ik}}{1 + \sum_k \theta_k^{(t)} t_{ik}}.
\end{eqnarray*}
It turns out this update based on quadratic minorization converges slower than the update \eqref{eqn:jensen-mm-update} based on Jensen's inequality.

\subsection{Additional Simulations for longitudinal model}\label{sec:supp_more_longitudinal_sims}

In each simulation scenario, the non-intercept entries of the predictor matrix $\bX_i$ are independent standard normal deviates. When simulating under our model for the CS and AR(1) covariance structures, the true regression coefficients $\bbeta_{\text{true}} \sim \text{Uniform}(-2, 2).$ When comparing estimates with {\rm MixedModels.jl} under the random intercept model for the Poisson, Bernoulli and negative binomial base, smaller regression coefficients $\bbeta_{\text{true}} \sim \text{Uniform}(-0.2, 0.2)$ hold. For Gaussian base, all precisions $\tau_{\text{true}} = 100.$ For the negative binomial base, all dispersion parameters are $r_{\text{true}} = 10$. Under both CS and AR(1) parameterizations of $\bGamma_i$, $\sigma^2_{\text{true}} = 0.5$ and $\rho_{\text{true}} = 0.5$. Each simulation scenario was run on 100 replicates for each sample size $n \in \{100, 1000, 10000\}$ and number of observations $d_i \in \{2, 5, 10, 15, 20, 25\}$ per independent sampling unit. By default, convergence tolerances are set to $10^{-6}.$

Under the VC parameterization of $\bGamma_i,$ the choice $\bGamma_{i,\text{true}} = \theta_{\text{true}} \times {\bf 1}_{d_i}{\bf 1}_{d_i}^\top$ allows us to compare to the random intercept GLMM fit using {\rm MixedModels.jl}. 
When the random effect term is a scalar, {\rm MixedModels.jl} uses Gaussian quadrature for parameter estimation. We compare estimates and run-times to the random intercept GLMM fit of {\rm MixedModels.jl} with 25 Gaussian quadrature points. We conduct simulation studies under two scenarios (simulation I and II). In simulation I, it is assumed that the data are generated by the quasi-copula model with $\theta_{\text{true}} = 0.1$, and in simulation II, it is assumed that the true distribution is the random intercept GLMM with $\theta_{\text{true}} = 0.01, 0.05$. 

Figures 1-4 summarize the performance of the MLEs using mean squared errors (MSE) under the AR(1) parameterization of $\bGamma_i$. Figures 5-8 summarize the same under the CS parameterization of $\bGamma_i$. Figures 9-10 help us assess estimation accuracy and how well the GLMM density approximates the quasi-copula density under simulation I for the Bernoulli and Gaussian base. Under simulation II, Figures 11-14 shed light on how well the quasi-copula density approximates the GLMM density under different magnitudes of variance components. Figure 11 shows that for the Bernoulli base distribution with $\theta_{\text{true}} = 0.05,$ the quasi-copula estimates of the variance component has average MSE of about $10^{-3}.$ Figure 12 shows that for the Bernoulli base distribution with $\theta_{\text{true}} = 0.01,$ the quasi-copula estimates of the variance component improves to an average MSE of about $10^{-4}.$ Figure 13 shows for the Gaussian base distribution with $\theta_{\text{true}} = 0.05,$ the quasi-copula estimates of the variance component has an average MSE around $10^{-4},$ and the quasi-copula estimates of the precision has an average MSE around $10^{-2}.$ Figure 14 shows that for the Gaussian base distribution with $\theta_{\text{true}} = 0.01,$ the quasi-copula estimates of the variance component improves to an average MSE of about $10^{-6},$ and the quasi-copula estimates of the precision improves to an average MSE of about $10^{-4}.$ The QC model accurately estimates the mean components even with large cluster sizes ($d_i = 25)$ and small sample sizes ($n = 100)$, even when the true density is that of the GLMM and LMM.

\begin{figure}
    \begin{minipage}[h]{0.43\linewidth}
        \begin{center}
        \includegraphics[width=1\linewidth]{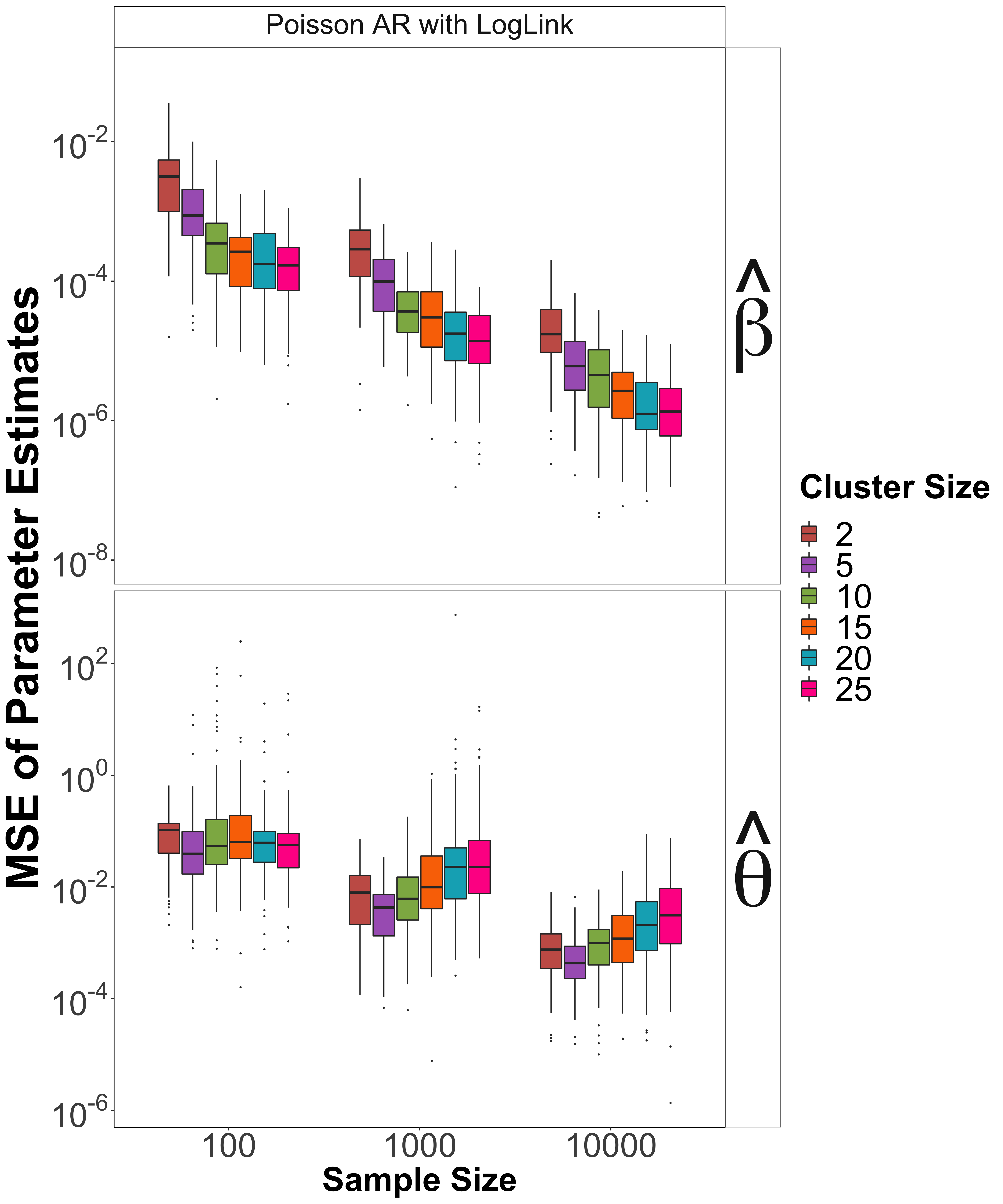} 
        \caption{MSE $\bbeta$ and $\btheta$ under the AR(1) covariance for the Poisson base.}
        \end{center} 
    \end{minipage}
    \hfill
    \begin{minipage}[h]{0.43\linewidth}
        \begin{center}
        \includegraphics[width=1\linewidth]{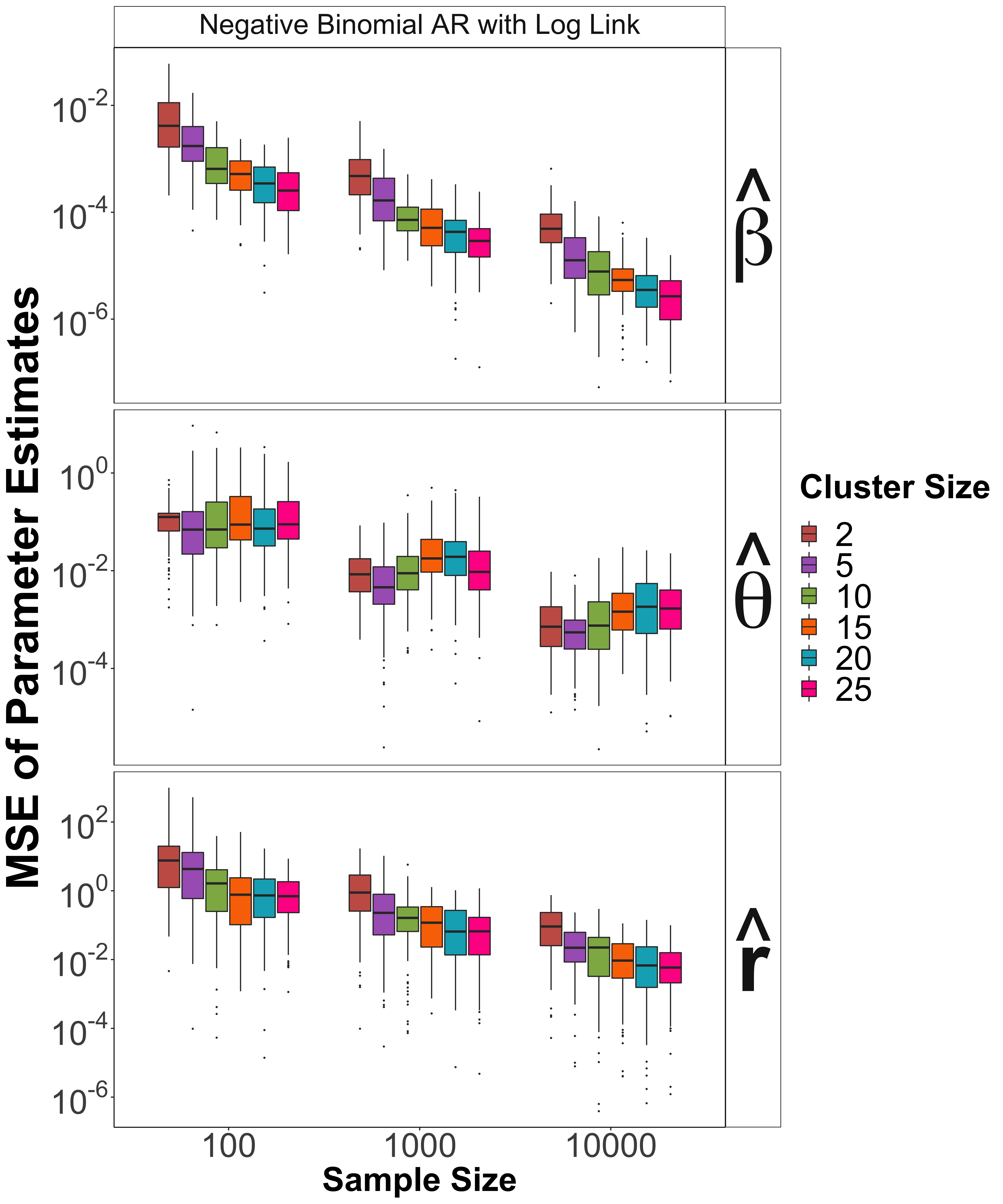} 
        \caption{MSE of $\bbeta$ and $\btheta$ under the AR(1) covariance for the negative binomial base.}
        \end{center}
    \end{minipage}
    \vfill
    \begin{minipage}[h]{0.43\linewidth}
        \begin{center}
        \includegraphics[width=1\linewidth]{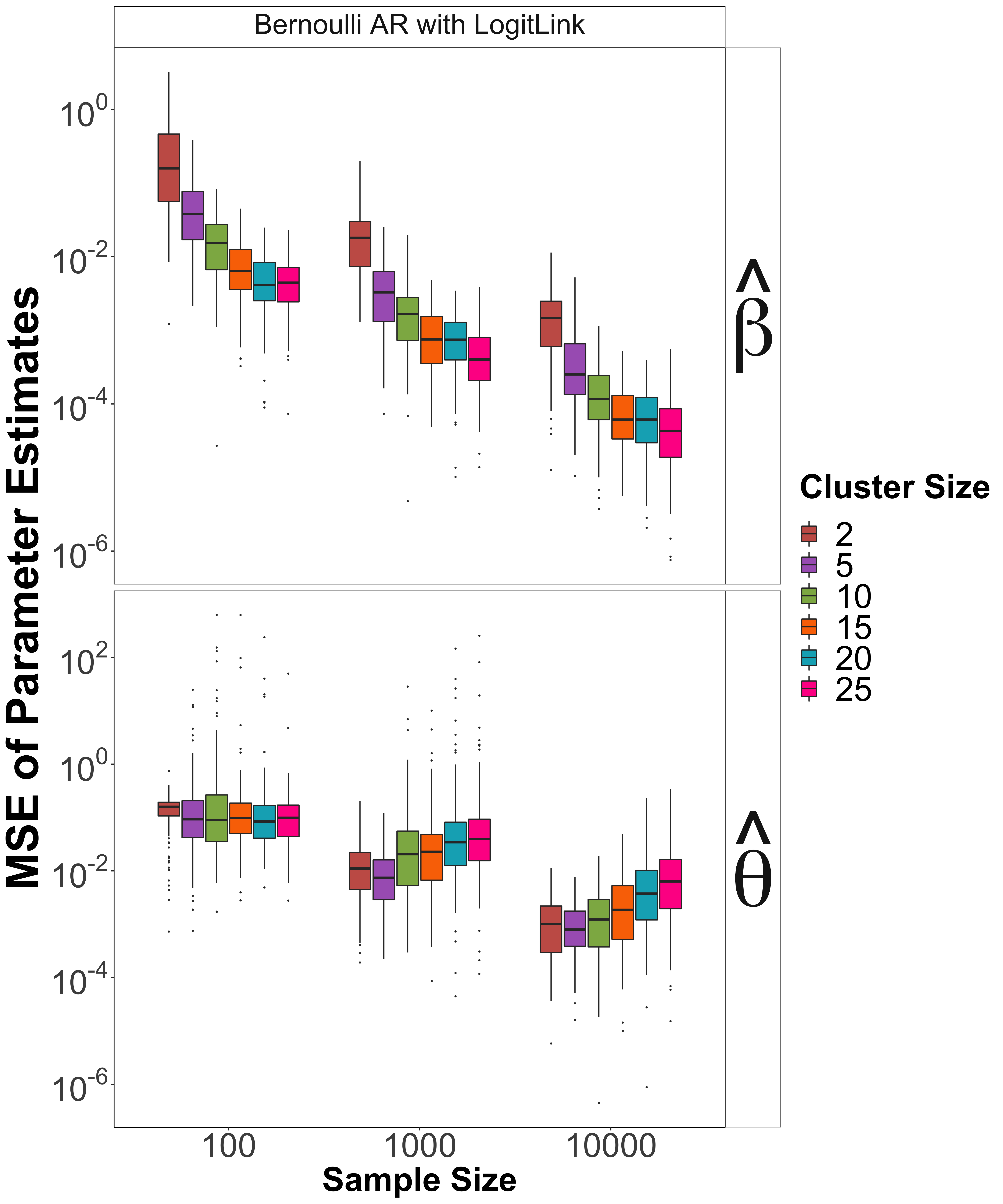} 
        \caption{MSE of $\bbeta$ and $\btheta$ under the AR(1) covariance for the Bernoulli base.}
        \end{center}
    \end{minipage}
    \hfill
    \begin{minipage}[h]{0.43\linewidth}
        \begin{center}
        \includegraphics[width=1\linewidth]{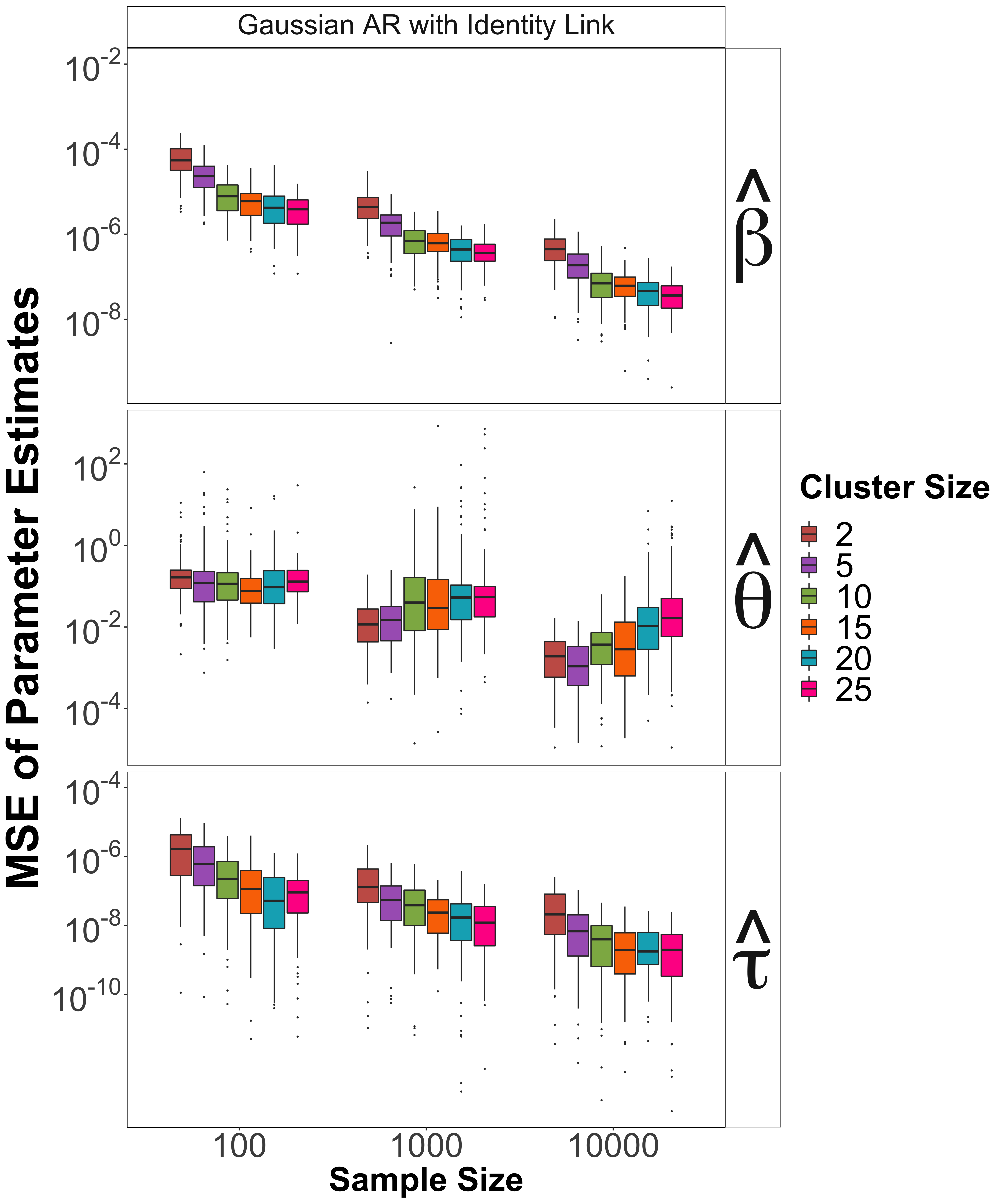} 
        \caption{MSE of $\bbeta$ and $\btheta$ under the AR(1) covariance for Gaussian base.}
        \end{center}
    \end{minipage}
\end{figure}

\begin{figure}
    \begin{minipage}[h]{0.43\linewidth}
        \begin{center}
        \includegraphics[width=1\linewidth]{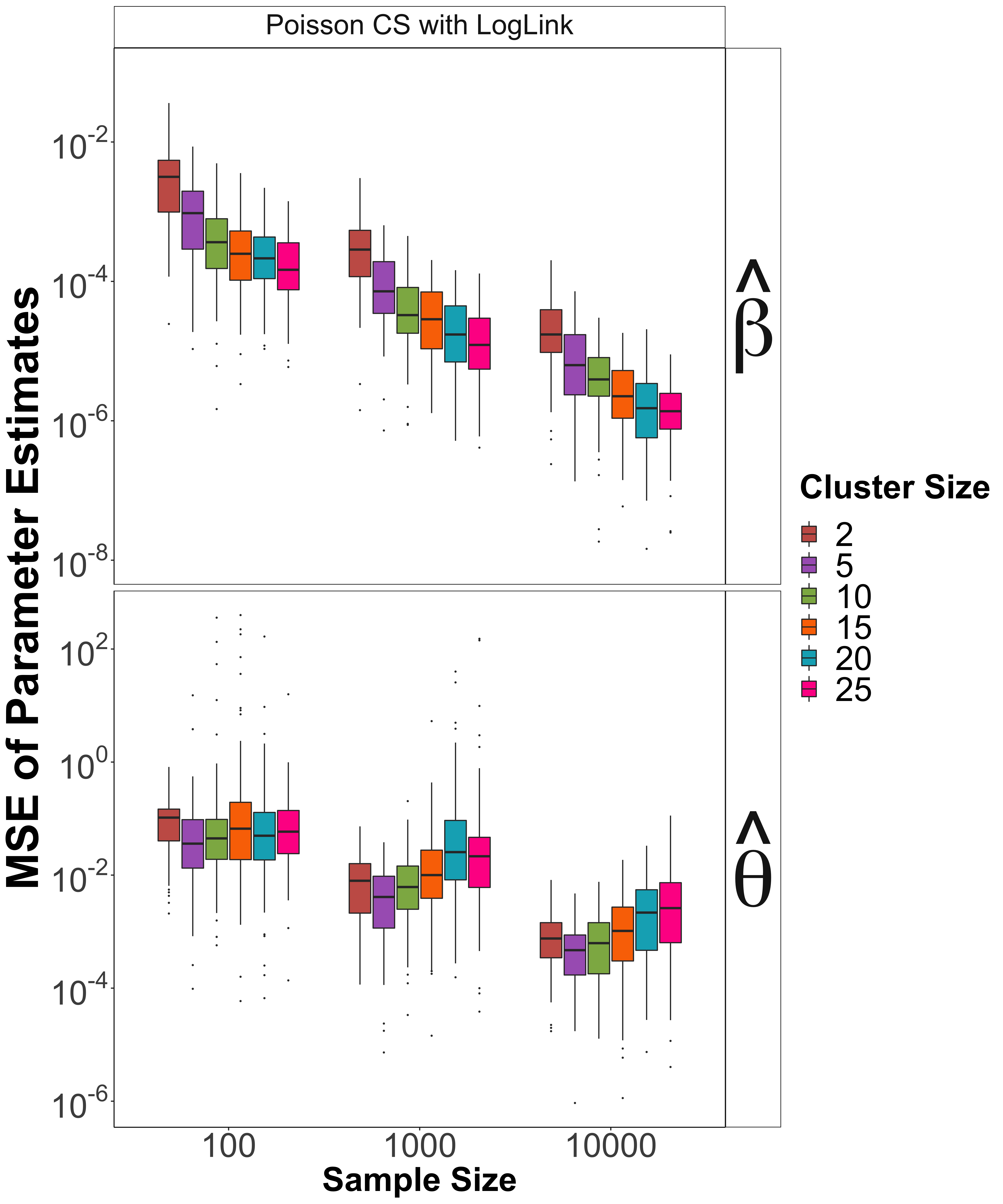} 
        \caption{MSE $\bbeta$ and $\btheta$ under the CS covariance for Poisson base.}
        \end{center} 
    \end{minipage}
    \hfill
    \begin{minipage}[h]{0.43\linewidth}
        \begin{center}
        \includegraphics[width=1\linewidth]{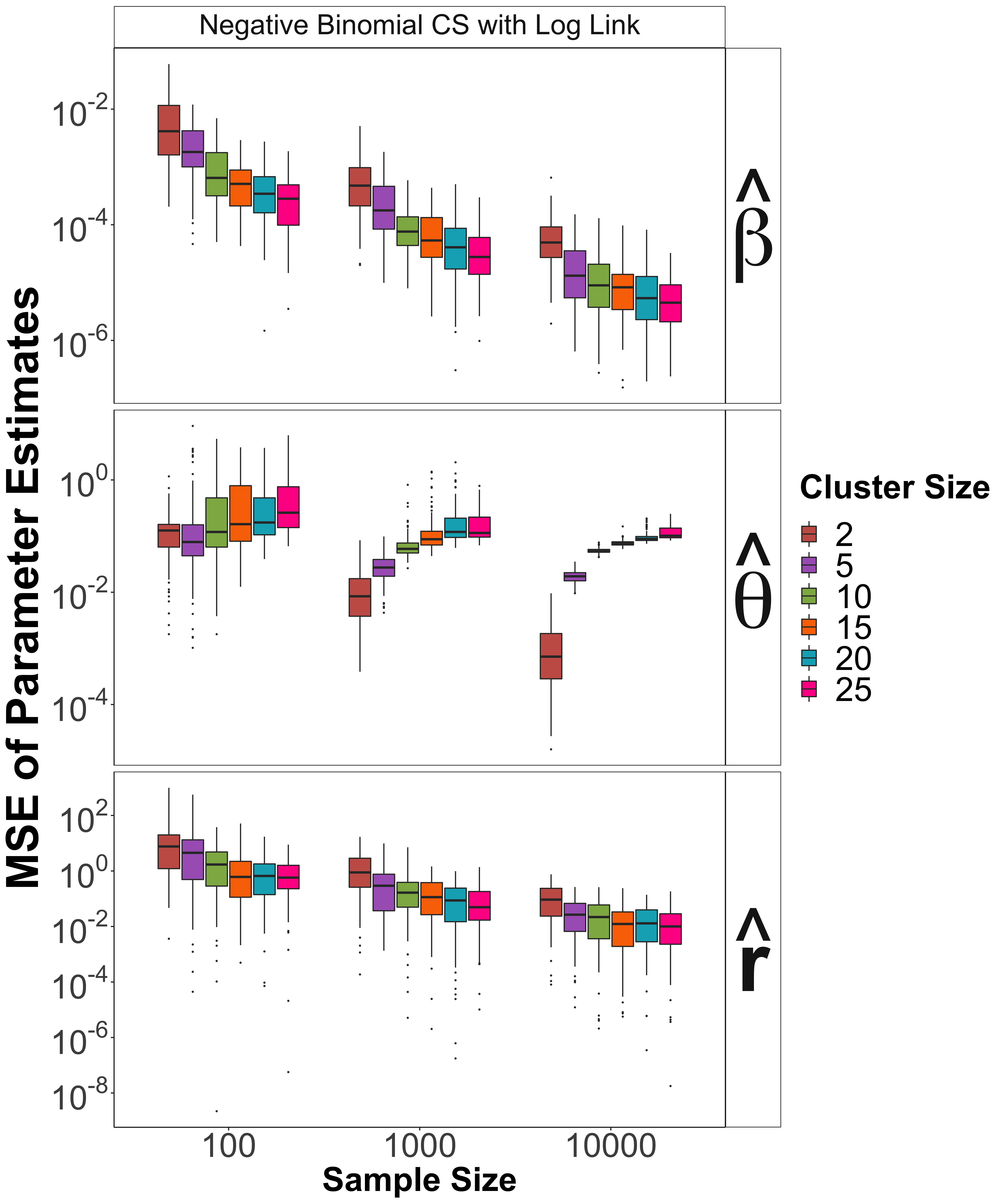} 
        \caption{MSE of $\bbeta$ and $\btheta$ under the CS covariance for negative binomial base.}
        \end{center}
    \end{minipage}
    \vfill
    \begin{minipage}[h]{0.43\linewidth}
        \begin{center}
        \includegraphics[width=1\linewidth]{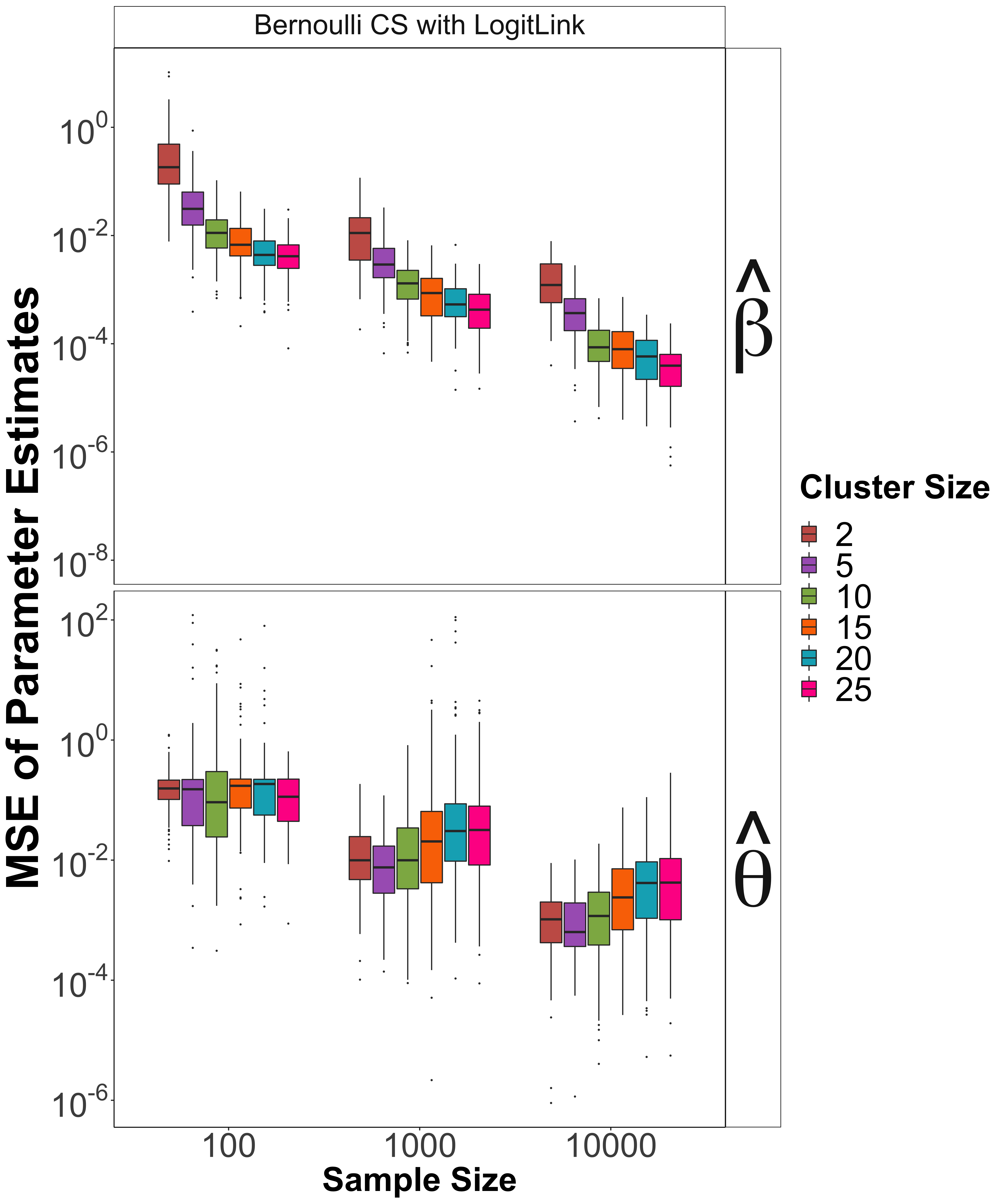} 
        \caption{MSE of $\bbeta$ and $\btheta$ under the CS covariance for Bernoulli base.}
        \end{center}
    \end{minipage}
    \hfill
    \begin{minipage}[h]{0.43\linewidth}
        \begin{center}
        \includegraphics[width=1\linewidth]{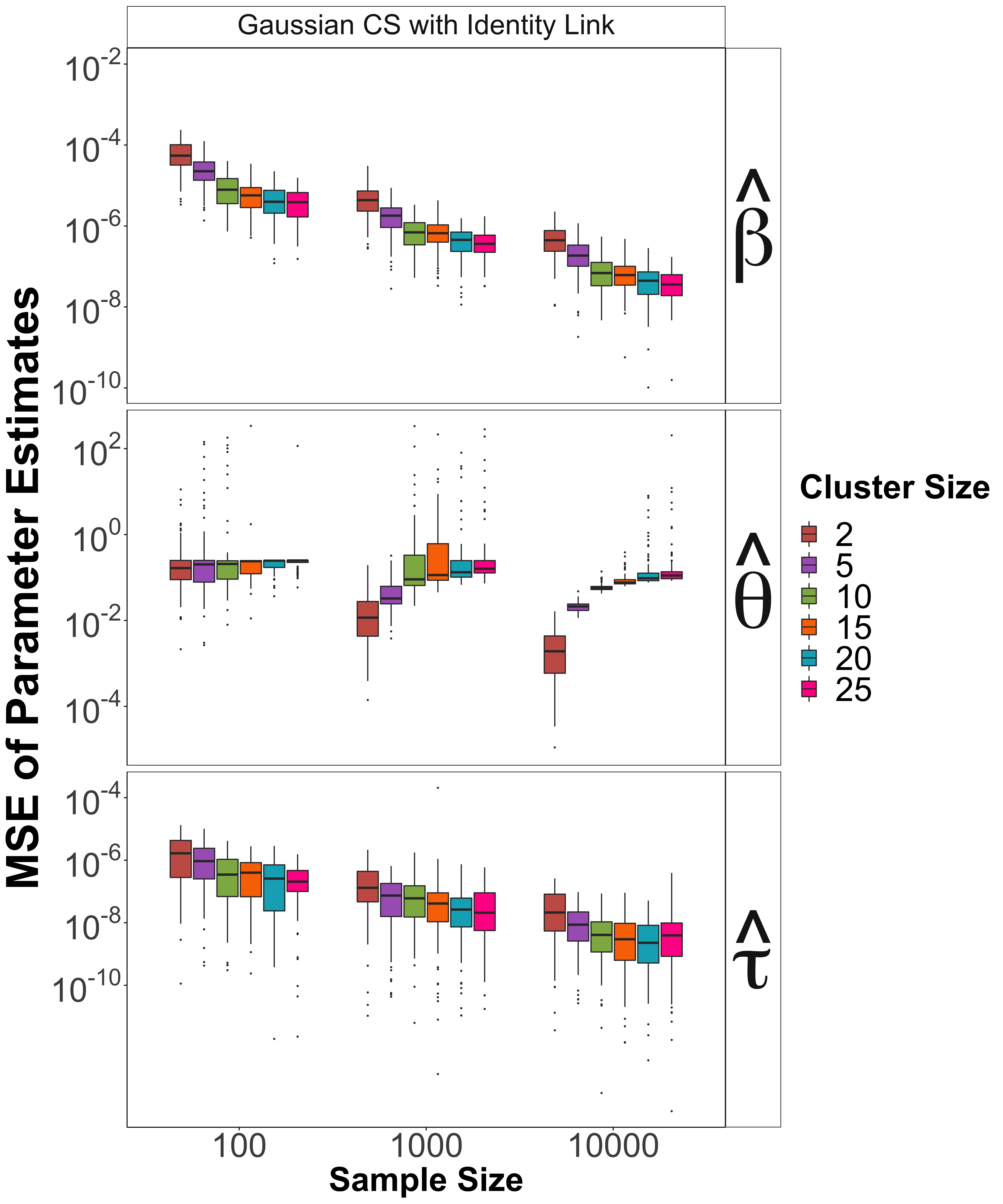} 
        \caption{MSE $\bbeta$ and $\btheta$ under the CS covariance for Normal base.}
        \end{center}
    \end{minipage}
\end{figure}

\begin{figure}
    \begin{minipage}[h]{0.47\linewidth}
        \begin{center}
        \includegraphics[width=1\linewidth]{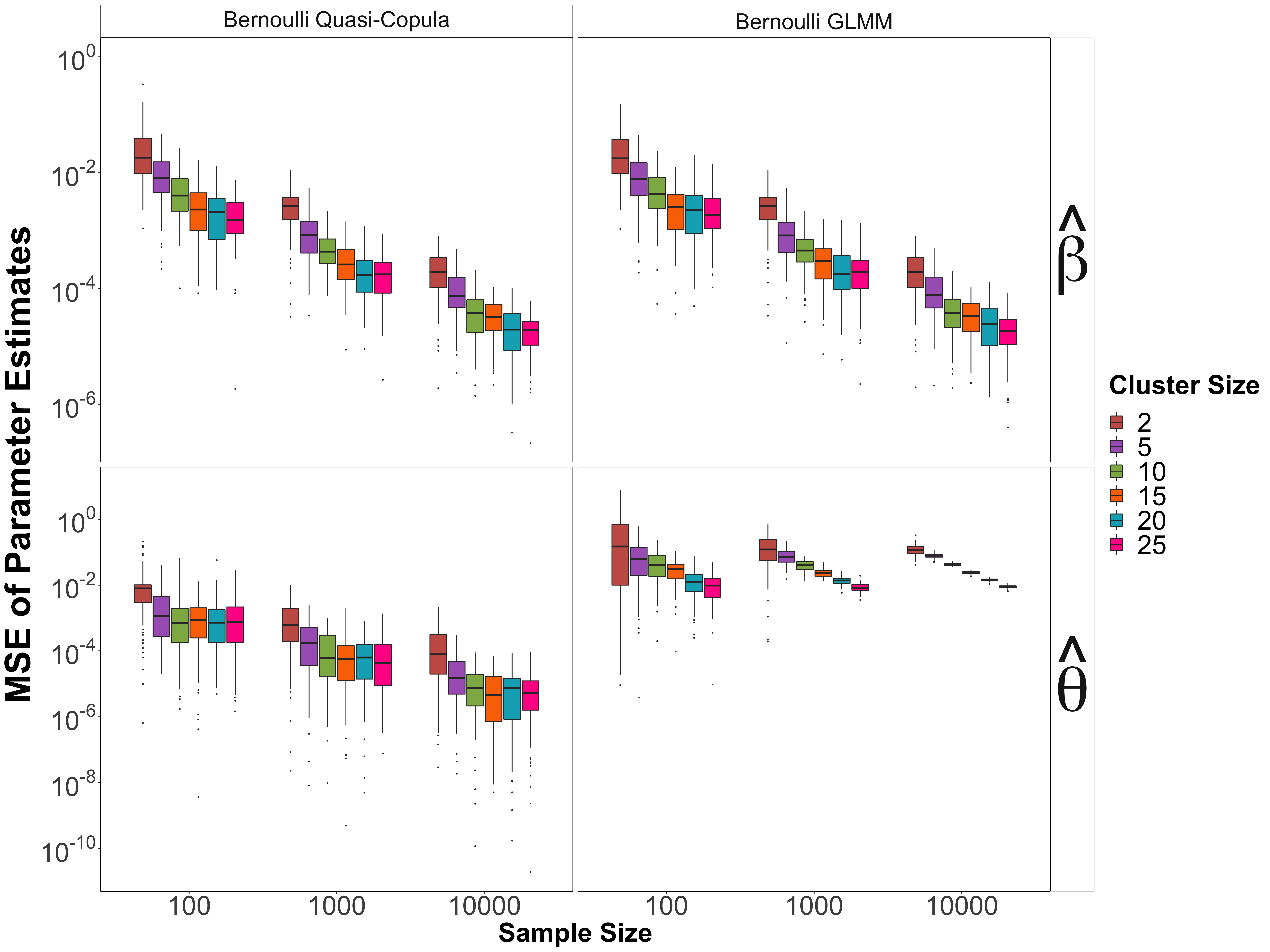} 
        \caption{Simulation I: MSE of $\bbeta$ and $\btheta$ under Bernoulli base and a single VC.}
        \end{center} 
    \end{minipage}
    \hfill
    \vspace{0.2 cm}
    \begin{minipage}[h]{0.47\linewidth}
        \begin{center}
        \includegraphics[width=1\linewidth]{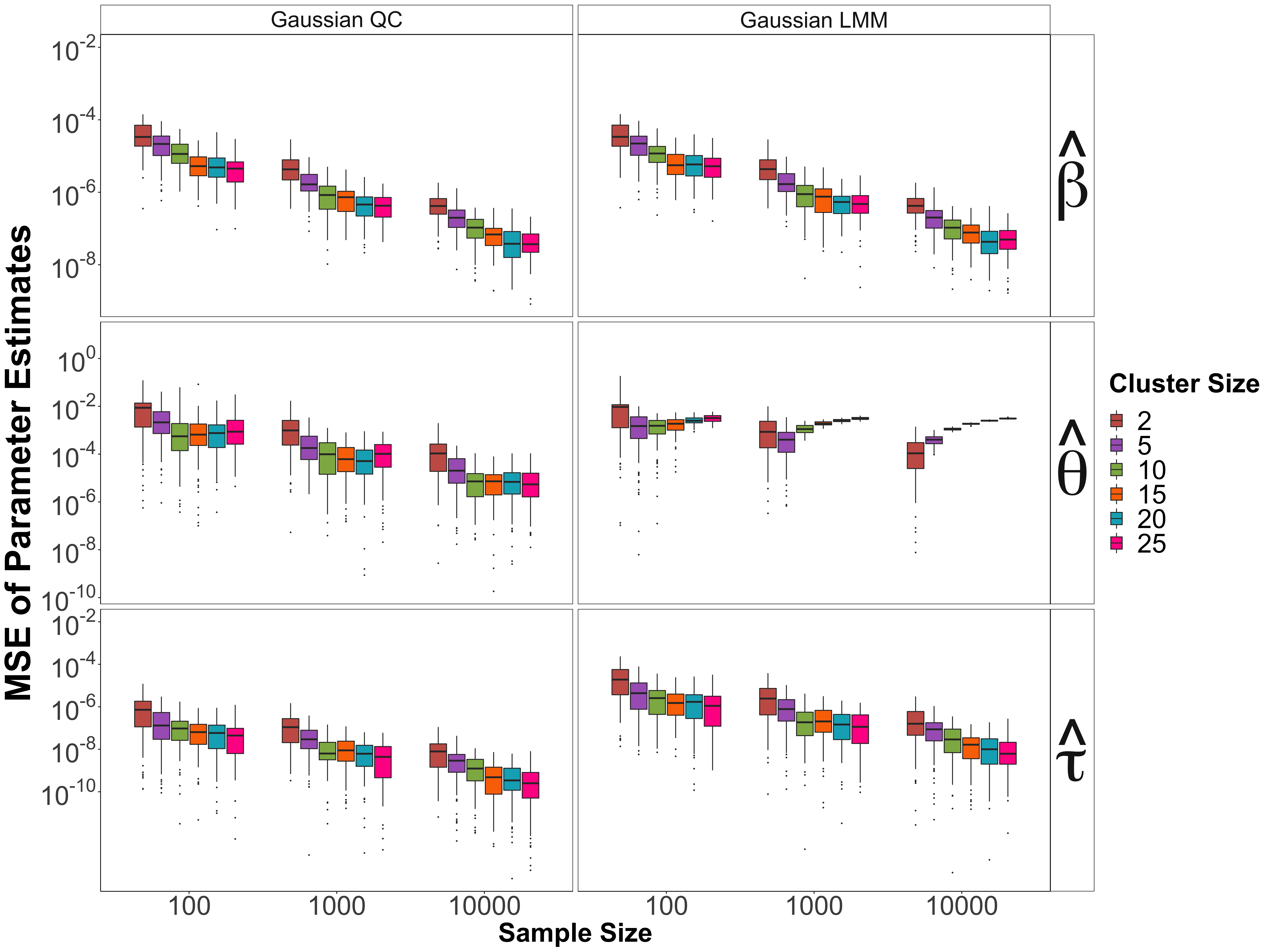} 
        \caption{Simulation I: MSE of $\bbeta$ and $\btheta$ under the Normal base and a single VC.}
        \end{center}
    \end{minipage}
    \vfill
    \vspace{0.2 cm}
    \begin{minipage}[h]{0.47\linewidth}
        \begin{center}
        \includegraphics[width=1\linewidth]{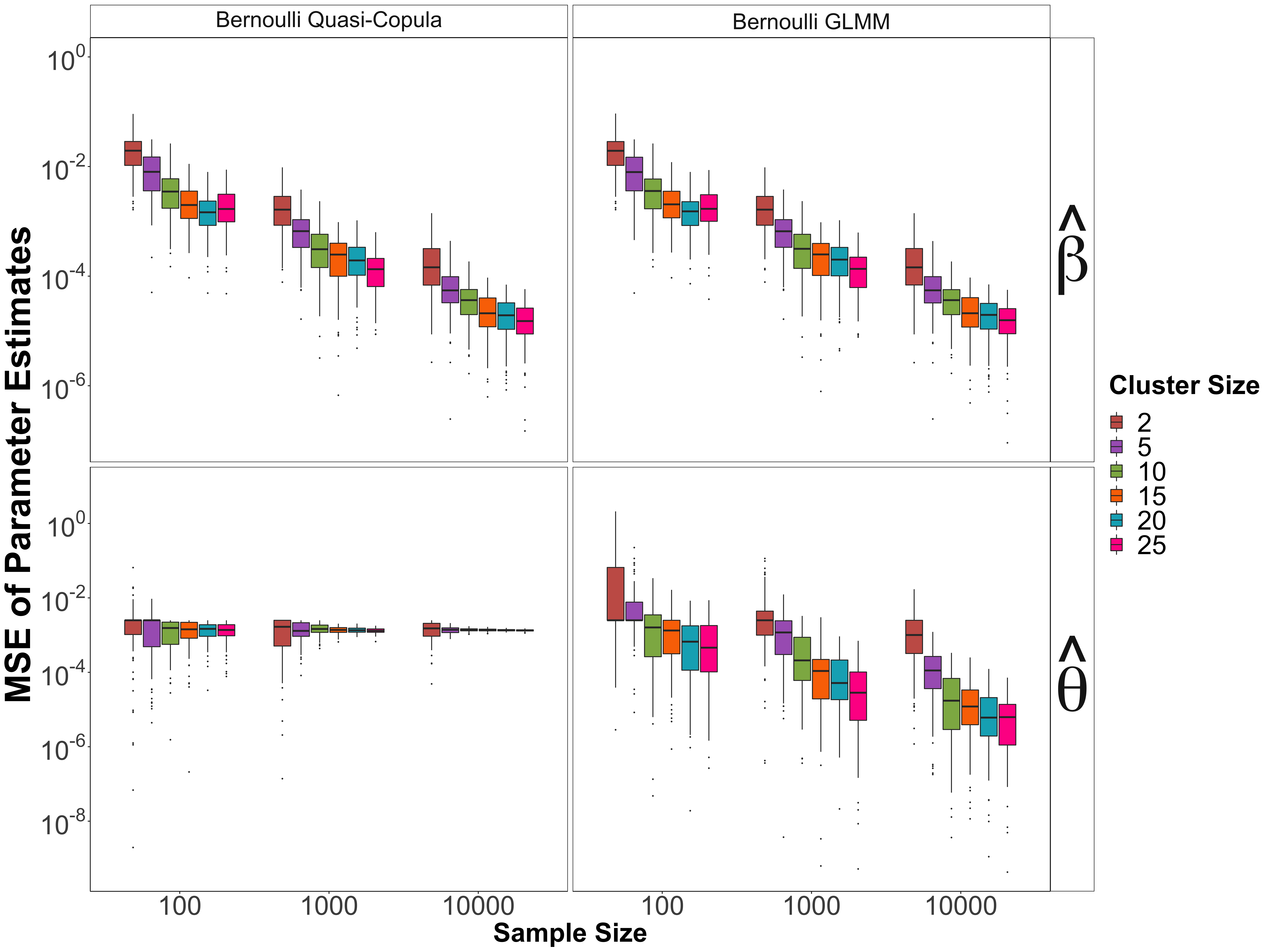} 
        \caption{Simulation I: MSE of $\bbeta$ and $\btheta$ under the Normal base with a single VC.}
        \end{center}
    \end{minipage}
    \hfill
    \begin{minipage}[h]{0.47\linewidth}
        \begin{center}
        \includegraphics[width=1\linewidth]{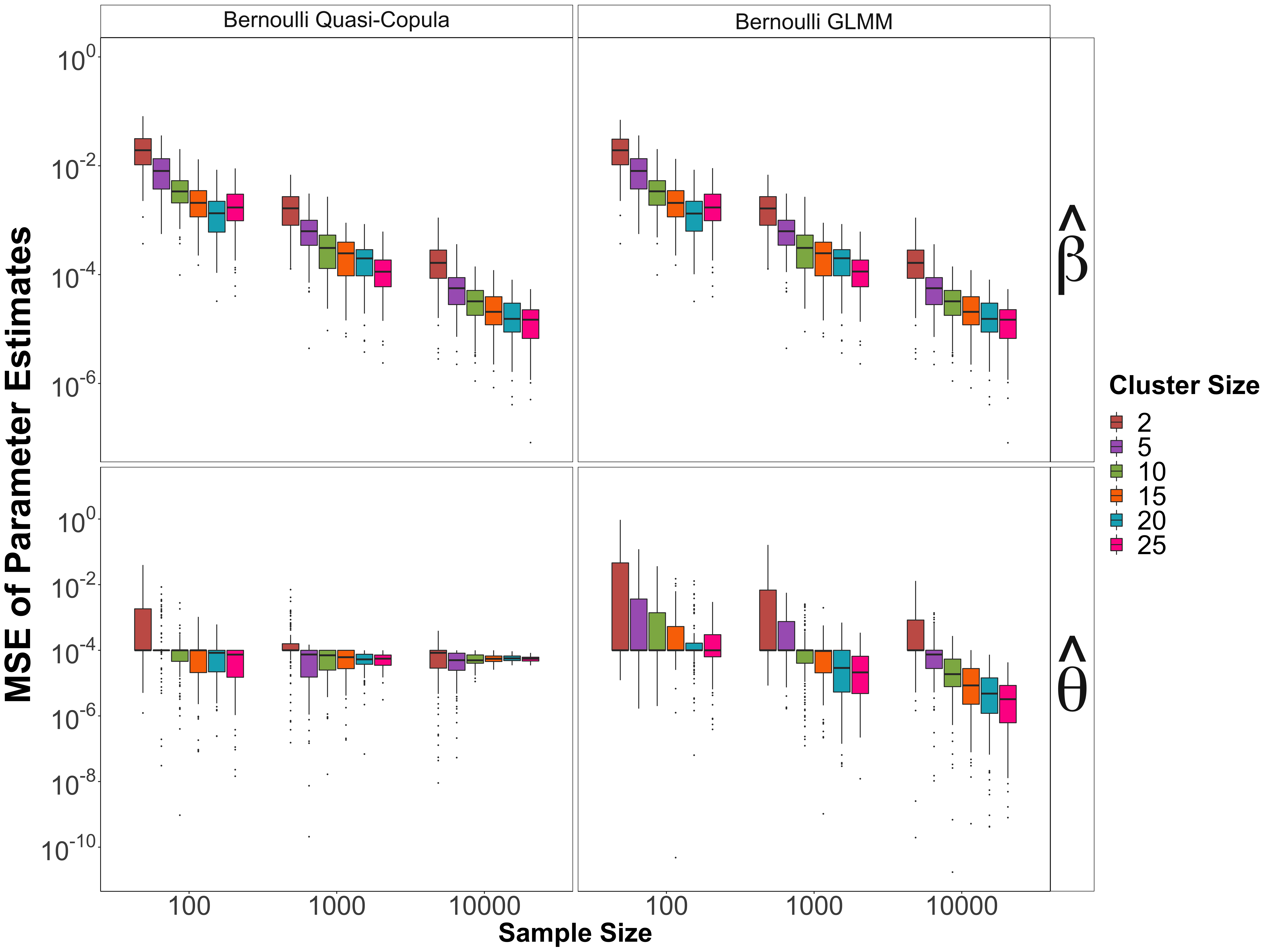} 
        \caption{Simulation II: MSE of $\bbeta$ and $\btheta = 0.01$ under Bernoulli base and a single VC.}
        \end{center}
    \end{minipage}
\end{figure}

\subsubsection{Additional Run Times}\label{sec:additional_runtimes}

Run times under simulation I and II are comparable. Tables 1-4 presents average run times and their standard errors in seconds, for $100$ replicates under the AR(1) and CS covariance structures. Tables 5-6 present average run times and their standard errors in seconds, for $100$ replicates under simulation II with  $\theta_{\text{true}} = 0.01$. All computer runs were performed on a standard 2.3 GHz Intel i9 CPU with 8 cores. Runtimes for the quasi-copula model are presented using multi-threading across 8 cores.

\begin{table}[H]
    \centering
    \small 
    \caption{Run times and (standard error of run times) in seconds based on $100$ replicates for Poisson Base under AR(1) and CS covariance structure with sampling unit size $d_i$ and sample size $n$.}
    \begin{tabular}{|c|c|c|c|}
    \hline
        \textbf{n} & $\mathbf{d_i}$ & \textbf{Poisson AR(1) time} & \textbf{Poisson CS time}  \\ \hline
    100 & 2 & 0.057 (0.001) & 0.065 (0.002) \\ \hline
    100 & 5 & 0.069 (0.002) & 0.079 (0.002) \\ \hline
    100 & 10 & 0.112 (0.008) & 0.133 (0.011) \\ \hline
    100 & 15 & 0.214 (0.021) & 0.212 (0.019) \\ \hline
    100 & 20 & 0.238 (0.023) & 0.234 (0.020) \\ \hline
    100 & 25 & 0.307 (0.025) & 0.289 (0.023) \\ \hline
    1000 & 2 & 0.060 (0.001) & 0.066 (0.001) \\ \hline
    1000 & 5 & 0.074 (0.001) & 0.081 (0.001) \\ \hline
    1000 & 10 & 0.096 (0.001) & 0.108 (0.002) \\ \hline
    1000 & 15 & 0.112 (0.002) & 0.125 (0.002) \\ \hline
    1000 & 20 & 0.153 (0.012) & 0.158 (0.008) \\ \hline
    1000 & 25 & 0.153 (0.003) & 0.180 (0.011) \\ \hline
    10000 & 2 & 0.201 (0.002) & 0.199 (0.002) \\ \hline
    10000 & 5 & 0.271 (0.002) & 0.302 (0.003) \\ \hline
    10000 & 10 & 0.358 (0.002) & 0.446 (0.004) \\ \hline
    10000 & 15 & 0.447 (0.004) & 0.564 (0.006) \\ \hline
    10000 & 20 & 0.543 (0.005) & 0.651 (0.006) \\ \hline
    10000 & 25 & 0.703 (0.008) & 0.757 (0.007) \\ \hline
    \end{tabular}
    \label{tab:poisson_AR_CS_time}
\end{table}

\begin{table}[H]
    \centering
    \small 
    \caption{Run times and (standard error of run times) in seconds based on $100$ replicates for negative binomial (NB) Base under AR(1) and CS covariance structure with sampling unit size $d_i$ and sample size $n$.}
    \begin{tabular}{|c|c|c|c|}
    \hline
  \textbf{n} & $\mathbf{d_i}$ & \textbf{NB AR(1) time} & \textbf{NB CS time}  \\ \hline
    100 & 2 & 0.323 (0.009) & 0.300 (0.009) \\ \hline
    100 & 5 & 0.339 (0.007) & 0.311 (0.008) \\ \hline
    100 & 10 & 0.320 (0.008) & 0.337 (0.012) \\ \hline
    100 & 15 & 0.334 (0.011) & 0.391 (0.016) \\ \hline
    100 & 20 & 0.364 (0.013) & 0.372 (0.015) \\ \hline
    100 & 25 & 0.376 (0.016) & 0.362 (0.016) \\ \hline
    1000 & 2 & 0.445 (0.004) & 0.381 (0.004) \\ \hline
    1000 & 5 & 0.499 (0.003) & 0.429 (0.004) \\ \hline
    1000 & 10 & 0.564 (0.004) & 0.520 (0.009) \\ \hline
    1000 & 15 & 0.654 (0.010) & 0.700 (0.021) \\ \hline
    1000 & 20 & 0.798 (0.019) & 0.864 (0.030) \\ \hline
    1000 & 25 & 0.938 (0.022) & 0.864 (0.030) \\ \hline
    10000 & 2 & 2.656 (0.012) & 2.297 (0.017) \\ \hline
    10000 & 5 & 3.161 (0.013) & 2.706 (0.012) \\ \hline
    10000 & 10 & 3.875 (0.015) & 4.001 (0.059) \\ \hline
    10000 & 15 & 4.924 (0.016) & 5.302 (0.140) \\ \hline
    10000 & 20 & 6.353 (0.028) & 6.073 (0.142) \\ \hline
    10000 & 25 & 7.449 (0.109) & 6.987 (0.144) \\ \hline
    \end{tabular}
\end{table}

\begin{table}[H]
    \centering
    \small 
    \caption{Run times and (standard error of run times) in seconds based on $100$ replicates for Bernoulli Base under AR(1) and CS covariance structure with sampling unit size $d_i$ and sample size $n$.}
    \begin{tabular}{|c|c|c|c|}
    \hline
  \textbf{n} & $\mathbf{d_i}$ & \textbf{Bernoulli AR(1) time} & \textbf{Bernoulli CS time}  \\ \hline
    100 & 2 & 0.052 (0.002) & 0.051 (0.002) \\ \hline
    100 & 5 & 0.062 (0.002) & 0.069 (0.003) \\ \hline
    100 & 10 & 0.176 (0.019) & 0.123 (0.012) \\ \hline
    100 & 15 & 0.218 (0.021) & 0.213 (0.017) \\ \hline
    100 & 20 & 0.253 (0.022) & 0.310 (0.021) \\ \hline
    100 & 25 & 0.299 (0.024) & 0.339 (0.021) \\ \hline
    1000 & 2 & 0.080 (0.002) & 0.056 (0.002) \\ \hline
    1000 & 5 & 0.081 (0.001) & 0.069 (0.002) \\ \hline
    1000 & 10 & 0.096 (0.006) & 0.088 (0.001) \\ \hline
    1000 & 15 & 0.121 (0.006) & 0.119 (0.007) \\ \hline
    1000 & 20 & 0.179 (0.016) & 0.179 (0.015) \\ \hline
    1000 & 25 & 0.226 (0.020) & 0.232 (0.020) \\ \hline
    10000 & 2 & 0.183 (0.002) & 0.171 (0.003) \\ \hline
    10000 & 5 & 0.256 (0.002) & 0.264 (0.003) \\ \hline
    10000 & 10 & 0.304 (0.002) & 0.356 (0.003) \\ \hline
    10000 & 15 & 0.432 (0.004) & 0.450 (0.004) \\ \hline
    10000 & 20 & 0.507 (0.005) & 0.535 (0.007) \\ \hline
    10000 & 25 & 0.614 (0.005) & 0.673 (0.007) \\ \hline
    \end{tabular}
\end{table}

\begin{table}[H]
    \centering
    \footnotesize 
    \caption{Run times and (standard error of run times) in seconds based on $100$ replicates for Gaussian Base under AR(1) and CS covariance structure with sampling unit size $d_i$ and sample size $n$.}
    \begin{adjustbox}{width=0.8\textwidth}
    \begin{tabular}{|c|c|c|c|}
    \hline
  \textbf{n} & $\mathbf{d_i}$ & \textbf{Gaussian AR(1) time} & \textbf{Gaussian CS time}  \\ \hline
    100 & 2 & 0.213 (0.008) & 0.214 (0.008) \\ \hline
    100 & 5 & 0.305 (0.021) & 0.338 (0.022) \\ \hline
    100 & 10 & 0.392 (0.025) & 0.432 (0.027) \\ \hline
    100 & 15 & 0.507 (0.028) & 0.441 (0.029) \\ \hline
    100 & 20 & 0.533 (0.027) & 0.448 (0.031) \\ \hline
    100 & 25 & 0.590 (0.027) & 0.429 (0.030) \\ \hline
    1000 & 2 & 0.236 (0.006) & 0.236 (0.006) \\ \hline
    1000 & 5 & 0.272 (0.005) & 0.309 (0.006) \\ \hline
    1000 & 10 & 0.365 (0.011) & 0.415 (0.010) \\ \hline
    1000 & 15 & 0.461 (0.024) & 0.547 (0.021) \\ \hline
    1000 & 20 & 0.548 (0.028) & 0.628 (0.026) \\ \hline
    1000 & 25 & 0.561 (0.030) & 0.669 (0.026) \\ \hline
    10000 & 2 & 0.604 (0.013) & 0.582 (0.011) \\ \hline
    10000 & 5 & 0.753 (0.016) & 0.793 (0.017) \\ \hline
    10000 & 10 & 0.871 (0.015) & 1.053 (0.015) \\ \hline
    10000 & 15 & 1.032 (0.018) & 1.300 (0.022) \\ \hline
    10000 & 20 & 1.233 (0.030) & 1.718 (0.025) \\ \hline
    10000 & 25 & 1.437 (0.033) & 2.191 (0.042) \\ \hline
    \end{tabular}
    \end{adjustbox}
\end{table}

\begin{table}[H]
    \centering
    \small 
    \caption{Run times and (standard error of run times) in seconds based on $100$ replicates under simulation II with Bernoulli Base, $\theta_{\text{true}} = 0.01,$ sampling unit size $d_i$ and sample size $n$.}
    \begin{adjustbox}{width=0.8\textwidth}
    \begin{tabular}{|c|c|c|c|}
    \hline
        \textbf{n} & $\mathbf{d_i}$ & \textbf{Bernoulli QC time} & \textbf{Bernoulli GLMM time}  \\ \hline
    100 & 2 & 0.048 ($<$0.001) & 0.022 (0.002) \\ \hline
    100 & 5 & 0.049 (0.001) & 0.041 (0.001) \\ \hline
    100 & 10 & 0.050 (0.001) & 0.086 (0.004) \\ \hline
    100 & 15 & 0.049 (0.001) & 0.125 (0.005) \\ \hline
    100 & 20 & 0.047 (0.001) & 0.167 (0.005) \\ \hline
    100 & 25 & 0.047 (0.001) & 0.203 (0.008) \\ \hline
    1000 & 2 & 0.045 (0.001) & 0.166 (0.003) \\ \hline
    1000 & 5 & 0.045 (0.001) & 0.446 (0.013) \\ \hline
    1000 & 10 & 0.043 (0.001) & 0.899 (0.022) \\ \hline
    1000 & 15 & 0.044 (0.001) & 1.435 (0.038) \\ \hline
    1000 & 20 & 0.054 (0.002) & 1.888 (0.041) \\ \hline
    1000 & 25 & 0.077 (0.002) & 2.461 (0.057) \\ \hline
    10000 & 2 & 0.138 (0.003) & 1.726 (0.034) \\ \hline
    10000 & 5 & 0.160 (0.003) & 4.711 (0.099) \\ \hline
    10000 & 10 & 0.189 (0.003) & 10.389 (0.221) \\ \hline
    10000 & 15 & 0.232 (0.003) & 15.958 (0.327) \\ \hline
    10000 & 20 & 0.276 (0.003) & 21.609 (0.313) \\ \hline
    10000 & 25 & 0.349 (0.003) & 28.723 (0.494) \\ \hline
    \end{tabular}
    \label{tab:bernoulli_time}
    \end{adjustbox}
\end{table}

\begin{table}[H]
    \centering
    \small 
    \caption{Run times and (standard error of run times) in seconds based on $100$ replicates under simulation II with Gaussian Base, $\theta_{\text{true}} = 0.01,$ sampling unit size $d_i$ and sample size $n$.}
    \begin{tabular}{|c|c|c|c|}
    \hline
      \textbf{n} & $\mathbf{d_i}$ & \textbf{Gaussian QC time} & \textbf{LMM time}  \\ \hline
    100 & 2 & 0.112 (0.002) & 0.003 (0.003) \\ \hline
    100 & 5 & 0.106 (0.003) & 0.001 ($<$0.001) \\ \hline
    100 & 10 & 0.097 (0.002) & 0.001 ($<$0.001) \\ \hline
    100 & 15 & 0.099 (0.004) & 0.001 ($<$0.001) \\ \hline
    100 & 20 & 0.105 (0.008) & 0.001 ($<$0.001) \\ \hline
    100 & 25 & 0.109 (0.008) & 0.003 (0.002) \\ \hline
    1000 & 2 & 0.110 (0.002) & 0.004 (0.002) \\ \hline
    1000 & 5 & 0.103 (0.002) & 0.002 ($<$0.001) \\ \hline
    1000 & 10 & 0.100 (0.002) & 0.006 (0.002) \\ \hline
    1000 & 15 & 0.095 (0.001) & 0.006 (0.001) \\ \hline
    1000 & 20 & 0.094 (0.002) & 0.008 (0.001) \\ \hline
    1000 & 25 & 0.099 (0.002) & 0.011 (0.002) \\ \hline
    10000 & 2 & 0.200 (0.005) & 0.018 (0.003) \\ \hline
    10000 & 5 & 0.192 (0.004) & 0.029 (0.003) \\ \hline
    10000 & 10 & 0.216 (0.004) & 0.050 (0.004) \\ \hline
    10000 & 15 & 0.219 (0.003) & 0.067 (0.003) \\ \hline
    10000 & 20 & 0.239 (0.003) & 0.091 (0.003) \\ \hline
    10000 & 25 & 0.258 (0.002) & 0.099 (0.003) \\ \hline
    \end{tabular}
    \label{tab:gaussian_time}
\end{table}

\subsection{Additional simulations for multivariate model} 

\subsubsection{Is the SNP screening procedure valid?}\label{sec:multivariate_gwas_extra_sims1}

\begin{figure}
    \centering
    \includegraphics[width=0.5\linewidth]{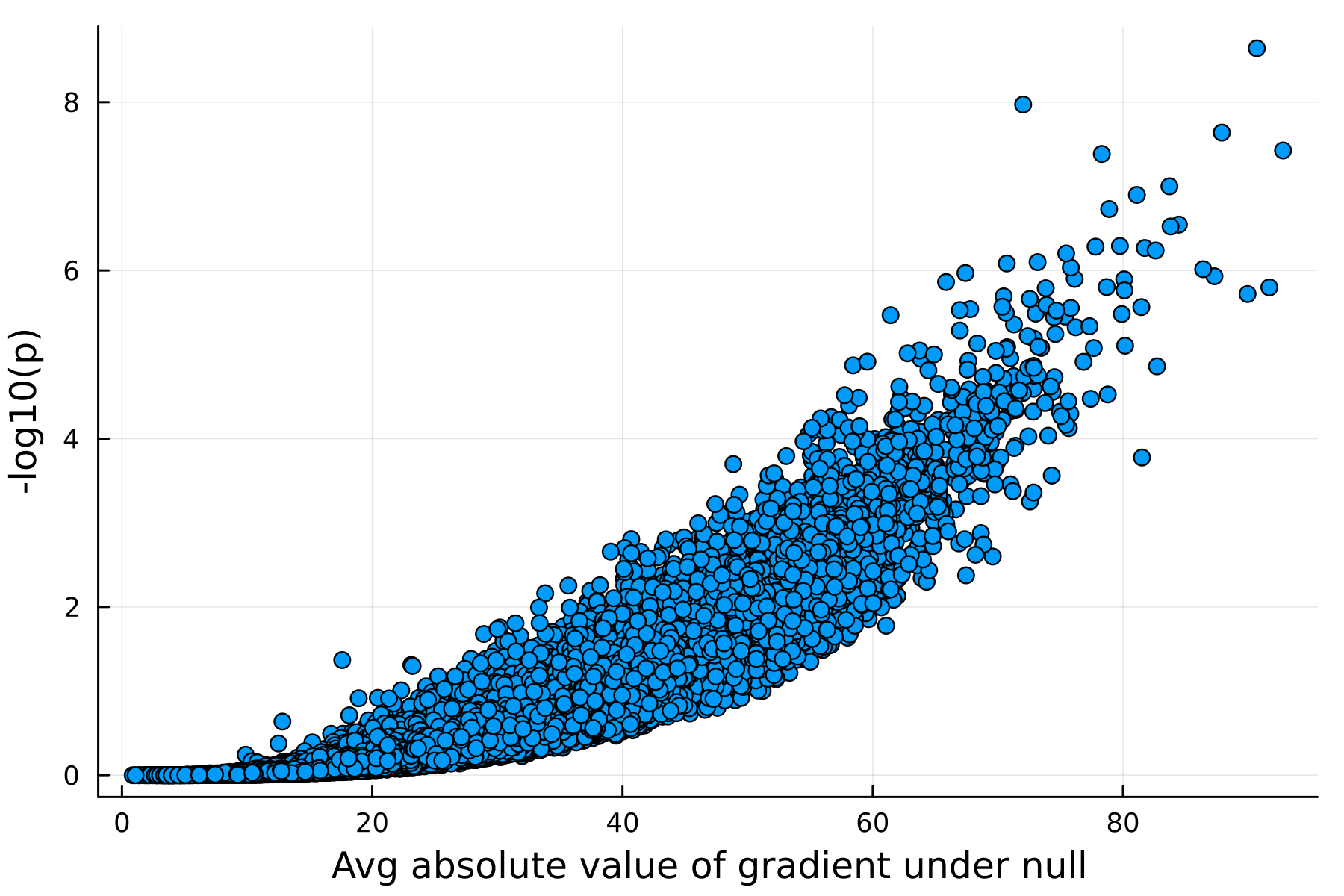}
    \caption{A comparison of the gradient of the SNP effect under the null model to their final p-values. Larger gradients tend to produce smaller p-values. }
    \label{fig:pval_ordering}
\end{figure}

Recall that Algorithm \eqref{alg:adhoc_LRT} only runs likelihood ratio tests on the most promising SNPs and employs an early stopping criteria when the current likelihood ratio p-value falls below some early termination threshold. The screening procedure relies on computing SNP gradients under the null model, given the intuition that a causal SNP should have large non-zero effect. Figure \ref{fig:pval_ordering} verifies this intuition in simulated data, where the simulation procedure follows Simulation III of the main paper except we choose $k=0$ causal SNPs. As shown in the plot, a ``larger" gradient (as measured by average absolute value) tend to ultimately produce a large $-\log_{10}(\text{p-value})$ i.e. be more significant under likelihood ratio test, with correlation coefficient $0.885$. 

\subsubsection{QQ plots}\label{sec:multivariate_gwas_extra_sims2}

To verify that p-values from testing the null \eqref{eq:qc_null_hypothesis} are valid, we conducted a simulation study with $n=5000$ samples, $d=4$ phenotypes (2 Gaussian, 1 Bernoulli, and 1 Poisson), $p=15$ non-genetic covariates, and $q=20000$ SNPs. The covariance matrix $\bGamma$ was defined as the AR1 model with $\rho = 0.5$ and $\sigma^2 = 1$. Genotype value $x_{ij}$ for sample $i$ at SNP $j$ were simulated as the sum of 2 independent $\text{Bernoulli}(0.3)$ draws. The non-genetic covariate were simulated from $N(0, 1)$ and the first column set to 1. Non-genetic effects were simulated from $\text{Uniform}(-0.5, 0.5)$. All genetic effects are 0. The mean of each sample $\bmu_i$ was formed by applying the inverse canonical link (identity, logit, log) for each base distribution to the linear predictors $\boldsymbol{\eta}_i = \bX_i\bbeta$. Finally, sample phenotypes $\by_i$ were simulated under the quasi-copula model for each subject $i$ independently. 

\begin{figure}
    \centering
    \includegraphics[width=0.5\linewidth]{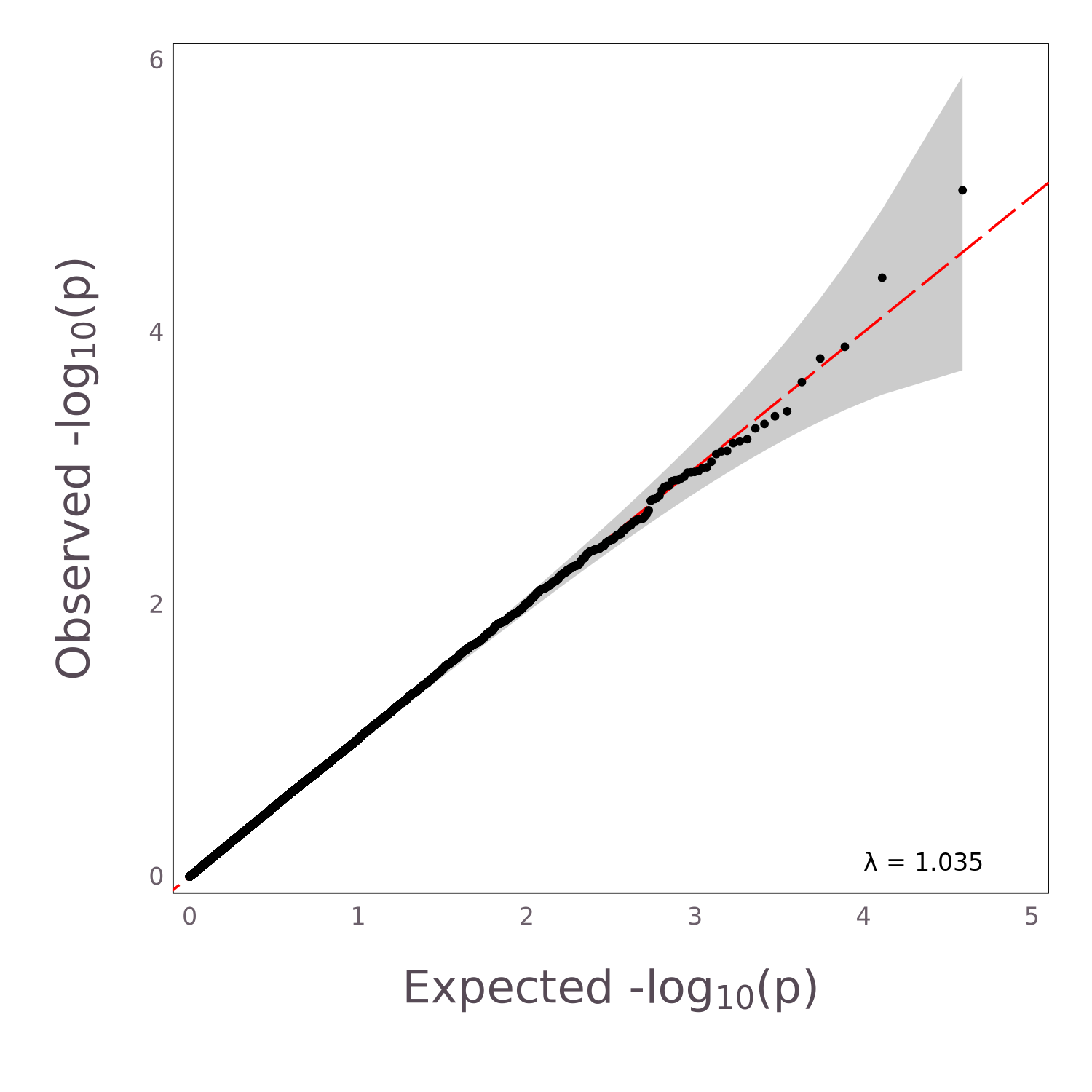}
    \caption{QQ plot of p-values resulting from Algorithm \ref{alg:adhoc_LRT} without early exiting. }
    \label{fig:QQ_plot}
\end{figure}

Under this setup, we ran Algorithm \ref{alg:adhoc_LRT} on the resulting SNPs with early stopping criteria set to infinity (i.e. no early stopping). The resulting p-values are displayed in a QQ plot in Figure \ref{fig:QQ_plot}. The plot meets our expectations and suggests that our estimation procedure is working properly.

\subsection{Quality control on UK-Biobank genotypes}\label{sec:UKB_qc}

We jointly analyzed BMI, SBP, and DBP on the UK-Biobank. For illustrative purposes, SBP and DBP were dichotomized as outlined in Section \ref{sec:gwas_example}. BMI was log-transformed to minimize the impact of outliers and then standardized to mean 0 and variance 1. 

Following our past analyses \citep{chu2020iterative,chu2023multivariate}, we filtered out subjects exhibiting sex discordance, high heterozygosity, or high SNP missingness according to recommendation by UK-Biobank. We then excluded subjects of non-European ancestry and first and second-degree relatives based on empirical kinship coefficients. We also excluded subjects on hypertension medicine at baseline. Finally, we excluded subjects with $<98\%$ genotyping success rate and SNPs with $<99\%$ genotyping success rate and imputed the remaining missing genotypes by the corresponding sample-mean genotypes. 

Given these reduced data and ignoring the Biobank's precomputed principal components, we computed afresh the top 5 principal components of the genotype matrix via \texttt{FlashPCA2} \citep{abraham2017flashpca2}. These principal components serve as predictors to adjust for hidden ancestry. We also designated sex and age as non-genetic predictors. The final data included $470,228$ SNPs and $185,656$ subjects. For faster computation time, we randomly selected $n=80000$ samples to perform the analysis. 

\subsection{Additional simulations for negative binomial longitudinal models}\label{sec:neg_bin_r_compare}

Finally, we also compared our negative binomial fits with those delivered by the three popular R packages for GLMM estimation in Tables \ref{table2}. On a single dataset with $d_i = 5,$ and $n = 10,000$ simulated under simulation II, the {\rm lme4} package \citep{BatesMachlerBolkerWalker15lme4} takes an inordinately long time to fit the model. Obtaining confidence intervals takes a significant amount of additional time, and inference of $r$ is impossible. The {\rm glmmTMB} package \citep{Brooks17glmmTMB} allows for inference of $r$ and takes much less time to form confidence intervals than {\rm lme4}, but it is still significantly slower than quasi-copula fitting. Both {\rm lme4} and {\rm glmmTMB} fit the negative binomial GLMM using Laplace Approximation, while the {\rm GLMMadaptive} package \citep{GLMMadaptive} uses adaptive Gaussian quadrature. In Tables \ref{table2} and \ref{table3}, we use {\rm GLMMadaptive} to fit the data with 25 Gaussian quadrature points. {\rm GLMMadaptive} allows for inference of $r$ and takes no additional time to form confidence intervals, but is still significantly slower than quasi-copula fitting. Run times in seconds for obtaining the estimates and confidence intervals in Table \ref{table2} appear in Table \ref{table3}.

\begin{table}[H]
    \centering
    
    \caption{MLE's and (confidence intervals) based on a single replicate under simulation II with negative binomial Base, $\theta_{\text{true}} = 0.01,$ sampling unit size $d_i = 5$ and sample size $n = 10000$.}
    \begin{adjustbox}{width=\textwidth}
    \begin{tabular}{|c|c|c|c|c|c|}
    \hline
        \textbf{Parameter} & \textbf{Truth} & \textbf{QC fit} & \textbf{lme4 fit} & \textbf{glmmTMB fit} & \textbf{GLMMadaptive fit}\\ \hline
    $\beta_1$ & 0.036 & 0.033 & 0.032 & 0.033 & 0.032\\ \hline
     ~ & ~ & (0.028, 0.037)  & (0.022, 0.042) & (0.023, 0.043) & (0.023, 0.042)\\ \hline
    $\beta_2$ & 0.107 & 0.106 & 0.106 & 0.106 & 0.106\\ \hline
     ~ & ~ & (0.101, 0.111)  & (0.097, 0.115) & (0.097, 0.115) & (0.097, 0.115) \\ \hline
    $\beta_3$ & 0.026 & 0.026 & 0.026 & 0.026 & 0.026\\ \hline
     ~ & ~ & (0.017, 0.035)  & (0.017, 0.035) & (0.017, 0.035) & (0.017, 0.035)\\ \hline
    $\theta$ & 0.01  & 0.007 & 0.009 & 0.008 & 0.009\\ \hline
     ~ & ~ & (0.003, 0.011)  & (0.002, 0.015) & (0.003, 0.019) & (0.005, 0.018)\\ \hline
    $r$ & 10 & 10.002 & 10.147 & 10.101 & 9.996\\ \hline
     ~ & ~ & (9.094, 10.910) & (NA, NA) & (8.640, 11.809) & (8.612, 11.602)\\ \hline
    \end{tabular}
    \label{table2}
    \end{adjustbox}
\end{table}

\begin{table}[H]
    \centering
    
    \caption{Run times and (confidence interval run times) in seconds based on a single replicate under simulation II with negative binomial Base, $\theta_{\text{true}} = 0.01,$ sampling unit size $d_i = 5$ and sample size $n = 10000$.}
    \begin{adjustbox}{width=\textwidth}
    \begin{tabular}{|c|c|c|c|c|c|}
    \hline
        \textbf{n} & $\mathbf{d_i}$ & \textbf{QC time} & \textbf{lme4 time} & \textbf{glmmTMB time} & \textbf{GLMMadaptive time} \\ \hline
        10000 & 5 & 1.247 (0.046) & 75.774 (158.034) & 98.944 (0.472) & 84.471 ($<$0.001) \\ \hline
    \end{tabular}
    \label{table3}
    \end{adjustbox}
\end{table}

\end{document}